\pdfoutput=1
\documentclass[aps,prd,twocolumn,amsmath,amssymb,amsfonts,nofootinbib,superscriptaddress,altaffilletter]{revtex4-1}

\usepackage{graphicx}
\usepackage[raggedright,hang]{subfigure}
\usepackage{hyperref}
\usepackage{bm}
\usepackage{amssymb}
\usepackage{amsmath}
\usepackage{longtable}
\usepackage{rotating}
\usepackage{color}
\usepackage{epsfig}
\usepackage{epsf}
\usepackage{fancyhdr}
\usepackage{dcolumn}
\usepackage{supertabular}

\begin{document}




\title{Einstein@Home all-sky search for periodic gravitational waves in LIGO S5 data}


\affiliation{LIGO - California Institute of Technology, Pasadena, CA  91125, USA}
\affiliation{California State University Fullerton, Fullerton CA 92831 USA}
\affiliation{SUPA, University of Glasgow, Glasgow, G12 8QQ, United Kingdom}
\affiliation{Laboratoire d'Annecy-le-Vieux de Physique des Particules (LAPP), Universit\'e de Savoie, CNRS/IN2P3, F-74941 Annecy-Le-Vieux, France}
\affiliation{INFN, Sezione di Napoli $^a$; Universit\`a di Napoli 'Federico II'$^b$ Complesso Universitario di Monte S.Angelo, I-80126 Napoli; Universit\`a di Salerno, Fisciano, I-84084 Salerno$^c$, Italy}
\affiliation{LIGO - Livingston Observatory, Livingston, LA  70754, USA}
\affiliation{Cardiff University, Cardiff, CF24 3AA, United Kingdom}
\affiliation{University of Sannio at Benevento, I-82100 Benevento, Italy and INFN (Sezione di Napoli), Italy}
\affiliation{Albert-Einstein-Institut, Max-Planck-Institut f\"ur Gravitationsphysik, D-30167 Hannover, Germany}
\affiliation{Leibniz Universit\"at Hannover, D-30167 Hannover, Germany}
\affiliation{Nikhef, Science Park, Amsterdam, the Netherlands$^a$; VU University Amsterdam, De Boelelaan 1081, 1081 HV Amsterdam, the Netherlands$^b$}
\affiliation{National Astronomical Observatory of Japan, Tokyo  181-8588, Japan}
\affiliation{University of Wisconsin--Milwaukee, Milwaukee, WI  53201, USA}
\affiliation{INFN, Sezione di Pisa$^a$; Universit\`a di Pisa$^b$; I-56127 Pisa; Universit\`a di Siena, I-53100 Siena$^c$, Italy}
\affiliation{University of Florida, Gainesville, FL  32611, USA}
\affiliation{INFN, Sezione di Roma$^a$; Universit\`a 'La Sapienza'$^b$, I-00185 Roma, Italy}
\affiliation{LIGO - Hanford Observatory, Richland, WA  99352, USA}
\affiliation{University of Birmingham, Birmingham, B15 2TT, United Kingdom}
\affiliation{Albert-Einstein-Institut, Max-Planck-Institut f\"ur Gravitationsphysik, D-14476 Golm, Germany}
\affiliation{Montana State University, Bozeman, MT 59717, USA}
\affiliation{European Gravitational Observatory (EGO), I-56021 Cascina (PI), Italy}
\affiliation{Syracuse University, Syracuse, NY  13244, USA}
\affiliation{LIGO - Massachusetts Institute of Technology, Cambridge, MA 02139, USA}
\affiliation{APC, AstroParticule et Cosmologie, Universit\'e Paris Diderot, CNRS/IN2P3, CEA/Irfu, Observatoire de Paris, Sorbonne Paris Cit\'e, 10, rue Alice Domon et L\'eonie Duquet, 75205 Paris Cedex 13, France}
\affiliation{Columbia University, New York, NY  10027, USA}
\affiliation{Stanford University, Stanford, CA  94305, USA}
\affiliation{IM-PAN 00-956 Warsaw$^a$; Astronomical Observatory Warsaw University 00-478 Warsaw$^b$; CAMK-PAN 00-716 Warsaw$^c$; Bia{\l}ystok University 15-424 Bia{\l}ystok$^d$; NCBJ 05-400 \'Swierk-Otwock$^e$; Institute of Astronomy 65-265 Zielona G\'ora$^f$,  Poland}
\affiliation{The University of Texas at Brownsville, Brownsville, TX 78520, USA}
\affiliation{San Jose State University, San Jose, CA 95192, USA}
\affiliation{Moscow State University, Moscow, 119992, Russia}
\affiliation{LAL, Universit\'e Paris-Sud, IN2P3/CNRS, F-91898 Orsay$^a$; ESPCI, CNRS,  F-75005 Paris$^b$, France}
\affiliation{NASA/Goddard Space Flight Center, Greenbelt, MD  20771, USA}
\affiliation{University of Western Australia, Crawley, WA 6009, Australia}
\affiliation{The Pennsylvania State University, University Park, PA  16802, USA}
\affiliation{Universit\'e Nice-Sophia-Antipolis, CNRS, Observatoire de la C\^ote d'Azur, F-06304 Nice$^a$; Institut de Physique de Rennes, CNRS, Universit\'e de Rennes 1, 35042 Rennes$^b$, France}
\affiliation{Laboratoire des Mat\'eriaux Avanc\'es (LMA), IN2P3/CNRS, F-69622 Villeurbanne, Lyon, France}
\affiliation{Washington State University, Pullman, WA 99164, USA}
\affiliation{INFN, Sezione di Perugia$^a$; Universit\`a di Perugia$^b$, I-06123 Perugia,Italy}
\affiliation{INFN, Sezione di Firenze, I-50019 Sesto Fiorentino$^a$; Universit\`a degli Studi di Urbino 'Carlo Bo', I-61029 Urbino$^b$, Italy}
\affiliation{University of Oregon, Eugene, OR  97403, USA}
\affiliation{Laboratoire Kastler Brossel, ENS, CNRS, UPMC, Universit\'e Pierre et Marie Curie, 4 Place Jussieu, F-75005 Paris, France}
\affiliation{University of Maryland, College Park, MD 20742 USA}
\affiliation{Universitat de les Illes Balears, E-07122 Palma de Mallorca, Spain}
\affiliation{University of Massachusetts - Amherst, Amherst, MA 01003, USA}
\affiliation{Canadian Institute for Theoretical Astrophysics, University of Toronto, Toronto, Ontario, M5S 3H8, Canada}
\affiliation{Tsinghua University, Beijing 100084 China}
\affiliation{University of Michigan, Ann Arbor, MI  48109, USA}
\affiliation{Louisiana State University, Baton Rouge, LA  70803, USA}
\affiliation{The University of Mississippi, University, MS 38677, USA}
\affiliation{Charles Sturt University, Wagga Wagga, NSW 2678, Australia}
\affiliation{Caltech-CaRT, Pasadena, CA  91125, USA}
\affiliation{INFN, Sezione di Genova;  I-16146  Genova, Italy}
\affiliation{Pusan National University, Busan 609-735, Korea}
\affiliation{Australian National University, Canberra, ACT 0200, Australia}
\affiliation{Carleton College, Northfield, MN  55057, USA}
\affiliation{The University of Melbourne, Parkville, VIC 3010, Australia}
\affiliation{INFN, Sezione di Roma Tor Vergata$^a$; Universit\`a di Roma Tor Vergata, I-00133 Roma$^b$; Universit\`a dell'Aquila, I-67100 L'Aquila$^c$, Italy}
\affiliation{University of Salerno, I-84084 Fisciano (Salerno), Italy and INFN (Sezione di Napoli), Italy}
\affiliation{Instituto Nacional de Pesquisas Espaciais,  12227-010 - S\~{a}o Jos\'{e} dos Campos, SP, Brazil}
\affiliation{The University of Sheffield, Sheffield S10 2TN, United Kingdom}
\affiliation{WIGNER RCP, RMKI, H-1121 Budapest, Konkoly Thege Mikl\'os \'ut 29-33, Hungary}
\affiliation{Inter-University Centre for Astronomy and Astrophysics, Pune - 411007, India}
\affiliation{University of Minnesota, Minneapolis, MN 55455, USA}
\affiliation{INFN, Gruppo Collegato di Trento$^a$ and Universit\`a di Trento$^b$,  I-38050 Povo, Trento, Italy;   INFN, Sezione di Padova$^c$ and Universit\`a di Padova$^d$, I-35131 Padova, Italy}
\affiliation{California Institute of Technology, Pasadena, CA  91125, USA}
\affiliation{Northwestern University, Evanston, IL  60208, USA}
\affiliation{Rochester Institute of Technology, Rochester, NY  14623, USA}
\affiliation{E\"otv\"os Lor\'and University, Budapest, 1117 Hungary}
\affiliation{University of Cambridge, Cambridge, CB2 1TN, United Kingdom}
\affiliation{University of Szeged, 6720 Szeged, D\'om t\'er 9, Hungary}
\affiliation{Rutherford Appleton Laboratory, HSIC, Chilton, Didcot, Oxon OX11 0QX United Kingdom}
\affiliation{Embry-Riddle Aeronautical University, Prescott, AZ   86301 USA}
\affiliation{Perimeter Institute for Theoretical Physics, Ontario, N2L 2Y5, Canada}
\affiliation{American University, Washington, DC 20016, USA}
\affiliation{University of New Hampshire, Durham, NH 03824, USA}
\affiliation{University of Southampton, Southampton, SO17 1BJ, United Kingdom}
\affiliation{Korea Institute of Science and Technology Information, Daejeon 305-806, Korea}
\affiliation{Hobart and William Smith Colleges, Geneva, NY  14456, USA}
\affiliation{Institute of Applied Physics, Nizhny Novgorod, 603950, Russia}
\affiliation{Lund Observatory, Box 43, SE-221 00, Lund, Sweden}
\affiliation{Hanyang University, Seoul 133-791, Korea}
\affiliation{Seoul National University, Seoul 151-742, Korea}
\affiliation{University of Strathclyde, Glasgow, G1 1XQ, United Kingdom}
\affiliation{The University of Texas at Austin, Austin, TX 78712, USA}
\affiliation{Southern University and A\&M College, Baton Rouge, LA  70813, USA}
\affiliation{University of Rochester, Rochester, NY  14627, USA}
\affiliation{University of Adelaide, Adelaide, SA 5005, Australia}
\affiliation{National Institute for Mathematical Sciences, Daejeon 305-390, Korea}
\affiliation{Louisiana Tech University, Ruston, LA  71272, USA}
\affiliation{McNeese State University, Lake Charles, LA 70609 USA}
\affiliation{Andrews University, Berrien Springs, MI 49104 USA}
\affiliation{Trinity University, San Antonio, TX  78212, USA}
\affiliation{University of Washington, Seattle, WA, 98195-4290, USA}
\affiliation{Southeastern Louisiana University, Hammond, LA  70402, USA}
\author{J.~Aasi$^\text{1}$}\noaffiliation\author{J.~Abadie$^\text{1}$}\noaffiliation\author{B.~P.~Abbott$^\text{1}$}\noaffiliation\author{R.~Abbott$^\text{1}$}\noaffiliation\author{T.~D.~Abbott$^\text{2}$}\noaffiliation\author{M.~Abernathy$^\text{3}$}\noaffiliation\author{T.~Accadia$^\text{4}$}\noaffiliation\author{F.~Acernese$^\text{5a,5c}$}\noaffiliation\author{C.~Adams$^\text{6}$}\noaffiliation\author{T.~Adams$^\text{7}$}\noaffiliation
% ...

 \collaboration{The LIGO Scientific Collaboration and the Virgo Collaboration}
\noaffiliation \author{and D.~P.~Anderson} \affiliation{University of California at Berkeley, Berkeley, CA 94720 USA}


\date{\today}

\begin{abstract}
This paper presents results of an all-sky search for periodic gravitational waves in the frequency range [50, 1\,190]~Hz and with frequency derivative range of $\sim [-20, 1.1]~\times~10^{-10}$~Hz~s$^{-1}$ for the fifth LIGO science run (S5). The search uses a non-coherent \textit{Hough-transform} method to combine the information from coherent searches on timescales of about one day. Because these searches are very computationally intensive, they have been carried out with the Einstein@Home volunteer distributed computing project. 
Post-processing identifies eight candidate signals; deeper follow-up studies rule them out.
Hence, since no gravitational wave signals have been found, we report upper limits on the intrinsic gravitational wave strain amplitude $h_0$.  For example, in the 0.5~Hz-wide band at 152.5~Hz, we can exclude the presence of signals with $h_0$ greater than $7.6~\times~10^{-25} $ at a $90\%$ confidence level.
This search is about a factor 3 more sensitive than the previous Einstein@Home search of early S5 LIGO data. 
\end{abstract}


\maketitle 

\section{\label{Intro}INTRODUCTION}

A promising class of sources for detectable gravitational wave signals
are rapidly rotating neutron stars with non-axisymmetric
deformations~\cite{NonAxNS1,NonAxNS2,NonAxNS3,NonAxNS4,NonAxNS5}. Such
objects are expected to emit long-lived continuous-wave
(CW) signals.  
In the rest frame of the neutron star, these waves have a constant
amplitude and are quasi-monochromatic with a slowly decreasing
intrinsic frequency. They are received at Earth-based detectors with a
Doppler modulation due to the relative motion between the source and
the detector. 
Consequently the observed phase evolution depends on the intrinsic
signal frequency, the first frequency time-derivative (also called
spindown), and the source sky position; these parameters shall
collectively be called the phase evolution parameters.  While using
higher order frequency derivatives could potentially improve the
astrophysical detection efficiency in a part of the parameter space
(see Sec.~\ref{sec:EatHset-up}), we shall not consider them in this
paper for computational reasons.  Finally, the received signal has a
time-dependent amplitude modulation due to the (time-dependent)
relative geometry of the wave and the detector.

The previous two decades have seen the construction and operation of
several kilometer-scale laser interferometric gravitational wave
detectors~\cite{ligoref,ligoref2,Virgo,virgo2,Geo,TAMA}.  The
detectors and the data analysis tools have steadily improved over this
period.  These have made it possible to search for various
gravitational wave signals with ever-improving sensitivity.
In this paper we focus on data from the fifth science run (S5) of the
LIGO (Laser Interferometer Gravitational wave Observatory) 
detectors, collected between the GPS times of 815\,155\,213 s (Fri Nov 04 16:00:00 UTC 2005) and  875\,145\,614  s (Sun Sep 30 00:00:00 UTC 2007).
The LIGO network~\cite{ligoref}
consists of three detectors: one at Livingston, Louisiana, USA, with
an arm length of 4 km (L) and two in the same vacuum envelope at
Hanford, Washington, USA, with arm lengths of 4 km (H) and 2 km, respectively. Only data from H and L detectors are used in this paper.
The Virgo and GEO\,600 detectors also collected data during the same
time interval but are not used in this analysis, which is optimized
for two detectors with similar sensitivities.

A coherent strategy for extracting faint CW signals buried in noisy
data using standard maximum-likelihood techniques in the presence of
``nuisance parameters'' was derived in~\cite{JKSpaper}.  The resulting
detection statistic is the so-called $\mathcal{F}$-statistic, which has
since been generalized to the case of multiple
detectors~\cite{MultiIfoFstat,MultiFstat2}.  The
$\mathcal{F}$-statistic has also been shown to arise as a special case
in a more general Bayesian framework \cite{Bstat}. Using the
$\mathcal{F}$-statistic means that we need to search explicitly only
over the phase evolution parameters.

Coherent wide-parameter-space searches utilizing the
$\mathcal{F}$-statistic have been carried out since the second LIGO
science run~\cite{S2ScoX1,S5CasA}.  The amplitude
sensitivity of this type of search improves as the square root of
the time
baseline. 
However, the template bank spacing decreases dramatically with the
time baseline, and the computational requirements of the search
increase rapidly.  Even with a coherent time baseline of just a few days, a
wide-frequency-band all-sky search is computationally extremely
challenging.  It becomes completely unfeasible if one considers instead
time baselines on the order of months.

As is often the case with computationally bound problems, hierarchical
approaches have been
proposed~\cite{HierarchP1,HierarchP2,HierarchP3}. In these strategies,
the entire data set is split into shorter segments.  Each segment is
analyzed coherently, and afterwards the information from the different
segments is combined incoherently (which means that the phase
information is lost). The amplitude sensitivity grows at best with the
fourth root of the number of segments.  Such methods have been used in
previous wide-parameter-space searches published by the LIGO and Virgo
collaborations
\cite{S2Hough,S4PSH,EatHS4R2,EatHS5R1,S5Powerflux2009,S5Powerflux2011}. 

A subset of these searches
\cite{S2Hough,S4PSH,S5Powerflux2009,S5Powerflux2011} used segments
sufficiently short (1\,800~s) that the signal remains within a single
Fourier frequency bin in each segment.  In these cases, a simple
Fourier transform suffices for each segment.  Three different variants
of such methods have been developed that combine the results from the
different short segments incoherently: the ``stack-slide'', the
``Hough-transform'' and the ``PowerFlux'' schemes. The stack-slide
procedure~\cite{HierarchP1,HierarchP3} averages the normalized power
from the Fourier transform of 30-minute segments (Short time baseline
Fourier Transform, SFT for short) of the calibrated detector strain
data. 
The PowerFlux scheme~\cite{S4PSH,S5Powerflux2009} can be seen as a variation
of the stack-slide method, where the power is weighted before
summing. The weights are chosen according to the detector noise and
antenna pattern to maximize the signal-to-noise ratio (SNR). The
Hough-transform method~\cite{HierarchP2,HoughP2} sums
weighted binary counts, depending upon whether the normalized power in
an SFT bin exceeds a certain threshold.

As the segment duration is increased, it becomes necessary to account
for signal modulations within each segment by computing the
$\mathcal{F}$-statistic over a grid in the space of phase evolution
parameters. This results in a significant increase in the
computational requirements of the search.  The distributed volunteer
computing project Einstein@Home~\cite{Einstweb} has been created to
address this need.  Two previous papers \cite{EatHS4R2,EatHS5R1}
report on results of such CW searches from the fourth LIGO science run
and from the first two months of S5,
respectively.  The method used was based on the computation of the
coherent $\mathcal{F}$-statistic on data segments from either the H or
L detectors separately, and only parameter space points with values of
$2\mathcal{F}$ larger than $25$ were returned back to the
Einstein@Home server for further inspection. The threshold value of
$25$ limited the ultimate sensitivity of that search: if a signal was
not loud enough to surpass that threshold on at least some of the
segments it would not be detected. The threshold value was set by
bandwidth constraints on the size of the results file uploaded back to
the server by the host, i.e.\ on the maximum number of significant
points that could be returned. These results were subsequently
combined by a coincidence scheme, performed offline in the
post-processing phase. In contrast, in the search presented here, the
combination of the results from the coherent searches takes place
directly on the host machines using a Hough-transform scheme. This
makes it possible to use a much lower threshold on $2\mathcal{F}$,
equal to $5.2$, that defines the parameter space points to be passed
on to the Hough-transform.  Moreover, in this search, data from the H
and L detectors are coherently
combined~\cite{MultiIfoFstat,MultiFstat2}.  Finally, more data was
searched in this analysis compared to any previous Einstein@Home
search.  The Einstein@Home runs presented here refer to searches based
on the first (S5R3) and second year of S5 LIGO data. This latter
search was run on Einstein@Home in two separate steps, called S5R5 and
S5R6. Since the S5R6 run used the same data as S5R5, but extended the
search region above 1~kHz, in this paper we simply refer to these two
runs as S5R5.

The paper is structured as follows. In Sec.~\ref{sec:lev3}, we briefly
review the Hough-transform method. Section~\ref{sec:EatHset-up}
describes the Einstein@Home distributed search used to analyze the data
set. Section~\ref{sec:S5R5pp} gives a detailed description of the S5R5
post-processing, which is based on the pioneer S5R3 post-processing (described
in Appendix~\ref{sec:S5R3pp}). Upper limit computations
from the more-sensitive S5R5 data are provided in
Sec.~\ref{sec:SensEst}. The study of some hardware-injected signals
is presented in Sec.~\ref{sec:HWInjS5R5}. In Sec.~\ref{sec:conclus} we make some concluding remarks.

\section{\label{sec:lev3} The data analysis method}
\subsection{The waveform model}

Let us begin by briefly describing the standard signal model for CW
signals.  In the rest frame of the neutron star, the gravitational
wave signal is elliptically polarized with constant amplitudes
$A_{+,\times}$ for the two polarizations $h_{+, \times}(t)$.  Thus, we
can find a frame such that
\begin{equation}
  \label{eq:h+x}
  h_+(t) = A_+ \cos\phi(t) \,,\qquad h_\times(t) = A_\times\sin\phi(t)\,. 
\end{equation}
The two amplitudes are related to an overall amplitude $h_0$ and the
inclination angle $\iota$ between the line of sight to the neutron
star and its rotation axis
\begin{equation}
  \label{eq:A+x}
  A_+ = \frac{1}{2}h_0(1+\cos^2\iota)\,, \qquad A_\times = h_0\cos\iota\,.
\end{equation}
The value of $h_0$ is model dependent.  For emission due to
non-axisymmetric distortions, the amplitude $h_0$ depends on the
ellipticity $\varepsilon$ of the star defined as
\begin{equation}
  \varepsilon = \frac{\left| I_{xx} - I_{yy}\right| }{I_{zz}}\,.
\end{equation}
Here $I_{zz}$ is the principal moment of inertia of the star, and
$I_{xx}$ and $I_{yy}$ are the moments of inertia about the other
axes.  The amplitude is given by
\begin{equation}
\label{eq:GWampl}
  h_0 = \frac{4\pi^2G}{c^4}\frac{I_{zz}f^2\varepsilon}{d},
\end{equation}
where $f$ is the frequency of the emitted GW signal (which is also
twice the rotational frequency of the star), $G$ is Newton's constant,
$c$ is the speed of light and $d$ is the distance to the star.  The
distribution of $\varepsilon$ for neutron stars is uncertain and model
dependent since the breaking strain for a neutron star crust is highly
uncertain (see e.g.\
\cite{NonAxNS2,Smoluchowski,Ruderman1991,Horowitz} for further
discussion).

Energy loss from the emission of gravitational and/or electromagnetic
waves, as well as possible local acceleration of the source, causes
the signal frequency arriving at the solar system to evolve.
To first order, it can be expressed as
\begin{equation}
  \label{eq:fhat}
  \hat{f}(\tau) = f_0 + \dot{f}(\tau-\tau_0),
\end{equation}
where $\tau$ is the arrival time of a wavefront at the solar system
barycenter (SSB), $f_0$ is the frequency at a fiducial reference time
$\tau_0$, and $\dot{f}$ denotes the first time derivative of the
frequency.  The astrophysical implications of ignoring higher order
derivatives in this Taylor expansion will be discussed later.  The
phase of the signal, $\phi(t)$, follows directly from the frequency
evolution with an initial phase $\phi_0$ at the reference time.

As the detector on the Earth moves relative to the SSB, the arrival time of a wavefront at the detector, $t$, differs from the SSB time $\tau$:
\begin{equation}
  \label{eq:tau(t)}
  \tau(t) = t + \frac{\vec{r}(t) \cdot \vec{n}}{c} +
  \Delta_{E\odot} - \Delta_{S\odot}\,. 
\end{equation}
Here $\vec{r}(t)$ is the position vector of the detector in the SSB
frame, and $\vec{n}$ is the unit vector pointing to the neutron star;
$\Delta_{E\odot}$ and $\Delta_{S\odot}$ are respectively the
relativistic Einstein and Shapiro time delays \cite{Taylor:1989sw}.
In standard equatorial coordinates with right ascension $\alpha$ and
declination $\delta$, the components of the unit vector $\vec{n}$ are
given by $(\mathrm{cos}\,\alpha \, \mathrm{cos}\,\delta,\
\mathrm{sin}\,\alpha \, \mathrm{cos}\,\delta,\ \mathrm{sin}
\,\delta)$.

Ignoring the relativistic
corrections, the instantaneous frequency $f(t)$ of a CW signal, as
observed at time $t$ by a detector on Earth, is described by the
well-known Doppler shift equation:
\begin{equation}\label{Eq:DopplerF}
f(t)  = \hat{f}(\tau) + \hat{f}(\tau) \frac{\vec{v}(\tau) \cdot \vec{n} }{c},
\end{equation}
where $\vec v(\tau)$ is the detector velocity with respect to the SSB frame;
$\vec v(\tau)$ is the sum of two components, from the yearly Earth motion
around the Sun ($\vec v_{y}$) and from the rotation of Earth around its
axis ($\vec v_{d}$). 

Finally, the received signal at the detector is
\begin{equation}\label{Eq:h(t)Signal}
h(t)  = F_+(t;\vec{n},\psi)h_+(t) +
F_\times(t;\vec{n},\psi)h_\times(t), 
\end{equation}
where $F_{+,\times}$ are the detector beam pattern functions which
depend on the sky position $\vec{n}$ and the relative polarization angle $\psi$
of the wave-frame \cite{JKSpaper,Bonazzola:1995rb}.  There are thus
altogether eight signal parameters, which include the four phase evolution
parameters $(f_0,\dot{f},\alpha,\delta)$, and four other parameters
$(h_0,\iota,\psi,\phi_0)$.

\subsection{\label{sec:HoughTA}The Hough-transform algorithm}

For completeness, we summarize the Hough detection
statistic in this section.  Further details of the method are given in
\cite{HierarchP2} and previous searches with this method applied to
short coherent times are reported in \cite{S2Hough,S4PSH}.

The Hough-transform is a well-known technique used mainly in digital
image processing for robust extraction of patterns. Such a procedure
is employed here for identifying points in the time-frequency plane
that match the pattern expected from a signal.  The time-frequency
data in our case is the $\mathcal{F}$-statistic computed as a function
of signal frequency, for each of the data segments of duration
$T_{\mathrm{seg}}$, over a grid of points in the space of
$(\alpha,\delta,\dot{f})$.  The grid in $(\alpha,\delta,\dot{f})$
space used for this $\mathcal{F}$-statistic computation is called the coarse
grid because its resolution is determined by the coherent
time baseline $T_{\mathrm{seg}}$, and makes no reference to
the
full observation time or the number of segments $N_{\mathrm{seg}}$.  The result of this
computation is thus a collection of $\mathcal{F}$-statistic values
$\mathcal{F}^i_{\alpha,\delta,\dot{f}}(k)$ where the integers $k$ and $i$ label a frequency bin (with spacing $\delta f$ as defined below)  and a data segment respectively.

The frequency and frequency derivative spacings for the coarse grid
are based on choosing the maximum allowed fractional loss in the
$\mathcal{F}$-statistic when the signal and template points are
slightly mismatched.  This leads naturally to the notion of a metric
in parameter space \cite{MetricBSD,MetricBen} and this has been
studied for the CW case in \cite{MultiFstat2}. The grid spacings in
$f,\dot{f}$ are given respectively by~\cite{ForMetricExprRes,S5CasA}
\begin{equation}\label{Eq:freqres}
\delta f = \frac{\sqrt{12 m}}{\pi T_{\mathrm{seg}}}
\end{equation}
and
\begin{equation}\label{eq:dfdotok}
\delta \dot{f} = \frac{\sqrt{720 m}}{\pi T_{\mathrm{seg}}^{2}},
\end{equation}
where $m$ represents the nominal single dimension mismatch value. For
all the Einstein@Home runs discussed here $m$ has been taken equal to $0.3$.  The
span $T_{\mathrm{seg}}$ of each segment has been set equal to
$25$~hours for all the runs. The frequency resolution, given by
Eq.~(\ref{Eq:freqres}), is $\delta f \sim 6.7 \; \mu $Hz for
all the Einstein@Home runs described in this paper.  As we shall see shortly, for
technical reasons it turned out to be necessary to use a finer spacing
for $\dot{f}$ than given by Eq.~\eqref{eq:dfdotok}.

In combining these $N_{\textrm{seg}}$ different
$\mathcal{F}$-statistic vectors, it is necessary to use a finer grid
in $(\alpha,\delta,\dot{f})$ centered around each coarse grid point.
Our implementation of the Hough-transform algorithm assumes that the
fine grid is a Cartesian product of a rectangular sky-grid and one
dimensional grids in $f$ and $\dot{f}$. Moreover the fine sky-grid is
assumed to be aligned with the $\alpha$ and $\delta$ directions in the
sky.  In order to completely cover the sky with the different fine
sky-grids, it is thus simplest to choose a rectangular coarse sky-grid
aligned with the $(\alpha,\delta)$ directions.  We choose a coarse
grid such that the spacing in $\delta$ is a constant and the spacing
in $\alpha$ is proportional to $(\cos\delta)^{-1}$.  This ensures that each
cell of the coarse grid covers a fixed solid angle.  

Since the coherent integration time is very close to a sidereal day, it is
reasonable to assume that the Hough sky-patch size
$d\theta$ (which in our case is the same as the coarse sky-grid
resolution) is determined by $v_{d}$, the Earth's rotation speed at the equator. 
At frequency $\hat{f}$ we
have~\cite{HierarchP2}:
\begin{equation}\label{eq:sky-patchSize} 
d\theta = \frac{c}{v_{d} \hat{f} T_{\mathrm{seg}}}\,.
\end{equation}
In practice, this estimate was verified by Monte-Carlo studies with
signal injections, and we used
\begin{equation}
\label{eq:coarseskyres}
d \theta_{\mathcal{F}} = \left\{ \begin{array}{ll}
 d \theta & \textrm{for S5R3}\\
 \mathcal{R} d \theta & \textrm{for S5R5},
  \end{array} \right.
\end{equation}
where the factor $\mathcal{R} = \sqrt{3}$ increases the size of the
sky-patch $d\theta$ for the S5R5 run.  The Monte-Carlo studies showed
that the grid spacing for S5R5 corresponded approximately to a
mismatch of $m\approx 0.3$ so that, for any other value of $m$, the spacing
would be approximately $\sqrt{m/0.3}\mathcal{R}d\theta$.

We also need to set the resolution of the refined sky-grid used by the
Hough algorithm.  Since the full observation time is of the order of a
year, the relevant scale here is set by the speed $v_y$ of Earth as it orbits the Sun.  Following~\cite{HierarchP2}, the resolution for the fine
sky-grid at a frequency $\hat{f}$ is given by
\begin{equation}\label{eq:HoughPixRes}
d \theta_{\mathrm{H}} = \frac{c \, \delta f}{\wp \hat{f} v_{y}}\,.
\end{equation}
The parameter $\wp$ scales the resolution compared to the conservative
estimate $c \, \delta f/(\hat{f} v_{y})$ and in practice,
again based on Monte-Carlo studies, we used $\wp = 0.5$.  One can see
that the increase in the number of sky position points from the coarse
$\mathcal{F}$-statistic grid to the fine Hough grid is
\begin{equation}
  \mathcal{N}_{\mathrm{sky}}^{\mathrm{ref}} = \left(d \theta_\mathcal{F}/d\theta_{\mathrm{H}}\right)^{2} =\left( {\pi\over \sqrt{12m}}\wp{ v_y\over v_d} \right)^2 \sim {\cal O}(10^4).
\label{eq:houghSkyRef}
\end{equation}
Taking $m=0.3$ and $\wp = 0.5$ yields
$\mathcal{N}_{\mathrm{sky}}^{\mathrm{ref}} \simeq 8\,444$ for the S5R5
run.

Finally, let us turn to the coarse and fine grids for $\dot{f}$.
Ideally, we should refine the coarse $\dot{f}$ grid spacing of
Eq.~\eqref{eq:dfdotok} by a factor $N_{\mathrm{seg}}$~\cite{HierarchP2}.  However, in our implementation of the Hough
transform, using this refinement turns out to increase the maximum memory
footprint of the different searches and would make it unsuitable for
Einstein@Home.  As a compromise, it was decided not to use
any refinement in $\dot{f}$ and instead to use a finer
resolution for the coarse grid.  Based on Monte-Carlo analyses, an
acceptable value for the $\dot{f}$ spacing turns out to be
\begin{equation}\label{eq:S5R5fdot}
  \left(\delta \dot{f} \right)_{\mathrm{S5R5}}= \frac{\sqrt{3.3 \, m}}{T_{\mathrm{seg}}^{2}} \sim 1.2 \times 10^{-10}~\mathrm{Hz}~\mathrm{s}^{-1}.
\end{equation}
This value was used for S5R5.  However, for S5R3, this was
incorrectly set to 
\begin{equation}\label{eq:wrongdfdot}
\left(\delta \dot{f}\right)_{\mathrm{S5R3}} = \frac{\sqrt{33 \, m}}{T_{\mathrm{seg}}^{2}} \sim 3.8 \times 10^{-10}~\mathrm{Hz}~\mathrm{s}^{-1},
\end{equation}
leading to a corresponding loss in sensitivity for S5R3.

\begin{figure}
\begin{center}
\includegraphics[width=9.2cm, angle=0, clip]{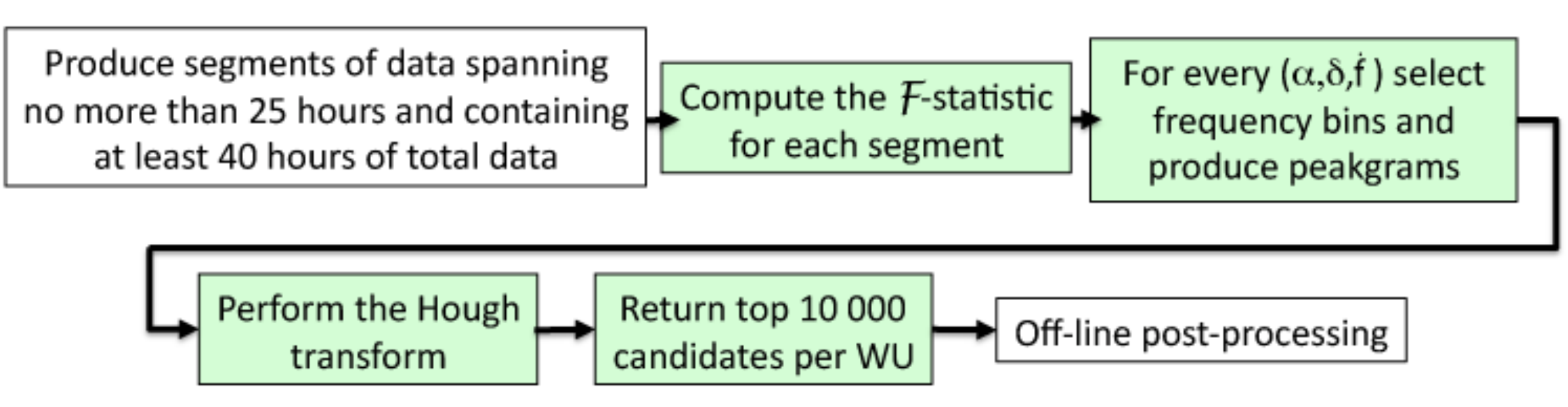} 
\caption{\label{Fig:DemodProc} High level schematic of the pipeline used in the searches.
For each data segment, the multi-detector $\mathcal{F}$-statistic is computed and frequency bins are selected setting a threshold on the $\mathcal{F}$-statistic. Such selected frequency bins are then used to create the Hough map. The output is a set of candidates in the parameter space. The color-filled boxes indicate the steps performed on the volunteer computers. The acronym ``WU'' is defined in Sec.~\ref{subsec:s5r5basics} and refers to the independent computing tasks into which we partition the computational work of the search.}
\end{center}
\end{figure}

The flow chart
of the search algorithm used for this search is depicted in
Fig.~\ref{Fig:DemodProc}. The input data set is composed of 30-minute
baseline SFTs. This set is partitioned in subsets such that no more
than 25 hours of data are spanned by each segment and such that there
is overall (including data from both detectors) at least 40 hours of
data in each segment. Let $T_{\mathrm{obs}}$ be the observation time
spanned by the $N_{\mathrm{seg}}$ segments constructed in this
way\footnote{Note that $T_{\mathrm{seg}} \neq
  T_{\mathrm{obs}}/N_{\mathrm{seg}}$ because of the gaps that are
  unavoidably present in the data stream, corresponding to times when
  the interferometers were not in lock, and to the selection of
  25 hour segments containing the requisite amount of data.}. The
multi-detector $\mathcal{F}$-statistic is computed for each segment at
each point of the search parameter space ($f$,$\dot{f}$, $\alpha$, $\delta$). The next step
consists of selecting parameter space points for which the $\mathcal{F}$-statistic
is above the fixed threshold. For every set
of~($f$,$\dot{f}$, $\alpha$, $\delta$), we assign a value
$n_{i}=1$ or $0$ in the $i^{\textrm{th}}$ segment depending on whether
the corresponding $\mathcal{F}$-statistic is above the threshold 5.2 or not;
this threshold turns out to be optimal \cite{HierarchP2}.
The values $n_{i}(f)$ are called a ``peakgram'', which is the input to the
Hough-transform.

The final statistic used by the Hough search is a weighted sum of
binary counts $n_{i}$, giving the so-called \textit{Hough number
  count} $n_{\mathrm{c}}$, expressed by \cite{S4PSH}
\begin{equation}
  \label{eq:nc}
  n_{\mathrm{c}} = \sum_{i=0}^{N_{\mathrm{seg}}-1} w_{i} n_{i}\,.
\end{equation}
The weight $w_i$ for a frequency $f$ and for a particular sky-location
is determined from the average antenna response and average detector
noise over the duration of the $i^{\mathrm{th}}$ segment.  Since the
input data in each segment consists of SFTs, we perform the averaging
over each SFT. Let $N_i$ be the number of SFTs in the
$i^{\textrm{th}}$ segment.  Let $S_{h}^{i,\gamma}$ be the single-sided
power spectral density (PSD) of the $\gamma^{\textrm{th}}$ SFT in the
$i^{\textrm{th}}$ segment, and averaged over a narrow frequency band
containing the search frequency.  The $w_i$ are given by
\begin{equation}
  \label{eq:weights}
  w_i \propto \sum_{\gamma=0}^{N_i-1} \left(F_{+(i,\gamma)}^2 + F_{\times (i,\gamma)}^2 \right) w_{i,\gamma}\,,
\end{equation}
where $F_{+(i,\gamma)}$ and $F_{\times(i,\gamma)}$ are the detector
antenna pattern functions for the $\gamma^{\mathrm{th}}$ SFT in the
$i^{\mathrm{th}}$ segment, and
\begin{equation}
  \label{eq:noisewt-harmonic}
    w_{i,\gamma}  = \frac{1}{S_h^{i,\gamma}} \times \left( \frac{1}{N_i}
    \sum_{\beta=0}^{N_i-1} \frac{1}{S_h^{i,\beta}}\right)^{-1} \,.
\end{equation}
It is easy to see that, in the
hypothetical case when the data is exactly stationary, so all the
$S_h^{i,\gamma}$ are identically equal to each other, then
$w_{i,\gamma} = 1$. More realistically, the $w_{i,\gamma}$ are
approximately unity for stationary data and the use of the harmonic
mean in Eq.~(\ref{eq:noisewt-harmonic}) ensures that the
$w_{i,\gamma}$ do not deviate too far from unity in the presence of
non-stationary noise.

The weight normalization is
\begin{equation}
  \label{eq:normW}
  \sum_{i=0}^{N_{\mathrm{seg}}-1} w_{i} = N_{\mathrm{seg}}\, ,
\end{equation}
which ensures that the Hough number count $n_{\mathrm{c}}$ lies within
the range $[0,N_{\mathrm{seg}}]$.  In~\cite{S4PSH} it is shown
that the weights $w_{i}$, first derived in~\cite{Palomba2005},
maximize the sensitivity, averaged over the orientation of the source (see Appendix \ref{sec:bugs} for a further discussion of the weights and some technical problems that were encountered in the search).

From the Hough number count $n_{\mathrm{c}}$ we define the \textit{significance} (or \textit{critical ratio}), CR:
\begin{equation}\label{eq:NcSign}
\mathrm{CR} = \frac{n_{\mathrm{c}}-\bar{n}_{\mathrm{c}}}{\sigma},
\end{equation}
that measures the significance of the measured $n_{\mathrm{c}}$ value
as the deviation from the expected
value $\bar{n}_{\mathrm{c}} = N_{\mathrm{seg}}p$ in absence of any
signal, in units of the expected noise fluctuations $\sigma$; $p$ is
the probability that a parameter space pixel is selected in the
absence of a signal. 
In case of unity weighting, the standard
deviation is simply that of the binomial distribution: $\sigma =
\sqrt{N_{\mathrm{seg}}p (1-p)}$. When the weights are used, the
standard deviation is given by
\begin{equation}
  \label{eq:sigmaW}
  \sigma = \sqrt{||\vec{w}||^2p(1-p)}\,,
\end{equation}
where $||\vec w||^2 = \sum_{i=0}^{N_{\mathrm{seg}}-1}w_{i}^2$~\cite{S4PSH}. The CR is the detection statistic returned by the hierarchical searches presented here. 

\section{\label{sec:EatHset-up} The Einstein@Home  distributed search}

The Einstein@Home project is built upon the BOINC (Berkeley Open
Infrastructure for Network Computing)
architecture~\cite{Boinc1,Boinc2,Boinc3}, a system that exploits the
idle time on volunteer computers to solve scientific problems that
require large amounts of computer power. During the S5R5 run,
Einstein@Home had approximately 225\,000 registered volunteers and
approximately 750\,000 registered host machines that contributed a
total of approximately 25\,000 CPU (Central Processing Unit) years.  

\subsection{Details of the S5R5 search}
\label{subsec:s5r5basics}

The computational load is partitioned in independent computing tasks,
called ``workunits'' (WUs), each of which is analyzed by a volunteer
machine.  In particular, 7\,369\,434 and 10\,945\,955 WUs have been
generated for the S5R3 and S5R5 runs, respectively.
The S5R5 run was launched on January 13, 2009 and ended on October 30, 2009. It
used 10\,560 S5 LIGO SFTs, collected between the GPS times of
852\,443\,819~s (Wed Jan 10 05:56:45 GMT 2007) and 875\,278\,812~s
(Mon Oct 01 12:59:58 GMT 2007). The analyzed data consists of 5\,550
and 5\,010 SFTs from the LIGO H and L interferometers, respectively.
The number of data segments used for the S5R5 run is 121, each
spanning no more than $T_{\mathrm{seg}}=25$~hours and with at least
$40$~hours of data, as already said.  Similar details for the S5R3 run can be found in
Appendix~\ref{sec:S5R3pp}.

The total search frequency range of the S5R5 search is $[50,
1\,190]$~Hz
, with a frequency resolution $\delta f \sim 6.7 \mathrm{\mu Hz}$ and
spindown resolution $\delta \dot{f} \sim0.12 $~ nHz~s$^{-1}$. 
Each WU analyzes a constant frequency band $B \simeq
20$~mHz, the full spindown interval, ranging roughly from $-2
$~nHz~s$^{-1}$ to $0.11$~nHz~s$^{-1}$, and a region of the sky, as we shall see in Sec.~\ref{sec:EatHWUdesign}. 

The original data contained instrumental artifacts in narrow frequency
bands that were known before the launch of the Einstein@Home
run. Those bands were identified and the corresponding frequency bins
in the SFTs were replaced with white Gaussian noise at the same level
as the neighboring frequencies. Table~\ref{tab:whiteNoiseBandsH} shows
which bands were treated in this manner and what instrumental artifact
they harbored. These control bands are useful to compare and contrast
the results obtained on real data against pure theoretical
noise. Measurements and studies after the Einstein@Home
run refined the frequencies and widths of these artifacts and
identified additional ones; the final lists of artifacts are given in
Appendix~\ref{sec:knownS5lines}, and were
used to discard candidates (as we shall see in
Sec.~\ref{sec:removeRSIL}).
The ``cleaning'' process affected $\sim 27$~Hz of search bandwidth,
in addition to bands that were eliminated later in post-processing.

\begin{table} [h!]
\begin{tabular}{lccccc} \hline \hline 
  Cause	      & $f_{\mathrm{L}}$~(Hz) 	&  Harmonics 			&  LFS~(Hz) 		& HFS~(Hz)  			& IFO\\
 \hline 
Calibration 	&     46.7		& 1		        			& 0.0		   	          & 0.0		& H\\ 
Calibration 	&     54.7		& 1		        			& 0.0		   	          & 0.0		& L\\ 
 Mains	&  60 	&  19 	&  1	&  1		& H, L\\
Wire	& 345	&  1	&  5	&  5		& L\\
 Wire	& 346	&  1	&  4	&  4		& H\\
Calibration 	&     393.1		& 1		        			& 0.0		   	          & 0.0		& H\\ 
Calibration 	&     396.7		& 1		        			& 0.0		   	          & 0.0		& L\\ 
Wire 	& 686.5     		& 1		        			& 1.0		   	          & 1.0		& L\\ 
Wire 	& 686.9     		& 1		        			& 0.3		   	          & 0.3		& H\\ 
Wire 	& 688.2    		& 1		        			& 0.3		   	          & 0.3		& H\\ 
Wire 	& 689.5    		& 1		        			& 0.5	   	          & 0.6		& H\\ 
Wire 	& 693.7     		& 1		        			& 0.7		   	          & 0.7		& L\\ 
Wire 	& 694.75     		& 1		        			& 1.25		   	          & 1.25		& H\\ 
Wire 	& 1029.5     		& 1		        			& 0.25		   	          & 0.25		& L\\ 
Wire 	&     1030.55 		& 1		        			& 0.1		   	          & 0.1		& H\\ 
Wire 	& 1031.0     		& 1		        			& 0.5		   	          & 0.5		& L\\ 
Wire 	&     1032.18 		& 1		        			& 0.04		   	          & 0.04		& H\\ 
Wire 	&     1032.58 		& 1		        			& 0.1	   	          & 0.1		& H\\ 
Wire 	& 1033.6     		& 1		        			& 0.2		   	          & 0.2		& L\\ 
Wire 	&     1033.7 		& 1		        			& 0.1	   	          & 0.1		& H\\ 
Wire 	&     1033.855		& 1		        			& 0.05	   	          & 0.05		& H\\ 
Wire 	&     1034.6 		& 1		        			& 0.4		   	          & 0.4		& H\\ 
Wire 	& 1041.0    		& 1		        			& 1.0		   	          & 1.0		& L\\ 
Wire 	&     1041.23		& 1		        			& 0.1		   	          & 0.1		& H\\ 
Wire 	&     1042.00		& 1		        			& 0.5		   	          & 0.2		& H\\ 
Wire 	&     1043.4		& 1		        			& 0.2		   	          & 0.2		& H\\ 
Calibration 	&     1144.3	& 1		        			& 0.0		   	          & 0.0		& H\\
Calibration  	&     1151.5	& 1		        			& 0.0		   	          & 0.0		& L\\
\hline \hline
\end{tabular}
\caption{\label{tab:whiteNoiseBandsH} Instrumental lines identified and ``cleaned'' before the Einstein@Home runs. The different columns represent: (I) the source of the line; (II) the central frequency of the instrumental line; (III) the number of harmonics; (IV) Low-Frequency-Side (LFS) of the knockout band; (V) High-Frequency-Side (HFS) of the knockout band; (VI) the interferometer where the instrumental lines were identified. Note that when there are higher harmonics, the knockout band width remains constant.}
\end{table}

The output data files from each WU are stored as ZIP-compressed ASCII
text files containing the 10\,000 most significant candidates ranked
according to the significance (as defined in Eq.~(\ref{eq:NcSign}))
over the parameter space searched by that WU.  The decision to keep
the top 10\,000 candidates was based on the maximum upload volume from
the hosts to the Einstein@Home servers. All in all, on the order of
$10^{11}$ candidates were returned to the Einstein@Home server from
each run, corresponding roughly to 2.3 TB of data.

The files contain nine quantities for each candidate.
The first four are the values (on the
coarse grids) of the frequency $f$, sky-position $(\alpha,\delta)$,
and spindown $\dot{f}$. The fifth is the significance of the candidate
as defined in Eq.~(\ref{eq:NcSign}). The remaining four quantities are
connected with the refined sky-grid centered on each coarse sky-grid
point (recall that refinement is performed only on the sky).  In
particular, the location of the most significant point on the fine sky-grid, and the mean and standard deviation of the Hough number count
values on all points of the fine grid are returned.  

\subsection{Validation of returned candidates}
\label{subsec:validation}

In order to eliminate potential errors, due to defective hardware
and/or software or to fraud, BOINC is configured so that each WU is
processed redundantly by computers owned by at least two different
volunteers. An automated validation process checks the consistency of
the results, ruling out those that are inconsistent, in which case new
WUs are generated to run again independently.  The first step of the
validation is to check that the file syntax is correct and that the
first four values, i.e.\ $(f,\alpha,\delta,\dot{f})$, are within the
appropriate ranges.  Next, for the pair of result files from each WU,
the validator checks that the values of $(f,\alpha,\delta,\dot{f})$
agree to within floating point accuracy (the frequency is in fact
checked to double precision).  Finally, the significance values
$\mathrm{CR}_1$ and $\mathrm{CR}_2$ from the two result files are
compared and are validated if
\begin{equation}
\Delta := |\mathrm{CR}_1-\mathrm{CR}_2|/(\mathrm{CR}_1+\mathrm{CR}_2) < 0.12\,.
\end{equation}
In S5R5, about $0.045\%$ of results that were processed by the
validator were marked as ``invalid'', including both syntax errors in
individual files and errors in comparisons of different results files.
Excluding the syntax errors in individual files, the error rate
arising from comparisons of pairs of distinct result files (most
likely due to differences in floating point arithmetic on different
computational platforms) was $\sim 0.015\%$.

Is it possible that two invalid results could agree with each other
and thus end up being marked as valid?  While it is difficult to
exclude this scenario with complete certainty, an upper limit for the
probability of this happening is $(0.015/100)^2 \approx 2.2\times
10^{-8}$ (only the $0.015\%$ error rate due to comparisons of distinct
result files is relevant here).  As mentioned earlier, there were a
total of $\sim 1.1 \times 10^7$ WUs. It is therefore unlikely that
even a single pair of result files would be incorrect and still pass
validation.

The threshold of 0.12 on the value of $\Delta$ defined above turns out
to be much looser than necessary.  The differences in the actual
observed values of $\Delta$ from a pair of matching result files are
usually much smaller.  The observed standard deviation of $\Delta$
turns out to be $\sim 0.012$, i.e.\ an order of magnitude smaller than
the threshold.  In addition, the standard deviation of the difference
$\mathrm{CR}_1-\mathrm{CR}_2$ is measured to be $\sim 0.15$, which
corresponds to a standard deviation of $\sim 0.7$ in terms of the
number count.  As we shall see later (see e.g.\
Fig.~\ref{Fig:NcFrS5R5R6}), the loudest events in every 0.5\,Hz band
have an average loudest number count $\gtrsim 70$. Thus, these
differences correspond to a $\lesssim 1$\% effect in the number
count at the 1-$\sigma$ level, and we expect this to have a negligible
effect on our analysis.  

\subsection{\label{sec:EatHWUdesign}Workunit design}

The design of the WUs must satisfy certain requirements.  The
first is that the WUs must be balanced, i.e.\ each WU must cover the
same number of parameter space points so that they can be completed in
roughly the same amount of time by a typical host machine.  Second,
the amount of data that must be downloaded by each host machine and
the maximum memory footprint of each job must be within appropriate limits.
Finally, one needs to choose the computational time for each WU on a
typical host machine and the total time that the project should run,
given the total computational power that is available.  To meet these
requirements, we need to understand how to split up the parameter
space, and to measure the CPU core time spent by the search
code on each part of the analysis.

We start with the basic parameters of the search, namely the total
observation time $T_{\mathrm{obs}}$, the coherent time baseline
$T_{\mathrm{seg}}$ and the number of segments $N_{\mathrm{seg}}$, the
resolution for the coarse and fine sky-grids,
$d\theta_{\mathcal{F}},d\theta_H$, given by
Eqs.~\eqref{eq:coarseskyres} and \eqref{eq:HoughPixRes}, respectively
and the frequency and spindown resolutions, $\delta f$ and $\delta
\dot{f}$, given by Eqs.~\eqref{Eq:freqres} and \eqref{eq:S5R5fdot},
respectively. Recall that the limit on the maximum
memory footprint of each job already forced us to forego any
refinement in $\dot{f}$.

\begin{figure}[h!]
\includegraphics[width=9.3cm, angle=0, clip]{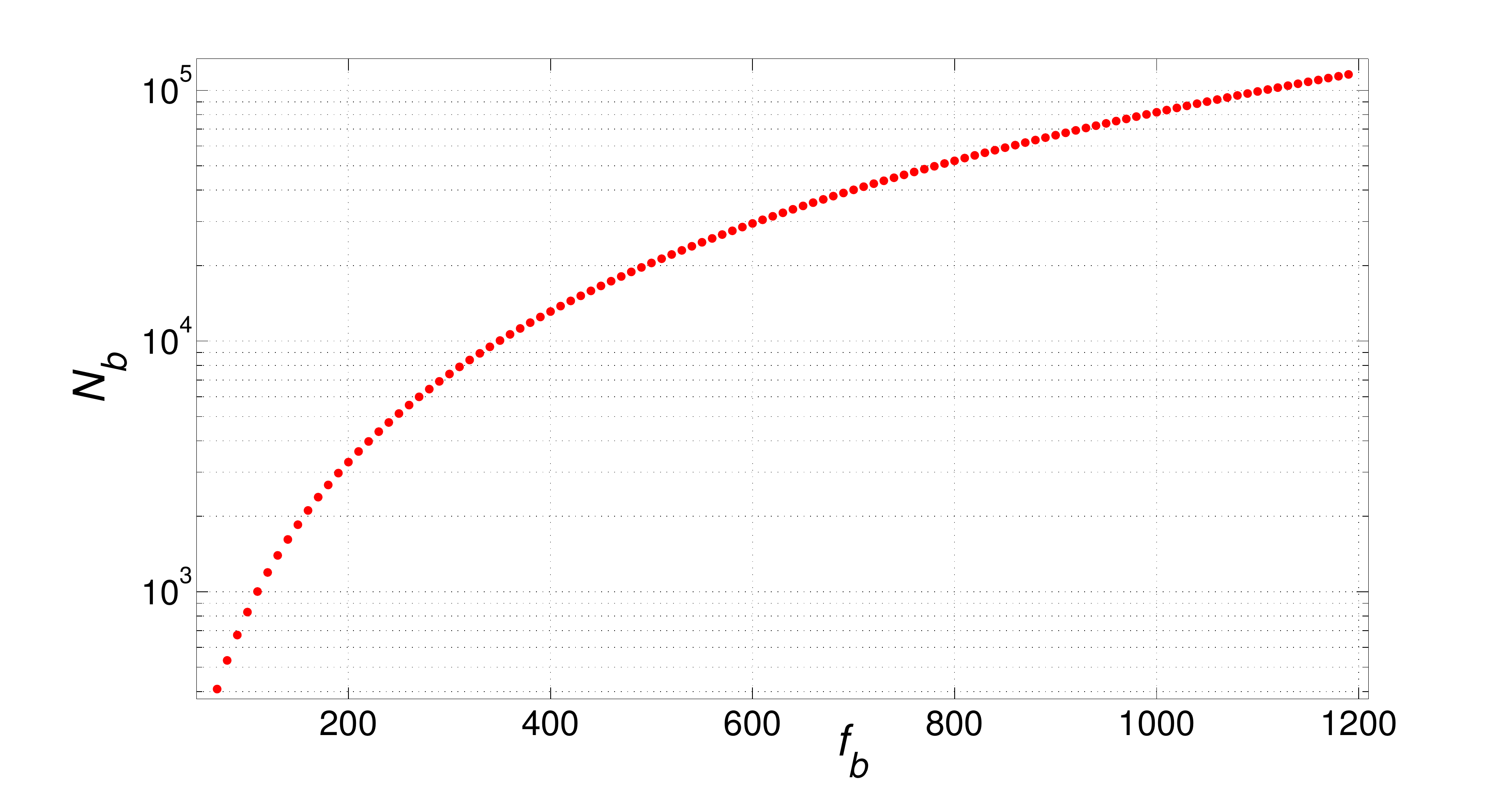}
\caption{\label{Fig:numSkyPoints} Total number of $\mathcal{F}$-sky-grid points, for the S5R5 run, as a function of the frequency, given by Eq.~(\ref{Eq:numSkyPoints}).}
\end{figure}

Unlike the previous Einstein@Home search~\cite{EatHS5R1},
where each WU searched the whole sky, here we choose each WU to
cover a fixed frequency bandwidth $B$, the entire spindown search range, and
a limited area of the sky. Let ${\hat f_b}$ be the highest frequency
in the $b^{\mathrm{th}}$ search band.
We want the computation time for every WU to be approximately the same, hence every WU must search the
same number of coarse sky-grid points.  Since the resolution in the
sky, $d \theta_{\mathcal{F}}$, is inversely proportional to the frequency of the
signal that we are searching for, WUs at higher frequencies
will be searching smaller portions of the sky.  Let $N_b$
be the number of coarse grid sky points over the whole sky for the
$b^{\rm th}$ band:
\begin{equation}
\label{Eq:numSkyPoints}
N_b={4\pi\over d\theta_{\mathcal{F}}^2}=4\pi\left({v_d\over c}\right)^2 {\hat f_b^2 T_{\mathrm{seg}}^2 \over \mathcal{R}^2},
\end{equation}
which is shown in Fig.~\ref{Fig:numSkyPoints}
for the
S5R5 run. At frequency $\hat f_b$, we then partition the sky in $P_b$
parts, each containing $N_{\mathrm{sky}}$ points:
\begin{equation}
\label{Eq:skyPartitions}
P_b={N_b\over N_\mathrm{sky}}=K \hat f_b^2,
\end{equation}
with $K={4\pi\left({v_d\over c}\right)^2 T_{\mathrm{seg}}^2}
\mathcal{R}^{-2}N_\mathrm{sky}^{-1}$.  In practice, to limit the
number of sky-grid files needing to be downloaded by the host machines,
the sky-grids are constant over $10$-Hz frequency bands and
are determined based on the highest frequency in each band.

We choose a constant range of $\dot{f}$ values over the entire
frequency band.  To see what this implies for potential sources, we
define the spindown age for a system emitting at a frequency $f$ and
spindown $\dot{f}$ to be $\tau = f/|\dot{f}|$ \footnote{The reader is
  warned that the spindown age as defined here can be rather
  different from the true chronological age of the star. For a neutron
  star with a present day frequency $f$ and spindown $\dot{f}$ due
  purely to the emission of GWs, the age would be $f/4|\dot{f}|$;
  more generally this depends on the physical mechanisms that are
  responsible for the spindown \cite{Ostriker:1969if}.}.  It is clear
that at a given search frequency, the minimum spindown age is
determined by the maximum value of $|\dot{f}|$ included in the search.
Our choice of a constant range of $\dot{f}$ means that the minimum
spindown age is frequency dependent:
\begin{equation}
  \label{eq:2}
  \tau_{\rm min}^b = \frac{f_b}{|\dot{f}|_{\rm max}}\,.
\end{equation}
We choose the minimum age at $50\,$Hz to be 800~years, and a fixed
range in $\dot{f}$: $\sim [-20, 1.1]~\times~10^{-10}$~Hz~s$^{-1}$.

The total number of WUs is simply the sum of the total number of sky-partitions
at each frequency $f_b$:
\begin{multline}\label{eq:Nwu}
N_{\mathrm{WU}} = \sum_{b} P_{b} = K \sum_{b} \hat f_{b}^{2} \approx \frac{K}{B} \int_{f_{\mathrm{min}}}^{f_{\mathrm{max}}} \hat f^{2} df \\
= \frac{K}{3 B} \left( f_{\mathrm{max}}^{3} - f_{\mathrm{min}}^{3} \right)\,. 
\end{multline}
In the third step we have replaced the sum over frequencies with an
integral between a minimum ($f_{\mathrm{min}}$) and maximum
($f_{\mathrm{max}}$) frequency.  As a consistency check, we shall see
later that $B \simeq 20\,$mHz, which is sufficiently small that this is a
good approximation.

The total computing time for one WU can be expressed as
\begin{equation}\label{Eq:tauWU}
\tau_{\mathrm{WU}} = N_{\mathrm{sky}} N_{\dot{f}} [(N_{f}+N_{\mathrm{sb}})\tau_{\mathcal{F}}+N_{f} \tau_{\mathrm{H}}], 
\end{equation}
where $N_{f}$, $N_{\dot{f}}$, and $N_{\mathrm{sky}}$
represent the number of coarse search frequency bins, spindown values,
and sky-grid-points respectively, $\tau_{\mathcal{F}}$ and
$\tau_{\mathrm{H}}$ are the times needed to compute the
$\mathcal{F}$-statistics and Hough number count for one point of the
coarse grid; $N_{\mathrm{sb}}$~\footnote{ The bins $N_{\mathrm{sb}}$ can be calculated by
  $\Delta f_{\mathrm{sb}}/\delta f$, where the average Hough ``sidebands''
  $\langle \Delta f_{\mathrm{sb}} \rangle$ can be estimated from the
  Hough master equation. The frequency offset corresponding to the
  half diagonal distance $\Delta \vec{n}$, over one Hough sky-patch,
  is $\langle \Delta f_{\mathrm{sb}} \rangle = \frac{\hat{f}}{c}
  \langle | \vec{v}_{y} \cdot \Delta \vec{n} | \rangle =
  \frac{v_{y}}{v_{d}}\frac{1}{\sqrt{2} \, T_{\mathrm{seg}}}$.  
By using $T_{\mathrm{seg}} = 25$~hours,
  we get for S5R3 $\langle
  \Delta f_{\mathrm{sb}} \rangle \sim 5 \times 10^{-4}$~Hz and
  $N_{\mathrm{sb}} \sim 75$~ bins, while for S5R5 $\langle \Delta
  f_{\mathrm{sb}} \rangle \sim 8.7 \times 10^{-4}$~Hz and
  $N_{\mathrm{sb}} \sim 130$~ bins. However, to be conservative, $(2 \times N_\mathrm{sb})$
  bins were used for safety reasons and to take into account for the
  changes of the velocity on the different segments. 
} represents additional ``sideband'' bins needed to
compute the Hough-transform.
The need for these sidebands can be understood by thinking about
computing the Hough number count for a frequency near the edge of a
given search band: that number count will involve summing the
peakgrams along a curved track that can extend a small distance to
either side of the target frequency.

The time $\tau_{\mathcal{F}}$ needed to compute the
$\mathcal{F}$-statistic for one point of the coarse grid can be
expressed as
\begin{equation}
\tau_{\mathcal{F}} =\tau_{\mathcal{F}}^{1} N_{\mathrm{SFT}},
\end{equation}
where $N_{SFT}$ is the total number of SFTs used in the search; $\tau_{\mathcal{F}}^{1}$ is the time needed to compute the
$\mathcal{F}$-statistic per single SFT. The time $\tau_{\mathrm{H}}$
in Eq.~(\ref{Eq:tauWU}), needed to compute the Hough number count
corresponding to one point of the coarse grid, can be written as
\begin{equation}
\tau_{\mathrm{H}} = \tau_{\mathrm{H}}^{1} N_{\mathrm{seg}} \mathcal{N}_{\mathrm{sky}}^{\mathrm{ref}},
\end{equation}
where $\tau_{\mathrm{H}}^{1}$ is the time to sum the Hough number
count per single data segment and per single point of the fine
grid. Here $\mathcal{N}_{\mathrm{sky}}^{\mathrm{ref}}$ is an overall refinement
factor, i.e.\ the number of grid points analyzed by the Hough algorithm
for each coarse grid point.  In our case, since the only refinement is
over the sky, $\mathcal{N}_{\mathrm{sky}}^{\mathrm{ref}}$ is given by
Eq.~\eqref{eq:houghSkyRef}.  The computational time is thus determined
by the two timing constants $\tau_{\mathcal{F}}^{1}$ and
$\tau_{\mathrm{H}}^{1}$.  For our implementation of the algorithm,
these constants were measured to be
\begin{equation}
  \label{eq:tauFHvalues}
\tau_{\mathcal{F}}^{1} = 180~\mathrm{ns}\,,\quad
\tau_{\mathrm{H}}^{1} = 1.1~\mathrm{ns}\,.
\end{equation}
These numbers are of course only average values for a typical host CPU core
available at the time of the Einstein@Home runs.

The presence of $N_{\mathrm{sb}}$ leads to an overhead for
the computation. We want to control this overhead and keep it
below some acceptable level.  Thus, we define the overhead $\epsilon$
to be the ratio between the time spent in computing the
$\mathcal{F}$-statistic for the ``sidebands'' and the total computational
time:
\begin{equation}
  \epsilon \equiv \frac{\Delta f_{\mathrm{sb}} \, \, \tau_{\mathcal{F}}}{B (\tau_{\mathcal{F}}+\tau_{\mathrm{H}}) + \Delta f_{\mathrm{sb}} \, \, \tau_{\mathcal{F}}}\,.
\end{equation}
The bandwidth B 
needs to be sufficiently large so that $\epsilon$ is
sufficiently small, but a too large value of $B$ can lead to
excessively high download volumes for the Einstein@Home clients.  From
the above equation, we see that the frequency bandwidth $B$ can be
determined by fixing $\epsilon$:
\begin{equation}
B = \frac{\Delta f_{\mathrm{sb}} \, \, \tau_{\mathcal{F}}}{\tau_{\mathcal{F}} + \tau_{\mathrm{H}}} \left(\frac{1-\epsilon}{\epsilon}\right).
\end{equation}
For all the Einstein@Home runs presented here we choose $\epsilon = 5\%$.  For the
S5R5 run, this leads to $B \simeq 20$~mHz.

The total run-time $\tau_{p}$ of the project is
\begin{equation}\label{eq:ProjDur}
\tau_{p} = \frac{\tau_{\mathrm{WU}} \, \, N_{\mathrm{WU}}}{N_{\mathrm{CPU}}},
\end{equation}
where $N_{\mathrm{CPU}}$ represents the number of volunteer CPU cores.
Given a certain number of CPU cores and having fixed $\tau_{p}$, the maximum search frequency $f_{\mathrm{max}}$ can be derived from the above equations to be
\begin{equation}
f_{\mathrm{max}}^{3} = f_{\mathrm{min}}^{3} + 3 \frac{\tau_{p} \, \, N_{\mathrm{CPU}}}{\tau_{\mathcal{F}} + \tau_{\mathrm{H}}} \frac{(1-\epsilon) \, \, \delta f}{\kappa \, \, N_{\dot{f}}},
\end{equation}
where $\kappa = {4\pi\left({v_d\over c}\right)^2 T_{\mathrm{seg}}^2} \mathcal{R}^{-2}$.
For the S5R3 and S5R5 runs, the nominal project duration was chosen to
be 6 months.  With the above choices, the
search frequency ranges for S5R3 and S5R5 turn out to be
respectively [50, 1\,200]~Hz and [50, 1\,190]~Hz.

\subsection{Accuracy of spindown model}
\label{subsec:spindownorder}

Let us briefly discuss the
second order spindown $\ddot{f}$ which, as mentioned previously, is
not a part of our search.  For our frequency resolution $\delta f$,
given by Eq.~\eqref{Eq:freqres}, and the full observation time
$T_{\mathrm{obs}}$, $\ddot{f}$ would have to be at least 
\begin{equation}
  \label{eq:fddotmin}
  \ddot{f}_{\rm min} = \delta f/T_{\mathrm{obs}}^2 \approx 1.3\times 10^{-20}\,\textrm{Hz~s}^{-2}
\end{equation}
in order for the signal to move by a single frequency bin over the full
observation time.  On the other hand, for a minimum spindown age
$\tau_{\rm min}$ at a frequency $f$, a useful estimate for the range
of $\ddot{f}$ that we should search is $f/\tau_{\rm min}^2 =
\dot{f}_{\rm max}^2/f$.  Thus, $\ddot{f}$ is potentially more
important at lower frequencies and higher spindown values.  Using the
maximum value of $|\dot{f}|$ in the search and the minimum value of
the search frequency gives us a value of $\ddot{f}$ that might be
potentially of astrophysical interest:
\begin{eqnarray}
  \label{eq:fddotrange} 
  \ddot{f}_{\rm ast} &=& 8\times 10^{-20} \,\textrm{Hz~s}{}^{-2} \nonumber \\
  &\times& \left( \frac{|\dot{f}|_{\rm max}}{2.0\times
      10^{-9}~\textrm{Hz/s} }\right)^2  \times \left( \frac{50~\textrm{Hz}}{f}\right) \,.
\end{eqnarray}
Comparing with the minimum value of $\ddot{f}$ obtained above, we see
that there is potentially a region in $(f,\dot{f})$ space where we
could improve our astrophysical detection efficiency by including
$\ddot{f}$; for our chosen range of $\dot{f}$ there is no effect of
$\ddot{f}$ above $\sim$$308\,$Hz.  It is important to note that the
calculation of Eq.~\eqref{eq:fddotmin} is too conservative because it
does not include any correlations between the phase evolution parameters.  On
the other hand, there is considerable uncertainty in the value of
$\ddot{f}_{\rm ast}$.  If the neutron star has a braking index $n \equiv
f\ddot{f}/\dot{f}^2$, then $\ddot{f}_{\rm ast}$ increases by a factor $n$.
If a star is spinning down purely due to gravitational wave emission,
then $n=5$.  On the other hand, for the Vela and Crab pulsars,
observed values of $n$ are $\sim 1.4$ and $\sim 2.5$ respectively~\cite{VelaBrakInd,CrabBrakInd}.  


The actual impact on our astrophysical reach is thus hard to quantify.
Let us consider as an example the extreme case when the spindown is
entirely due to gravitational radiation so that the braking index is
$n=5$. For this case, a conservative estimate of the part of the
spindown range $|\dot{f}|_{\rm cons}$ that is included in our search
is:
\begin{equation}
  \label{eq:fdotcons}
  \frac{-\dot{f}_{\rm cons}}{2\times 10^{-9}\,\textrm{Hz}~\textrm{s}^{-1}} \leq
  0.18 \sqrt{\frac{f}{50\,\textrm{Hz}}}\,.
\end{equation}
Here we have used Eqs.~\eqref{eq:fddotmin} and \eqref{eq:fddotrange}
modified by the braking index factor.  This corresponds to a minimum
spindown value of $-1.8$~nHz~s$^{-1}$ at the upper frequency of
1190\,Hz and $-3.6\times 10^{-10}\,$Hz~s$^{-1}$ at $50\,$Hz.

\section{\label{sec:S5R5pp} S5R5 Post-Processing}
As said earlier, roughly $10^{11}$ candidates from the S5R5 run
were returned to the Einstein@Home server. They were then transferred to the 6720-CPU-core Atlas Computing Cluster~\cite{atlas} at the Albert Einstein Institute in Hannover, and post-processed. The goal is to filter the set of $10^{11}$ candidates, excluding false candidate events. The post-processing strategy consists of the following steps:
\begin{itemize}
\item selection of 100 most significant candidates in 0.5~Hz frequency bands;
\item removal of known instrumental noise artifacts;
\item removal of unknown data artifacts through the $\mathcal{F}$-statistic consistency veto;
\item follow-up of the most significant candidates with S5R3 data;
\item fully-coherent follow-up of the surviving candidates.
\end{itemize}
The items outlined above are described in the next subsections.

\subsection{\label{sec:PostProcessing} Selecting the top candidates
  in frequency bands}

As is commonly done in CW searches, the results are examined
separately in fixed-size search frequency bands; here we choose to
perform the analysis in 0.5\,Hz bands.  As described earlier, in
designing the WUs, we have previously been led to break up the
frequency range in $\sim 20$~mHz bands and the sky has been
partitioned as well.  This was however done for purely technical
reasons to make the search on Einstein@Home feasible.  The choice of
frequency bands for the post-processing is based on different
requirements.  First, we would like the detector to have roughly
constant sensitivity within each frequency band.  Furthermore, as we
shall see, the search does not result in a convincing detection
candidate, and upper limits will be set over each of these frequency
bands.  Having a large number of very narrow bands would make the
calculation of the upper limit very computationally intensive.  The
choice of $0.5\,$Hz is a compromise between these two requirements.
This choice is in fact comparable to previous CW searches and will
make comparisons straightforward.  Finally, we note that all other
things being equal, having a larger frequency band will in principle
also lead to a decrease in sensitivity simply because of having a
larger number of templates.  This is however a relatively minor effect
in the present case.

For each of the 0.5\,Hz-wide frequency bands, we select the 100 most
significant candidates for further analysis, leading to a set of
228\,000 loudest S5R5 candidates. This choice
was dictated by the available computational and human
resources for the post-processing.  As will be illustrated in the
following, at the end of the automated post-processing procedure there
will remain of order 10 candidates which survive all selection
criteria. This is about the number that we can afford to follow-up
manually with further investigations.  As our follow-up procedures are
further automated and optimized, it will become possible to consider
lower thresholds and to inspect a correspondingly larger number of
candidates.

\begin{figure*}
\begin{center} 
\includegraphics[width=15cm, angle=0, clip]{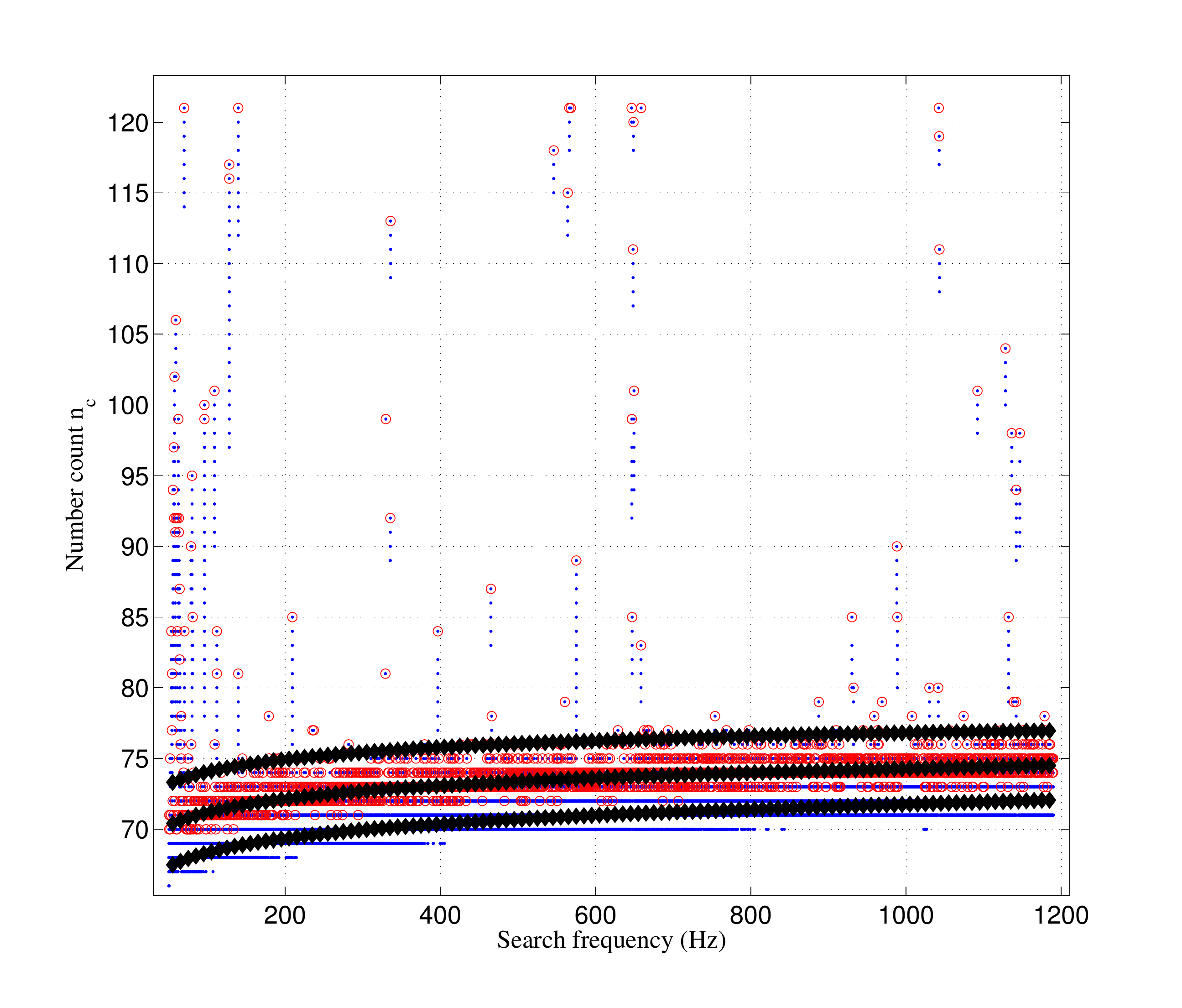} 
\caption{\label{Fig:NcFrS5R5R6} 
Number count of top 100 loudest candidates (blue dots) selected in 0.5~Hz-wide frequency bands as a function of the search frequency, across the entire S5R5 search frequency range. The loudest (most significant) candidate in every 0.5~Hz band is indicated by a red circle.  The expected values of the loudest candidates for Gaussian noise alone are shown by the central black curve. The lower and upper black curves show $\pm 3$ standard deviations from the expected value. }
\end{center}
\end{figure*}

The number count of such candidates is plotted in
Fig.~\ref{Fig:NcFrS5R5R6} as a function of the frequency.
For most bands, it is consistent with expectations.
The number count generally increases with
increasing frequency because the number of sky points
searched over increases (see
Eq.~(\ref{eq:sky-patchSize})) and the maximum expected value of a
random variable, over repeated trials, grows with the number of
independent trials.  Let us ignore the effect of the weights, and take
the Hough number count $n_c$ (defined in Eq.~(\ref{eq:nc})) to be an
integer random variable following a binomial distribution.  The
cumulative probability for obtaining $n_c$ or lower is
\begin{equation}
\label{Eq:cumulativeBinomial}
F(n_c)=\sum_{n=0}^{n_c} \left( {\begin{array}{*{20}c} N_{\mathrm{seg}} \\ n \\ \end{array}} \right) p_{\mathrm{seg}}^n~(1-p_{\mathrm{seg}})^{N_{\mathrm{seg}}-n}\,.
\end{equation}
The binomial parameters are
$N_{\mathrm{seg}}=121$ and $p_{\mathrm{seg}}=0.267$ (consistent with a
$2\mathcal{F}$ threshold of 5.2).
The probability $p_{\mathrm{max}}$ that the maximum number count is
$n_c$ over a set of $N_{\mathrm{trials}}$ independent trials is 
\begin{equation}
  \label{Eq:ProbMaxRepeatedBinomial}
  p_{\mathrm{max}}(n_c, N_{\mathrm{trials}} | N_{\mathrm{seg}},p_{\mathrm{seg}}) = 
  F(n_c)^{N_{\mathrm{trials}}} -F(n_c-1)^{N_{\mathrm{trials}}}.
\end{equation}
We compute Eq.~(\ref{Eq:ProbMaxRepeatedBinomial}) as a function of
$N_{\mathrm{trials}}$, which we take to be the number of templates
searched to cover $0.5$ Hz bands, i.e.\ $7.5\times 10^4$ frequency
values $\times$ 18 spindown values $\times$ 8\,444 Hough pixels per $\mathcal{F}$ sky point $\times$ $N_b$, the total number of  $\mathcal{F}$ sky points shown in Fig.~\ref{Fig:numSkyPoints}. 
The number of Hough pixels has been computed using Eq.~(\ref{eq:houghSkyRef}) with $m=0.3$ and $\wp =0.5$. Figure~\ref{Fig:NcFrS5R5R6} shows the expected value of the maximum (central black curve), computed using the probability function given by Eq.~(\ref{Eq:ProbMaxRepeatedBinomial}), superimposed on our measurements (red circles) and confirming that there is broad agreement, in most bands, between our results and the expectations for Gaussian noise.

\subsection{\label{sec:removeRSIL}Removing known data artifacts}

As a first step of the post-processing pipeline, we eliminated from
the list of top candidates any candidate whose frequency was too close
to that of either a known artifact or to the cleaned noise bands
described in section~\ref{subsec:s5r5basics}.  Specifically, we
discarded those candidates whose detection statistic could have been
constructed with contributions either from:
\begin{itemize}
\item data polluted in either of the two instruments by spurious
  disturbances; details of such detector disturbances are given in
  Appendix~\ref{sec:knownS5lines} and, in particular, a list of known
  spectral disturbances for the H and L instruments are listed in
  Tables~\ref{tab:HS5linesRS} and~\ref{tab:LS5linesRS}.
\item from fake noise that had been inserted by the cleaning process and
  hence could not host a CW signal (see
  Table~\ref{tab:whiteNoiseBandsH} in Sec.~\ref{subsec:s5r5basics}).
\end{itemize}
After this veto, about $25 \%$ of the candidates were eliminated from
the original set of 228\,000 loudest S5R5 candidates. More precisely,
a total of $172\,038$ S5R5 candidates survived this veto.
The bandwidth removed due to the lines listed in
Appendix~\ref{sec:knownS5lines} amounted to $\sim$$243\,$Hz; an
additional $27\,$Hz was removed due to the cleaned noise bands.

\subsection{The $\mathcal{F}$-statistic consistency veto}
\label{subsec:fstatconsistency}

We have thus far considered only known instrumental disturbances for
vetoing candidates.  However, we expect there to be more such
disturbances present in the data that have not yet been explicitly
identified.  The idea is to discriminate between disturbances in a
single detector and signals, which should produce consistent values of
the $\mathcal{F}$-statistic in both detectors~\cite{PaolaMAPpaper}.
We refer to this method as the $\mathcal{F}$-statistic consistency
veto.

For each of the 172\,038 S5R5 surviving candidates, the
single-detector and multi-detector $\mathcal{F}$-statistic was
computed for each of the 121 data segments and then averaged over the
segments.  We refer to these averaged $2\mathcal{F}$ values as
$\left<2\mathcal{F}_{\mathrm{H}} \right>$ and
$\left<2\mathcal{F}_{\mathrm{L}} \right>$ for the H and L detectors
respectively, and $\left<2\mathcal{F}_{\mathrm{HL}} \right>$ for the
coherent combination of the data from the two detectors. Candidates
were discarded if either $\left<2\mathcal{F}_{\mathrm{H}} \right>$ or
$\left<2\mathcal{F}_{\mathrm{L}} \right>$ were greater than
$\left<2\mathcal{F}_{\mathrm{HL}} \right>$. Using this veto, a small
fraction ($4.1\%$) of candidates was eliminated, leaving $164\,971$
surviving S5R5 candidates.

\begin{figure}[h]
\begin{center}
\includegraphics[width=8.5cm, angle=0, clip]{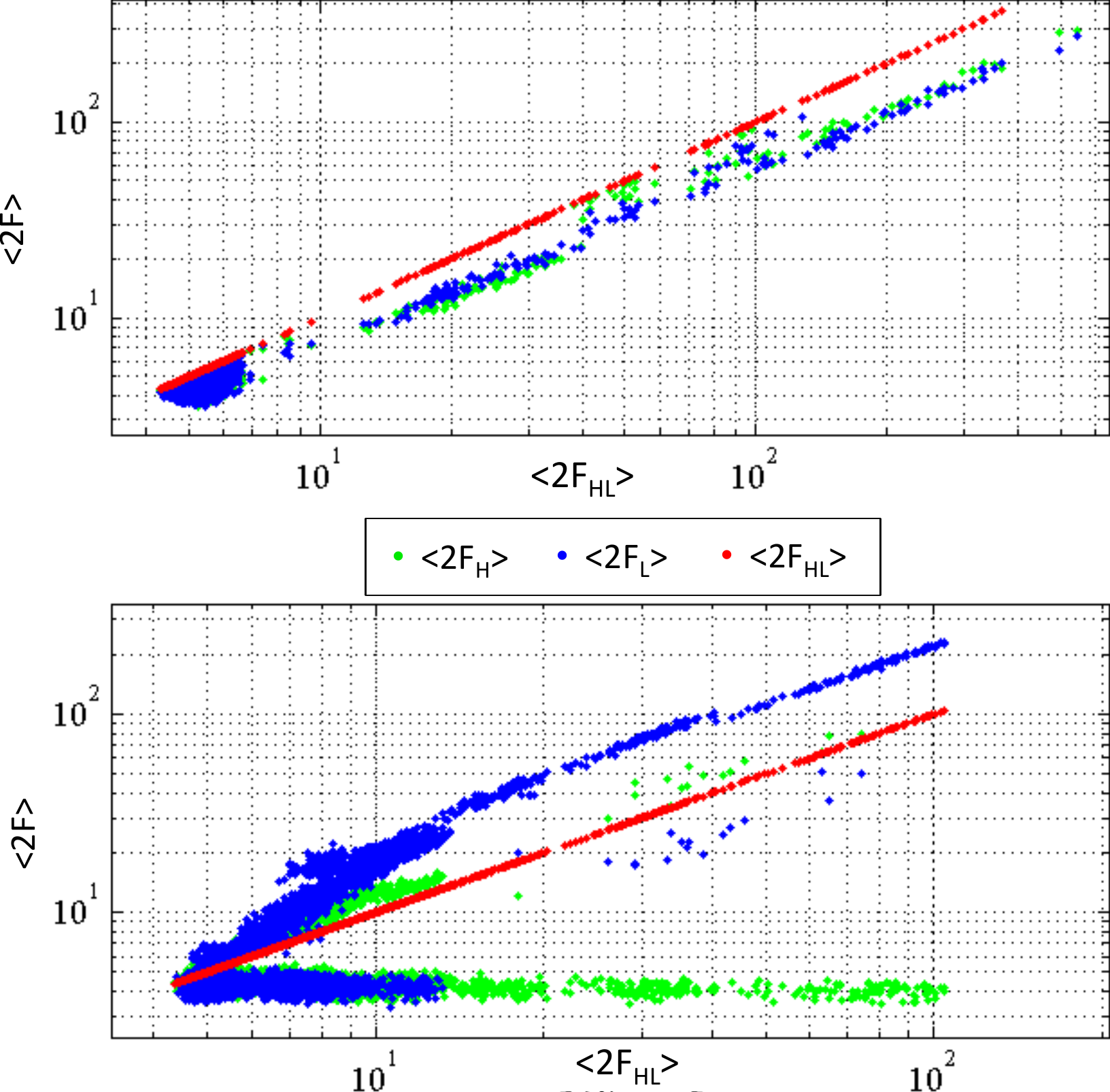}
\caption{\label{Fig:AveTwoFS5R5R6survVet} Values of $2\mathcal{F}$ averaged over $121$ data segments for the single-detector case, $\left<2\mathcal{F}_{\mathrm{H}} \right>$ (green dots), $\left<2\mathcal{F}_{\mathrm{L}} \right>$ (blue dots) and the multi-detector case, $\left<2\mathcal{F}_{\mathrm{HL}} \right>$ (red dots), against those for the combined multi-detector statistic. The top (bottom) plot shows such values for $164\,971$ ($7\,067$) surviving (vetoed) S5R5 candidates such that $\left<2\mathcal{F}_{\mathrm{H}} \right>$ and (or) $\left<2\mathcal{F}_{\mathrm{L}} \right>$ is less (greater) than $\left<2\mathcal{F}_{\mathrm{HL}} \right>$.}
\end{center}
\end{figure}

The impact of this veto is limited in this case due to the prior
removal of the bulk of instrumental artifacts. However, the
$\mathcal{F}$-statistic consistency veto represents an efficient
method to remove disturbances that clearly stand out of the noise in
the absence of independent instrumental
evidence. Figure~\ref{Fig:AveTwoFS5R5R6survVet} shows the average $2
\mathcal{F}$-values for $164\,971$ surviving (top plot) and $7\,067$
vetoed (bottom plot) S5R5 candidates as a function of the
multi-detector average $2 \mathcal{F}$-values. By construction, all
the surviving candidates in the top panel of
Fig.~\ref{Fig:AveTwoFS5R5R6survVet} lie below the red dotted line,
which defines the veto criterion.

\subsection{Distribution of candidates}
\label{subsec:distribution}

We have now applied all of our vetoes that try to remove instrumental
artifacts.
While there will of course remain other low amplitude
instrumental spectral lines and hardware signal injections (described
later in Sec.~\ref{sec:HWInjS5R5}), we now need to deal with the
possibility that, say, even Gaussian noise can mimic a signal in some
cases.  All remaining candidates will need to undergo detailed
individual inspection and we will only be able to afford this for a
few candidates.  As our follow-up techniques become more refined,
optimized and automated, we will be able to improve this part of the
pipeline and dig deeper into the noise.

\begin{figure}[h]
\begin{center}
\includegraphics[width=8.5cm, angle=0, clip]{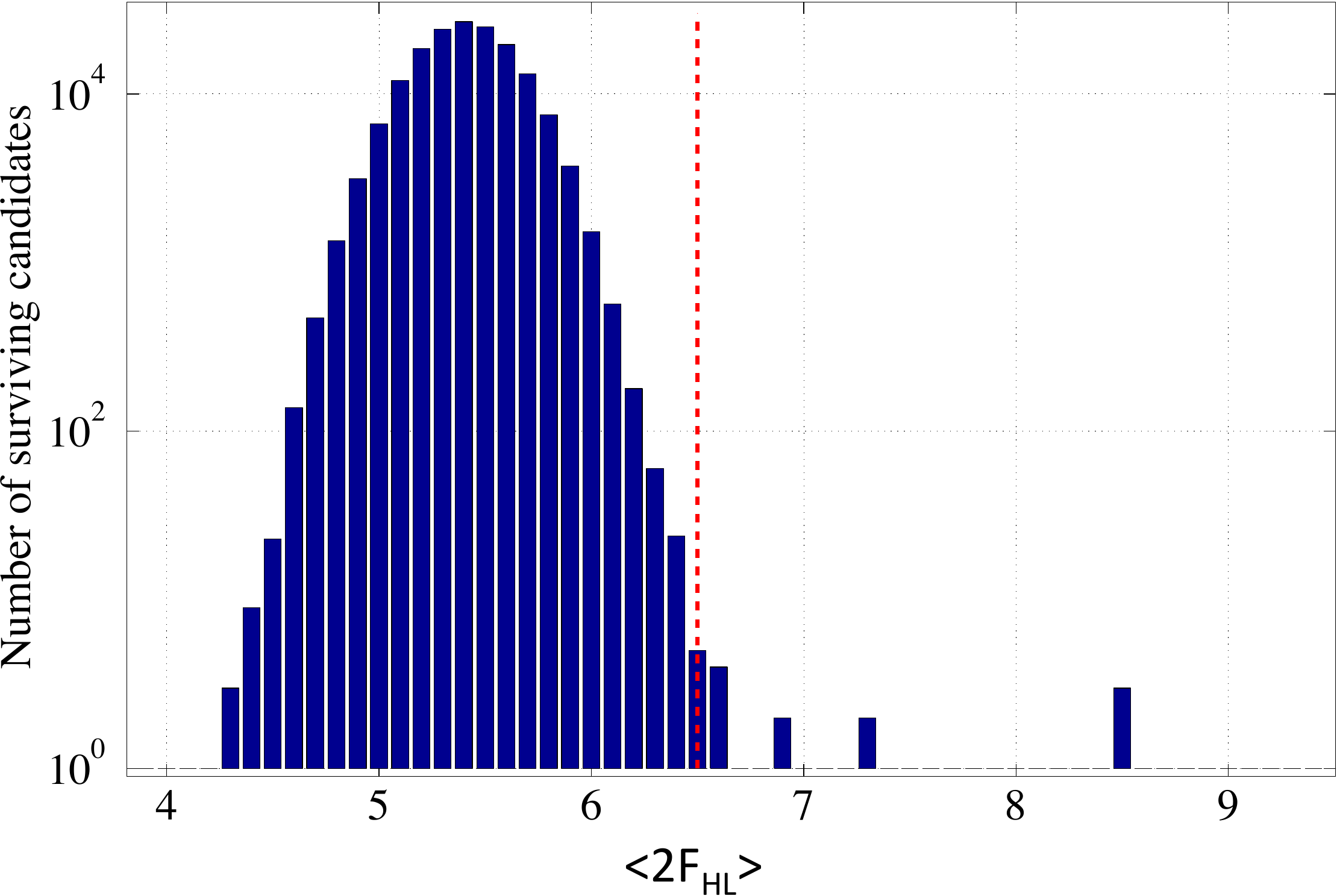}  
\caption{\label{Fig:averageFstatHist} Histogram of average multi-detector $2 \mathcal{F}$-values for 164\,971 S5R5 surviving candidates. The red dotted line draws the boundaries of the bulk of candidates due to instrumental noise, and corresponds to the threshold $\left<2\mathcal{F}_{\mathrm{HL}} \right> = 6.5$.} 
\end{center}
\end{figure}

Figure~\ref{Fig:averageFstatHist} shows the histogram of
$\left<2\mathcal{F}_{\mathrm{HL}} \right>$ for the 164\,971 surviving
candidates up to $\left<2\mathcal{F}_{\mathrm{HL}} \right>$-values of
$9$.  The distribution actually extends up to
$\left<2\mathcal{F}_{\mathrm{HL}} \right> \sim 542.8$ but we show only
the low $\left<2\mathcal{F}_{\mathrm{HL}} \right>$-values distribution
in order to explain our next choice of threshold. There are $166$
candidates with $\left<2\mathcal{F}_{\mathrm{HL}} \right> > 9$ and
they are all clustered at the two frequencies of $\sim$$108.9$ Hz and
$\sim$$575.2$ Hz, corresponding to two simulated signals injected in the data
stream, as discussed in Sec.~\ref{sec:HWInjS5R5}.  The region
$\left<2\mathcal{F}_{\mathrm{HL}} \right> < 6.5$ contains well over
$99\%$ of the candidates and, as seen in
Figure~\ref{Fig:averageFstatHist}, below $6.5$ the density of
candidates increases very sharply.  We will take $6.5$ as a threshold
for the next step in our follow-up procedure.  

\begin{figure}
\begin{center}
\includegraphics[width=9cm, angle=0, clip]{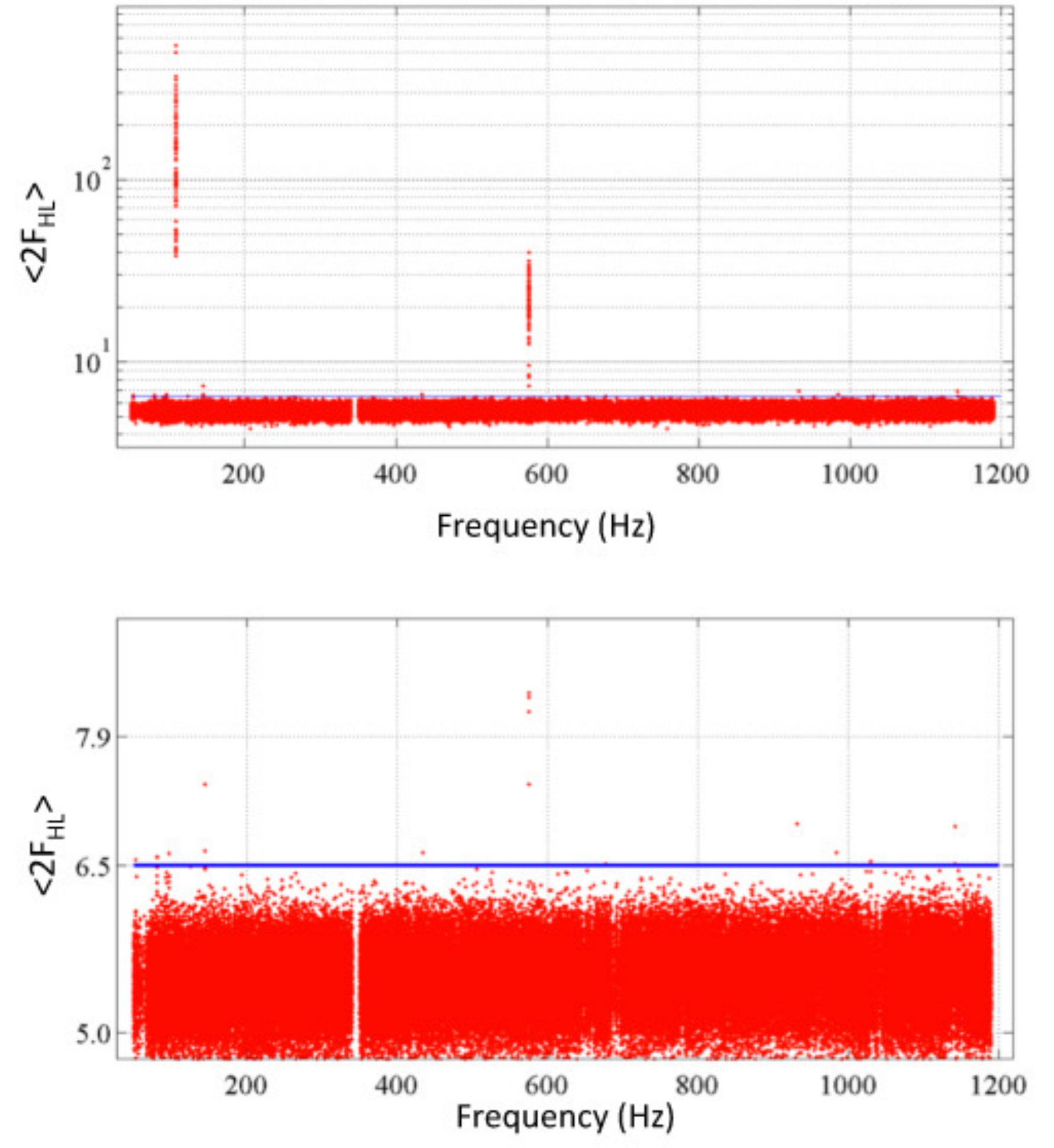}
\caption{\label{Fig:AverageFstatVsFreq} Average multi-detector $2\mathcal{F}$-values of the S5R5 candidates surviving the $\mathcal{F}$-statistic consistency veto as a function of the frequency. The horizontal line represents the threshold value of $\left<2\mathcal{F}_{\mathrm{HL}} \right>=6.5$. The bottom plot shows the top plot in the region close to the threshold.}
\end{center}
\end{figure}

The number of candidates expected to survive the $6.5$ cut on
$\left<2\mathcal{F}_{\mathrm{HL}} \right>$ in fixed $0.5\,$Hz
bands increases with frequency because the number of sky locations
searched scales with the square of the searched frequency. In order to
compute the false alarm probability corresponding to this threshold in
different frequency bands, note that in the absence of a signal, the
value of $2\mathcal{F}_\mathrm{HL}$ in the $i^{\mathrm{th}}$ segment,
$2\mathcal{F}_\mathrm{HL}^{(i)}$, follows a $\chi^2$ distribution with 4
degrees of freedom.  Furthermore, since
\begin{equation}
  \left<2\mathcal{F}_{\mathrm{HL}} \right> \times 121 =
  \sum_{i=1}^{121} 2\mathcal{F}_{\mathrm{HL}}^{(i)}\,,
\end{equation}
it is clear that $\left<2\mathcal{F}_{\mathrm{HL}} \right> \times 121$
is a $\chi^{2}$ random variable with $(4\times121)$ degrees of freedom. The false alarm probability corresponding to a threshold at
$(6.5\times 121)$ for such a random variable is $\sim$$10^{-16}$. This
corresponds to expected false alarm rates of about $0.1\%$, $0.6\%$,
$2.6\%$ and $10\%$ for searches in Gaussian noise over over $0.5$ Hz bands at $100$~Hz,
$250$~Hz, $500$~Hz and $1\,000$~Hz, respectively, considering the
number of independent trials given by the number of searched templates
in the respective bands.  Disregarding the non-Gaussian line features evident in Fig.~\ref{Fig:AverageFstatVsFreq}, 
the ratio of the number of candidates observed above the $6.5$ threshold at lower frequencies (say below $800$ Hz) to that at higher frequencies (say above $800$ Hz) is not inconsistent\footnote{Due to the low number statistics it is hard to make a sharper statement.} with the ratios of the false alarm rates computed above. We note that the false alarm probability given above overestimates the number of expected candidates above the $6.5$ threshold because that threshold is not the only cut applied to the data. The previous cuts, discussed in the preceding sections, lower the actual false alarm probability of the surviving candidates.

\begin{figure}[h]
\begin{center}
\includegraphics[width=8.5cm, angle=0, clip]{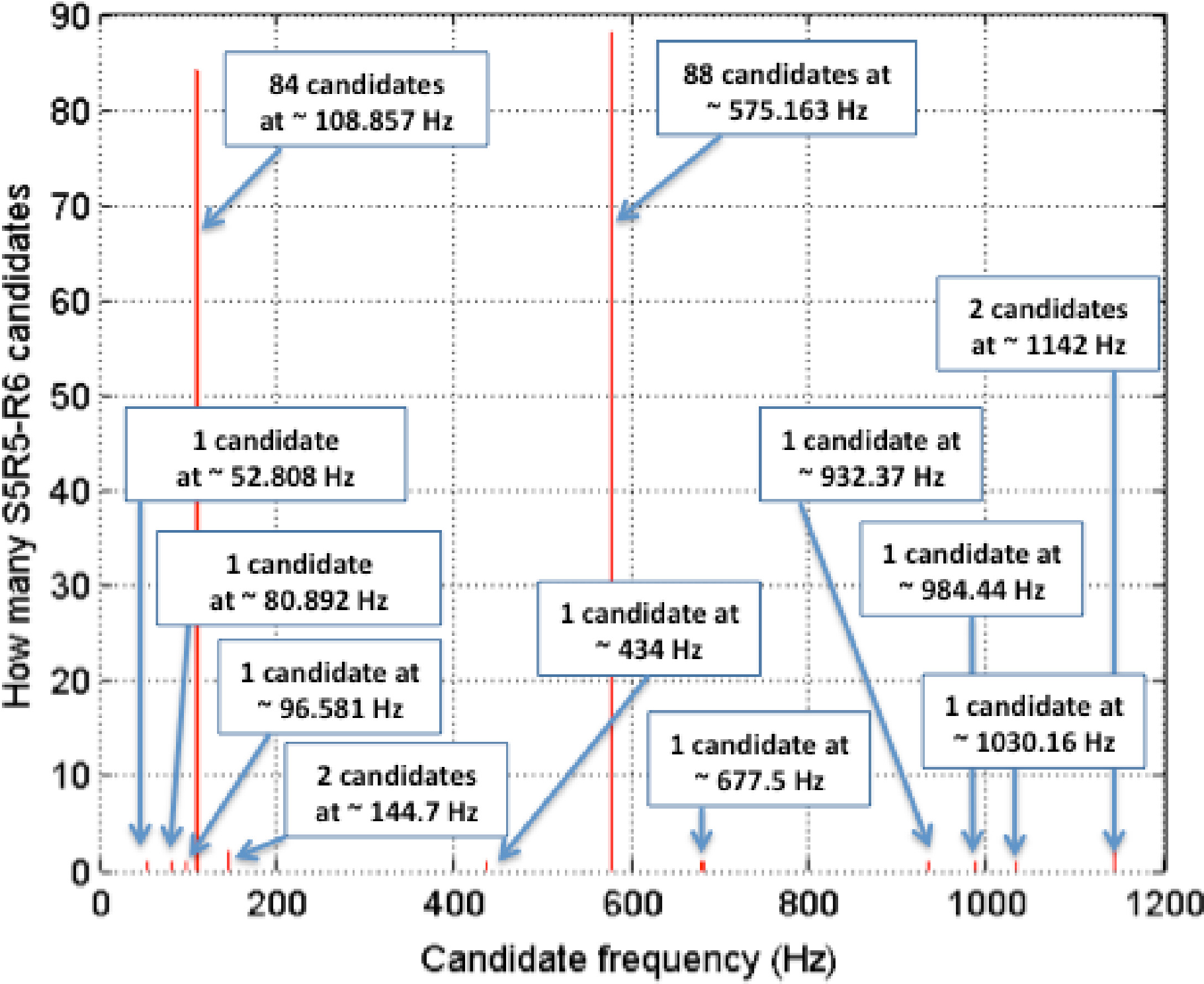} 
\caption{\label{Fig:S5R5FUcands} Histogram of the S5R5 candidates that have been further investigated through deep follow-up study with the S5R3 data set.}
\end{center}
\end{figure}

There are 184 remaining S5R5 candidates, for which
$\left<2\mathcal{F}_{\mathrm{HL}} \right> > 6.5$ and they are shown in
Fig.~\ref{Fig:S5R5FUcands} as a function of the frequency. They are
clustered at twelve frequencies and only the most significant
candidate from every cluster has been followed up.

\subsection{Following up candidates with S5R3 data} 
\label{subsec:s5r3followup}

For the next step in our post-processing pipeline, recall that our
underlying signal model of Eq.~(\ref{eq:h+x}) assumes that the signal
is long-lived; thus its amplitude is constant in time and its
intrinsic frequency evolves smoothly according to Eq.~(\ref{eq:fhat})
with a constant spindown.
This is an idealized model: although pulsars are the most stable
clocks in the Universe, neutron stars are nonetheless known to glitch,
to be perturbed by external agents, and in some cases to be affected
by significant timing noise.
Furthermore, for sufficiently long observation times, the spin-down 
evolution model that we use here, including only the first spindown order, may not be adequate to 
describe the actual signal model (see also the discussion in Section \ref{subsec:spindownorder}). 
However, since the data set used in the S5R3
run ends just about a week before the S5R5 data set, it is reasonable
to assume that any putative signals should be present in both data
sets.  Moreover, the average noise floor level in the detectors
turns out to be approximately stable between S5R3 and S5R5.  Thus, we
might expect that a detectable signal should be visible in both
searches.  

The next follow-up step for each candidate then consists of a
hierarchical search carried out on the same WU as done for S5R5 (i.e.\
over the same parameter space), but using the S5R3 data set. The
closest\footnote{The distance used to judge closeness between
  candidates is a Euclidean distance expressed in bins in the four
  dimensions $(f,\dot{f},\alpha,\delta)$.} candidate to the original one
from such a search was identified and the value of its detection
statistic was compared with what one would expect if the S5R5
candidate were due to a signal.  In particular, the expected number
counts in S5R3 and S5R5 should be related according to
\begin{equation}
  E[N_c^{\mathrm{S5R3}}]=N_c^{\mathrm{S5R5}}\times 84 /121 \,,
\label{Eq:expectedS5R3nc} 
\end{equation}
where $84$ is the number of data segments used in S5R3, as we shall see in Appendix~\ref{sec:S5R3pp}.
Possible reasons for this not to be a good approximation would be if
the detector noise floor were to vary significantly between S5R3 and
S5R5, or the fact that the relative geometry between the detector and
source varies in time.  We have already remarked that the noise floor
is, on the average, stable between S5R3 and S5R5.  Furthermore, albeit
in each segment the expected $2\mathcal{F}$-values for a given signal might be
different, if we average this expected value over many non-overlapping
segments we expect this to converge within a few tens of segments;
recall here that each segment spans a duration of 25 hours, while the
antenna pattern function of the detectors has a periodicity of 24
hours.  Thus, it is reasonable to assume that
Eq.~(\ref{Eq:expectedS5R3nc}) is valid.

Candidates for which the measured value $N_c^{\mathrm{S5R3}}$ was more
than $3 \, \sigma$ less significant than the expected
$E[N_c^{\mathrm{S5R3}}]$ were discarded as not being consistent with a
CW signal, where $\sigma$ was computed using Eq.~(\ref{eq:sigmaW}).
As shown in Table~\ref{Tab:FUs5r5r6cands}, two
candidates were discarded by this follow-up test, at $\sim$$80.9$~Hz
and $\sim$$108.9$~Hz.  However, the second of these, as well as the
candidates at $\sim$$52.8$~Hz and $\sim$$575.2$~Hz,
represent three simulated signals
injected only part of the time during S5, as discussed in
Sec.~\ref{sec:HWInjS5R5}. 

\begin{table}
\begin{tabular}{lcccc} \hline \hline 
Frequency~(Hz)  & CR	&  $N_c^{\mathrm{S5R5}}$	& $E[N_c^{\mathrm{S5R3}}]$ 	& ${N_c}^{\mathrm{S5R3}}$  \\
 \hline 
52.808297682 &  7.7   & 70   & 49 & 37 \\
80.891816186 &  10.1 & 82   &  57 & 34 \\
96.581099597 & 8.1 & 72   &  50 & 37 \\
108.85716501 & 13.9 &  101  & 70 & 55 \\
144.74321811 & 7.9 & 71   & 49 & 42 \\
434.09886421 & 7.7	& 70    & 49 & 46 \\
575.16357663 & 11.6 & 89  & 62 & 53 \\
677.47882796	& 8.0	& 71 & 49 & 46 \\
932.36948703	&   8.5 &  74  &  51 	& 46 \\
984.44286823	& 8.2 & 73 &  51 & 47 \\
1030.1650892	& 9.1 &  77  & 53 &  53 \\
1141.9926498 	& 9.4 & 78  & 54 &  48 \\
\hline\hline
\end{tabular}
\caption{\label{Tab:FUs5r5r6cands} S5R5 candidates followed-up using the S5R3 data set. The different columns represent: (I) the candidate frequency (shown also in Fig.~\ref{Fig:S5R5FUcands}), (II) its significance (CR), (III) the S5R5 number count ($N_c^{\mathrm{S5R5}}$), (V) the expected and (VI) observed number count values after following the candidate up ($E[N_c^{\mathrm{S5R3}}]$ and $N_c^{\mathrm{S5R3}}$, respectively).}
\end{table}
As we can see from Fig.~\ref{Fig:S5R5FUcands}, the bulk of the
candidates arise from the strong hardware-injected pulsar~2 and~3 signals (see
Sec.~\ref{sec:HWInjS5R5}). In particular, 84 and 88 candidates are
clustered near the frequencies of these two injected signals
respectively.

\subsection{Fully-coherent follow-up}

Excluding the hardware-injected simulated signals,
the post-processing up to this point has left us with 8 surviving
candidates.  These have been significant enough to pass our thresholds
and have not been clearly identified as instrumental artifacts, or eliminated by
inconsistency between the H and L detectors, or by the follow-up with the S5R3 data set.  We
therefore need to consider other more sensitive methods.  If these
candidates are real signals, then their SNRs and
significance should increase if the parameter space grids are made
finer, or as the coherent integration time becomes larger.  

We use a three-step procedure consisting of a grid-based semi-coherent
Hough search, followed by a semi-coherent and a final fully-coherent
$\mathcal{F}$-statistic search, using the Mesh Adaptive Direct Search
(MADS) algorithm for constrained optimization. The reference
implementation of the MADS algorithm is publicly available through the
NOMAD library \cite{Audet04meshadaptive, LeDigabel2011}. Hence, in the
following, we refer to such searches simply as NOMAD
searches. Contrary to the traditional grid-based methods, a mesh
adaptive search constructs the trial points as the search evolves
aiming to find the maximum of the statistic.  

The three steps of the follow-up procedure are the following:
\begin{enumerate}
\item Re-run the Hough search around a given candidate, but with
  a finer grid to reduce the mismatch with a putative signal.  The
  search region includes 5 frequency bins on either side around the
  candidate and 16 neighboring coarse sky-grid points.  The fine Hough
  sky-grid is refined by a factor of 2 in each direction by using
  $\wp=1$ (see Eq.~\ref{eq:HoughPixRes}) instead of $0.5$ as in the
  original search.  Furthermore, we refine the coarse $\dot{f}$ grid
  spacing of Eq.~\eqref{eq:dfdotok} by a factor
  $N_{\mathrm{seg}}=121$~\cite{HierarchP2}.

\item The loudest candidate from the first step is used as a starting
  point for the semi-coherent $\mathcal{F}$-statistic NOMAD
  optimization.  The detection statistic in this step is the sum
  of the $\mathcal{F}$-statistic values from each segment.
  This search has been performed in
  a fixed parameter space box around the starting point.  The
  dimensions of the box are $\Delta f = 10\time10^{-4}$ Hz, $\Delta
  \alpha = 0.10$ rad, $\Delta\delta = 0.24$ rad and
  $\Delta\dot{f}=10^{-10}$ Hz~s$^{-1}$. The loudest candidate found in this
  semi-coherent $\mathcal{F}$-statistic NOMAD search is passed on to
  the next step.  

\item In the third step, the loudest candidate from the previous step
  is used as a starting point for the fully-coherent
  $\mathcal{F}$-statistic NOMAD search. This search spans the entire
  duration of the S5R5 data set and has been carried out in a
  parameter space box defined by using the diagonal elements of the
  inverse Fisher matrix in each dimension around the starting
  point. These elements are described by Eq.~(16) in~\cite{MirosProc} and have been computed from the inverse of the
  semi-coherent parameter space metric computed at the candidate
  point, re-scaled by the measured SNR at the same point; for more
  details we refer the reader to~\cite{MirosProc}.

\end{enumerate}

In both the semi-coherent and fully-coherent NOMAD searches, we ran
multiple instances of the algorithm iterating over the mesh coarsening
exponent using both deterministic \cite{Audet04meshadaptive} and
stochastic~\cite{AbramsonADD09} methods for the choice of search
directions. Based on Monte-Carlo studies, the false-dismissal
probability of the follow-up procedure is found to be less than 10\%.

As said earlier, in the presence of a real signal we expect the significance of a
candidate to increase as the template grid becomes finer because there
will be a template with a smaller mismatch with respect to the real
signal. At that template the signature of the signal should be more
evident and all the consistency tests should continue to hold. If the
candidate signal detected on the finer grid does not pass a
consistency test, this indicates that it is not behaving as we would
expect from the signals that we are targeting.
The candidates at $\sim$$96.6$~Hz, $\sim$$144.7$~Hz, $\sim$$932.4$~Hz,
$\sim$$1030.2$~Hz and $\sim$$1142$~Hz 
fail a multi-detector versus single-detector
$\mathcal{F}$-statistic consistency test (see 
Sec.~\ref{subsec:fstatconsistency}) after performing the semi-coherent
NOMAD search;
therefore, they cannot be considered defensible CW signals.
Moreover, line artifacts appear in the average power spectrum of S5 H data at\
$\sim$$932.4$~Hz, $\sim$$1030.2$~Hz and $\sim$$1142$~Hz.

The remaining candidates, namely at $\sim$$434.1$~Hz, $\sim$$677.5$~Hz and $\sim$$984.4$~Hz,
survive the $\mathcal{F}$-statistic consistency test on the
finer grid and are followed-up with the fully-coherent
$\mathcal{F}$-statistic NOMAD search.
However, for each of them, the maximum value of the detection statistic over the
parameter space searched is much lower than would be expected based on
the original candidate parameters, and in fact is 
consistent with the expectation for Gaussian noise. Hence, also these
three candidates do not survive a more sensitive inspection and cannot
be considered viable detection candidates.  Thus, we see that all the
candidates listed in Table~\ref{Tab:FUs5r5r6cands} are inconsistent
with the properties of a true CW signal.

\section{\label{sec:SensEst} Upper limit estimation} 

The analysis of the Einstein@Home searches presented here has not
identified any convincing CW signal. Hence, we proceed to set upper limits on
the maximum intrinsic gravitational wave strain $h_0$ that is
consistent with our observations for a population of CW signals
described by Eq.~(\ref{Eq:h(t)Signal}), from random positions in the
sky, in the gravitational wave frequency range $[50.5, 1\,190]$~Hz, and with
spindown values in the range of $\sim [-20,
1.1]~\times~10^{-10}$~Hz~s$^{-1}$.  The nuisance parameters $\cos\iota$,
$\phi_0$, and $\psi$ are assumed to be uniformly
distributed. 
As commonly done in all-sky, all-frequency searches, the upper limits
are given in different frequency sub-bands and here we have chosen
these to be 0.5 Hz wide. Each upper limit is based on the most
significant event from the S5R5 search in its 0.5 Hz
band.

\subsection{Monte-Carlo upper-limit estimates}

Our procedure for setting upper limits uses Monte-Carlo signal
injection studies using the same search and post-processing pipeline (except for the S5R3 and fully-coherent follow-ups)
that we have described above.  In every $0.5\,$Hz band, our goal is to
find the value of $h_0$ (denoted $h_0^{90\%}$) such that $90\%$ of the
signal injections at this amplitude would be recovered by our search
and are more significant than the most significant candidate from our
actual search in that band.  We can thus exclude, with $90\%$
confidence, the existence of sources (from our specific population)
that have an amplitude $h_0 > h_0^{90\%}$.

In order to estimate $h_0^{90\%}$, for each injection at a randomly
chosen parameter space point, a hierarchical search is performed over
a small parameter space region, which consists of:
\begin{itemize}
\item a 0.8~mHz frequency band centered at the S5R5 frequency
  grid-point closest to the randomly chosen source frequency;
\item 4 spindown values around the S5R5 frequency derivative
  grid-point closest to the randomly chosen spindown;
\item a sky-patch consisting of 10 S5R5 coarse sky-grid-points
  closest to the randomly chosen sky location. 
\end{itemize}
At the end of each hierarchical search, the most significant candidate
is selected and post-processed as described in the previous
section. The vetoes for excluding known instrumental lines are not
required because we have already excluded them from the upper-limit
analysis.  We first check if this candidate is significant enough to
be part of the 100 Einstein@Home loudest candidates originally
selected in its corresponding 0.5~Hz band. Then we perform the
$\mathcal{F}$-statistic consistency check and finally we compare the
computed average multi-interferometer $2\mathcal{F}$-value
($\left<2\mathcal{F}_{\mathrm{HL}}^{\mathrm{Cand}}\right>$) with the maximum
$\left<2\mathcal{F}_{\mathrm{HL}} \right>$ value we have in the
corresponding 0.5~Hz band. If
$\left<2\mathcal{F}_{\mathrm{HL}}^{\mathrm{Cand}}\right>$ is greater than
the maximum $\left<2\mathcal{F}_{\mathrm{HL}} \right>$,
then the simulated source is considered to be
recovered and more significant than the most significant
candidate of the search.  The confidence level is defined as
$C={n_{\rm rec}/ n_{\rm tot}}$, where $n_{\rm rec}$ is the number of
recovered candidates, and $n_{\rm tot}=100$ is the total number of
injections performed.

\begin{figure}[h]
\begin{center}
\includegraphics[width=9cm, angle=0, clip]{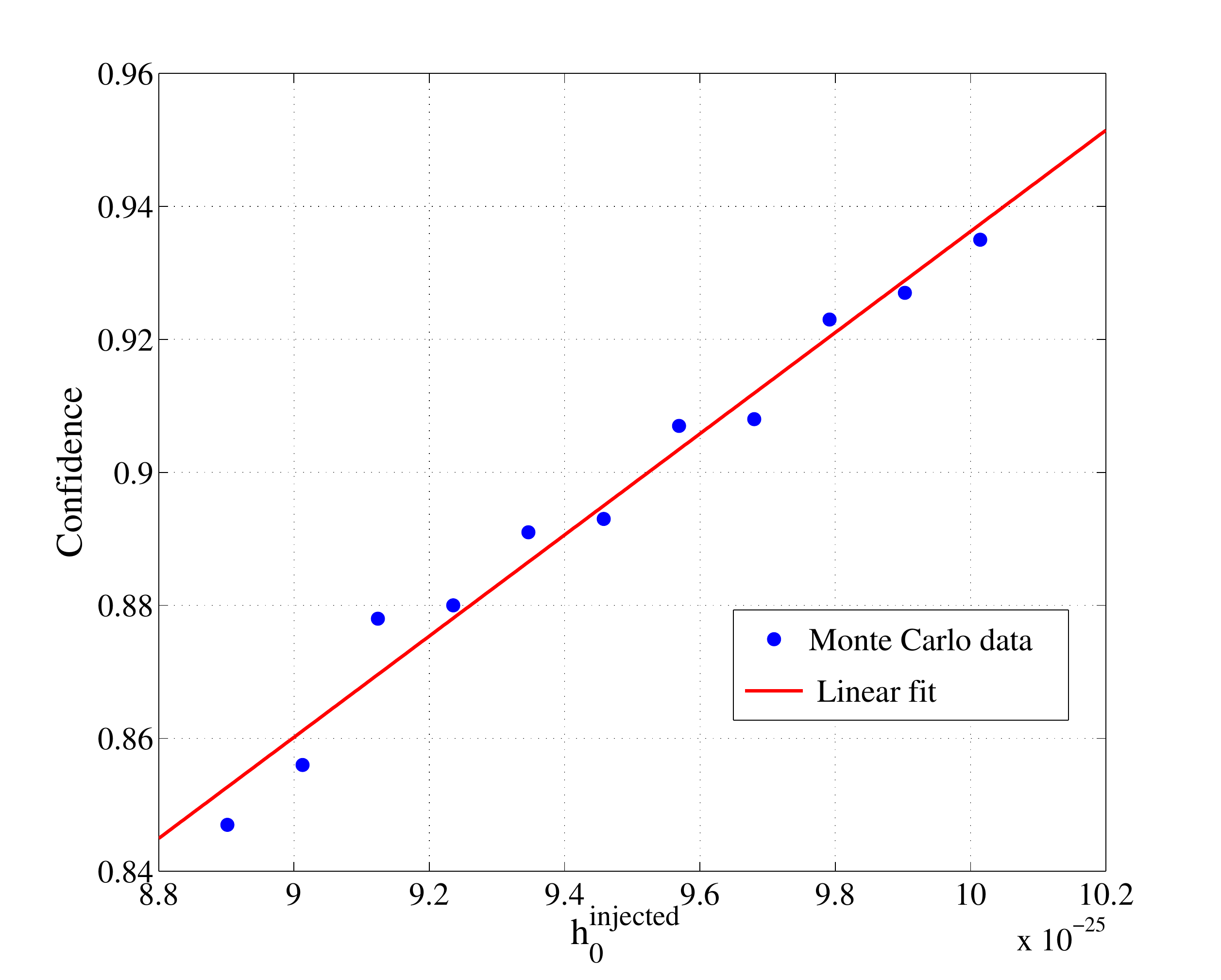}
\caption{\label{Fig:Cversush0injected} Confidence versus the injected $h_0$ values for sets of 1\,000 injections in the band $[216, 216.5]$~Hz. This plot illustrates how the uncertainty on the confidence level affects the uncertainty on the value of $h_0^{90\%}$.}
\end{center}
\end{figure}

After some preliminary tuning to determine a range of $h_0$ values
close to the $90\%$ confidence level, we use an iterative procedure to
determine the confidence as a function of the injected population
$h_0$ until we hit a confidence value close to $90\%$, within the
expected $1~\sigma$ fluctuations. Since we use 100 injections, from a
binomial statistic we estimate the $1~\sigma$ fluctuation to be $3\%$
and hence we associate the $h_0^{90\%}$ value to any measured
confidence in the range $87\%-93\%$. The $3\%$ uncertainty in
confidence translates to an uncertainty in $h_0^{90\%}$ smaller than
$5\%$, as can be seen from Fig.~\ref{Fig:Cversush0injected}, which
shows a typical confidence versus injected $h_0$ behavior. Each point
in Fig.~\ref{Fig:Cversush0injected} was derived with 1\,000 injections
and hence is affected by fluctuations smaller than
$1\%$. 

\begin{figure*}[]
\begin{center}
\includegraphics[width=18cm, angle=0, clip]{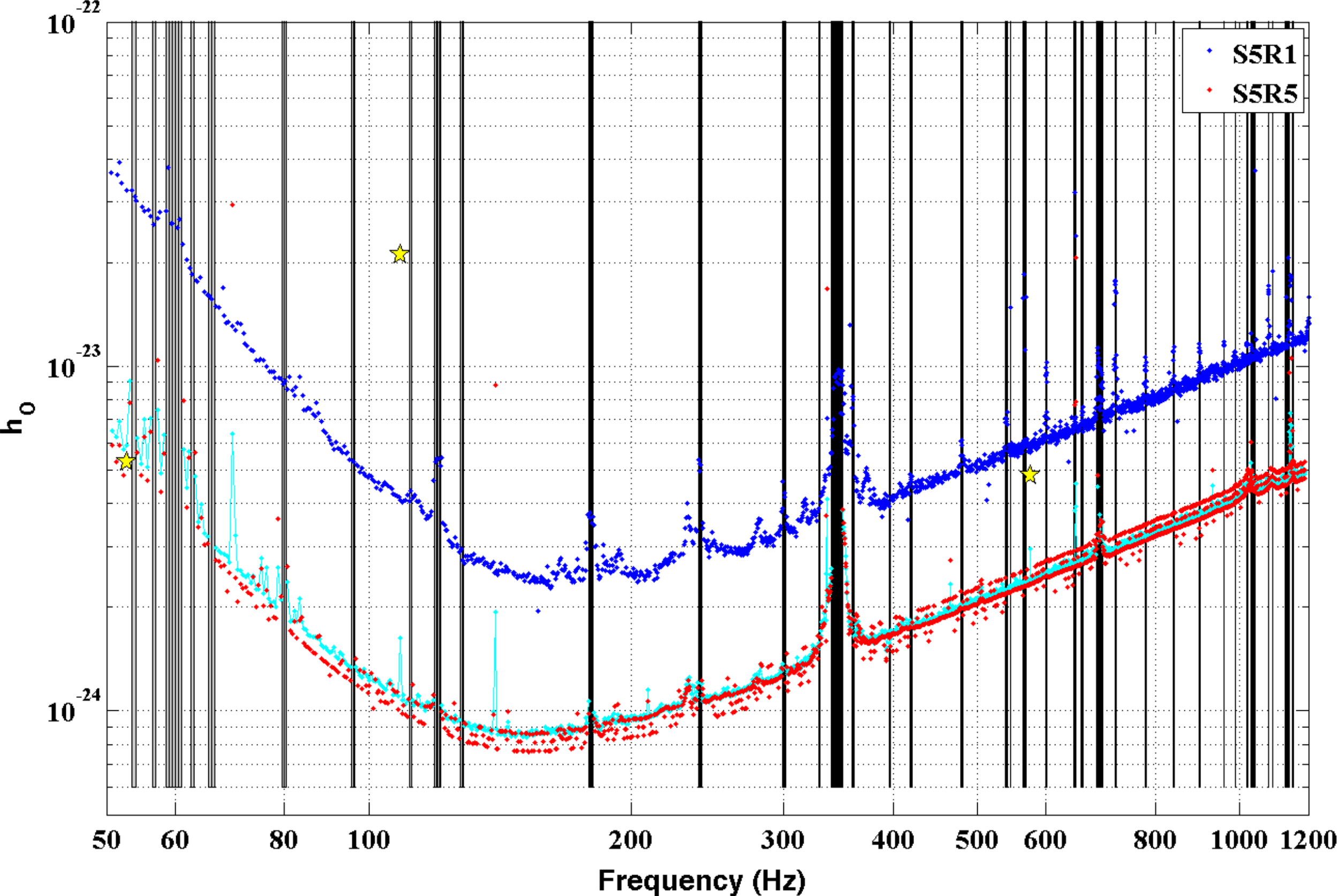}
\caption{\label{Fig:SensEstimS5R5R6} Upper limits for the S5R5 Einstein@Home
search (red dots) as well as the previous Einstein@Home search, called S5R1,
which used early S5 data (blue dots)~\cite{EatHS5R1}.
The three stars correspond to hardware-injected simulated pulsars which were recovered in the S5R5 search.
The curves represent the source strain amplitude $h_{0}$ at which 90\% of simulated signals would be detected. The vertical bars represent 156 half-Hz frequency bands contaminated by instrumental disturbances for which no upper limits are provided. The upper limits for the 0.5~Hz-wide bands starting at 69.5~Hz, 139.5~Hz, 335.5~Hz and 648~Hz are fairly high due to significant \textit{partial} contamination in these bands by lines listed in Tables~\ref{tab:HS5linesRS} and~\ref{tab:LS5linesRS}. Note that the broadness of the red curve is due to the 5\% steps used to vary the injected population $h_0$ values in the Monte-Carlo signal software-injections until a confidence value close to $90\%$ is reached.
In addition, less than 1/4 of the spectral range shown was excluded in many narrow bands because of known instrumental artifacts, as described in Sec.~\ref{sec:removeRSIL}.
 The cyan curve shows the predicted $h_0^{90\%}$ upper limits according to Eq.~(\ref{eq:AnalyticalFitULs}).}
\end{center}
\end{figure*}

The lower (red) curve in Fig.~\ref{Fig:SensEstimS5R5R6} shows the resulting upper limits as a function of the frequency. The upper (blue) curve shows the upper limit values from the previous Einstein@Home search in early S5 data~\cite{EatHS5R1}. The current upper limit values are about a factor 3 more constraining than the previous Einstein@Home ones. In particular, the most constraining upper limit falls in the 0.5~Hz-wide band at 152.5~Hz, where we can exclude the presence of signals with $h_0$ greater than $7.6~\times~10^{-25} $. The three stars shown in Fig.~\ref{Fig:SensEstimS5R5R6} correspond to the simulated pulsars~2,~3 and~5, i.e.\ the hardware injections recovered in the S5R5 search (discussed in Sec.~\ref{sec:HWInjS5R5}). 

The numerical data for the plot in Fig.~\ref{Fig:SensEstimS5R5R6} can be obtained separately \footnote{See Supplemental Material at \textcolor{red}{[Replace PRD link instead than ID LIGO-P1200026]} for numerical values of upper limits.}.
A conservative estimate of the overall uncertainty on the $h_{0}^{90\%}$ values shown in
Fig.~\ref{Fig:SensEstimS5R5R6} is $15\%$, having added to the
$1~\sigma$ statistical upper limit estimation procedure uncertainty
the $10\%$ amplitude calibration uncertainties for the data used in
this Einstein@Home run~\cite{Calib}.

As we have excluded from the search those frequency bands hosting
spectral artifacts (Tables~\ref{tab:HS5linesRS}
and~\ref{tab:LS5linesRS}) and the cleaned noise bands
(Table~\ref{tab:whiteNoiseBandsH}), we therefore also exclude these
frequency intervals from the upper limit statements. Vertical bars in
Fig.~\ref{Fig:SensEstimS5R5R6} represent 156 half-Hz frequency
bands for which no upper limits are provided because the entire half-Hz band
has been excluded. 

As shown in Fig.~\ref{Fig:SensEstimS5R5R6}, the upper limits on $h_{0}$ provided in the 0.5~Hz-wide bands starting at 69.5~Hz, 139.5~Hz, 335.5~Hz and 648~Hz are fairly high, roughly equal to $3 \times 10^{-23}$, $8.8 \times 10^{-24}$, $1.7 \times 10^{-23}$ and $2 \times 10^{-23}$, respectively. This is due to significant \textit{partial} contamination in these bands by lines listed in Tables~\ref{tab:HS5linesRS} and~\ref{tab:LS5linesRS}; the upper limit is given for the remaining, clean part of the band, but loud candidates from the disturbed part make up the loudest 100
candidates selected in the processing, so a simulated signal must be especially loud to surpass those. Note that, for the same reason, if we had set upper limits in the 156 half-Hz bands shown in Fig.~\ref{Fig:SensEstimS5R5R6} we would have obtained similarly high upper limits on $h_{0}$.

\subsection{Analytic sensitivity estimates}  

The $h_0^{90\%}$ upper limits can be independently predicted using the
method in \cite{Wette.2012}, adapted to the
Hough-on-$\mathcal{F}$-statistic search method (see~\cite{Prix.Wette.2012} for details).  The upper limit procedure
described above is modelled by a simple threshold on the number count,
where the thresholds are the largest number counts observed in each
0.5~Hz upper limit band.  The probability that, in the neighbourhood
of an injected signal, the number count $n_{\mathrm{c}}$ will exceed a
threshold $n_{\mathrm{c,th}}$ is denoted by $P[ n_{\mathrm{c}} >
n_{\mathrm{c,th}} | \rho(h_0,\cos\iota,\phi_0,\psi,m) ]$; this
probability can be calculated analytically from the known distribution of
$n_{\mathrm{c}}$.  The recovered SNR, $\rho$, is a function of the
nuisance parameters and of the mismatch $m$ between the injected signal
and the nearest template.  In addition, note
that in the presence of a signal, $2\mathcal{F}$ follows a non-central
$\chi^2$ distribution with 4 degrees of freedom; $\rho^2$ is the
non-centrality parameter of this distribution.  In the presence of a signal, averaging $P$
over the parameters of $\rho$ (except $h_0$) gives $\langle P(
n_{\mathrm{c}} > n_{\mathrm{c,th}} | \rho(h_0) )\rangle$, which equals
the confidence of recovering a population of injections with amplitude
$h_0$.  In each 0.5~Hz band, we determine the value of $h_0$ such that
$\langle P( n_{\mathrm{c}} > n_{\mathrm{c,th}} | \rho(h_0) )\rangle =
90\%$; this value is then the predicted value of $h_0^{90\%}$, given by
\begin{equation}
\label{eq:AnalyticalFitULs}
h_0^{90\%} = H \sqrt{ \frac{ S_h }{ T_{\mathrm{data}} } } \,
\end{equation}
as a function of the detector noise $S_h$
and the total data volume $T_{\mathrm{data}} = N_{\mathrm{seg}} T_{\mathrm{seg}}$.
The factor $H$ varies from $\sim$$141$ to $\sim 150$ over the range of
search frequencies, and is plotted in \cite{Prix.Wette.2012}.  It is
given by $H = 2.5 \hat{\rho}^{90\%} \sqrt{N_{\mathrm{seg}}}$, where
$\hat{\rho}^{90\%}$ is the mean injected SNR per segment of a population of
signals as described above, and is itself a function of the false
alarm and false dismissal probabilities, and $N_{\mathrm{seg}}$
\cite{Wette.2012,Prix.Wette.2012}.  The variation of $H$ as a function
of frequency arises from the variation of $\hat{\rho}^{90\%}$ as a
function of the false alarm probability in each upper limit band,
which are calculated from the largest number counts $n_{\mathrm{c}}$
plotted in Fig.~\ref{Fig:NcFrS5R5R6}.

The predicted values are shown by the cyan curve in Fig.~\ref{Fig:SensEstimS5R5R6}. The root-mean-square error between the Monte-Carlo estimated and predicted $h_0^{90\%}$ values ($\sim 7\%$ over all frequencies) is comparable with the uncertainties due to calibration ($10\%$) and the finite stepping of the Monte Carlo procedure ($5\%$).
This demonstrates that the sensitivity of the Hough-on-$\mathcal{F}$-statistic search method, as a function of the search parameters, is well understood.

\subsection{Astrophysical reach}
\begin{figure}[h!]
\begin{center}
\includegraphics[width=9cm, angle=0, clip]{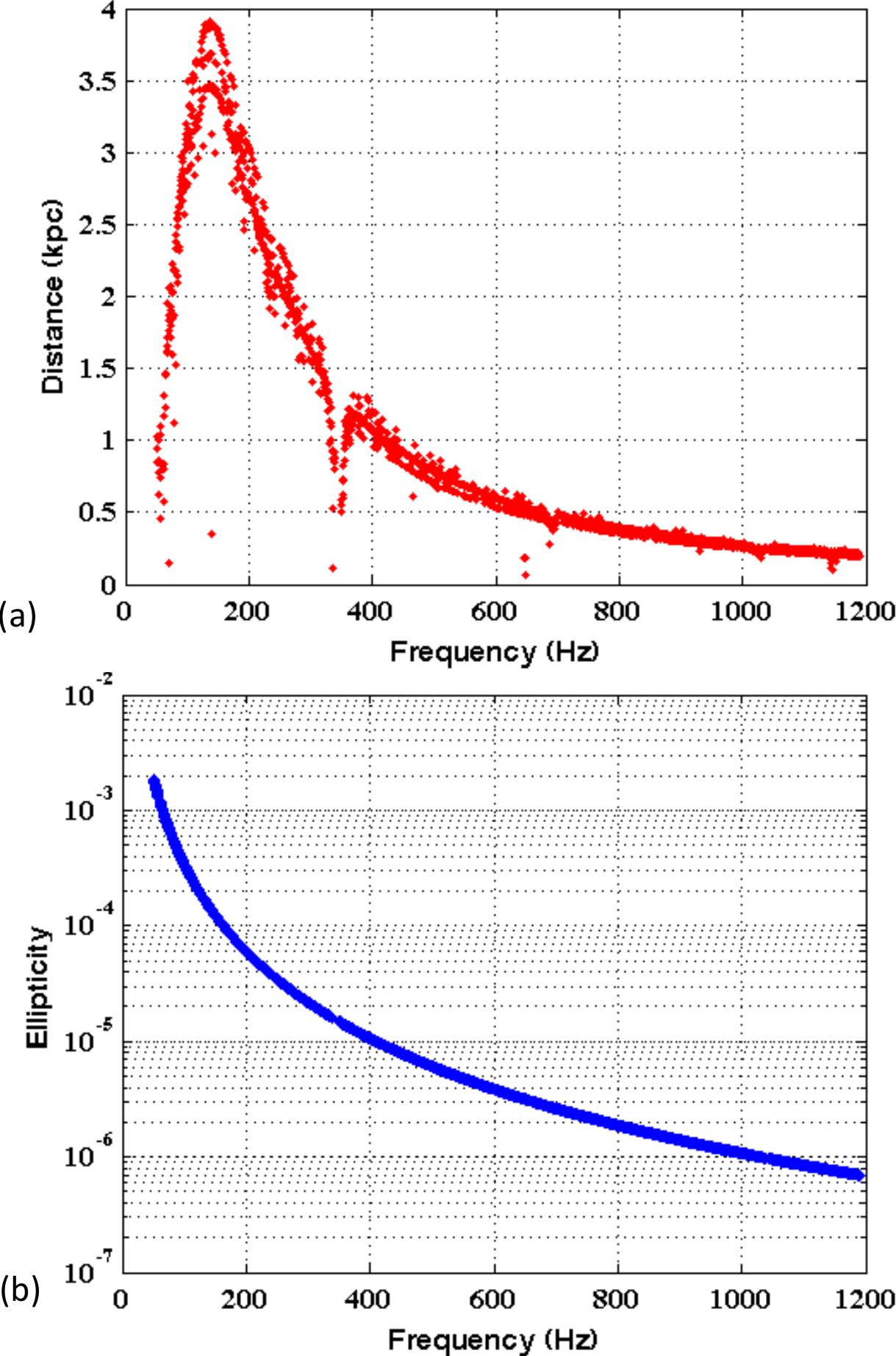} 
\caption{\label{DistEllip} Panel~(a) and (b) represent the distance range (in kpc) and the maximum ellipticity, respectively, as a function of the frequency. Both the panels are valid for neutron stars spinning down solely due to gravitational radiation and assuming a spindown value of $\sim -2$~nHz~s$^{-1}$. In these plots, the 156 half-Hz frequency bands for which no upper limits are provided have not been considered.}
\end{center}
\end{figure}

Figure~\ref{DistEllip} shows the maximum reach of our search. 
The top panel shows the maximum distance at which we could have detected a source emitting a CW signal with
strain amplitude $h_0^{90\%}$. The source is assumed to
be spinning down at the maximum rate considered in the search, 
$\sim -2$~nHz~s$^{-1}$, and emitting at the spindown limit, i.e.\ with all of the
lost rotational energy going into gravitational waves. The intrinsic
gravitational wave strain from a source at a distance $d$, with frequency $f$, frequency derivative $\dot{f}$, and emitting at the spindown limit is
 \begin{equation}
 h_0^{sd}=4.54\times 10^{-24} \left(1~\mathrm{kpc} \over d\right) \left(250~\mathrm{yr} \over -f/(4 \dot{f})\right)^{1/2},
 \label{Eq:spindownLimit}
 \end{equation}
where the canonical value of $10^{38}$ kg $\text{m}^2$ is assumed for $I_{zz}$ in Eq.~(\ref{eq:GWampl}).
The bottom panel of the figure does not depend on any result from the
search. It shows the spindown ellipticity values as a function of the
frequency for sources emitting in gravitational waves all the energy
lost while spinning down at a rate of $\sim -2$~nHz~s$^{-1}$. This is
obtained by setting $\dot{f}=-2\,$nHz~s$^{-1}$ in the following
equation:
\begin{equation}
 \varepsilon^{sd}=7.63\times 10^{-5} \left(-\dot{f} \over 10^{-10}~\mathrm{Hz}~\mathrm{s}^{-1}\right)^{1/2} \left(100~\mathrm{Hz} \over f\right)^{5/2}.
 \label{Eq:EpsilonAtSpindownLimit}
 \end{equation}
Around the frequency of greatest sensitivity, 152.5~Hz, we are sensitive to objects as far as 3.8~kpc and with an ellipticity $\varepsilon \sim 10^{-4}$. Normal neutron stars are expected to have $\varepsilon$ less than a few times $10^{-6}$ based on theoretical predictions~\cite{Horowitz}. A plausible value of  $\varepsilon \sim 3.5 \times 10^{-6}$ could be detectable by a search like this if the object were emitting at 625~Hz and at a distance no further than 500~pc.

\section{\label{sec:HWInjS5R5} Study of hardware-injected signals}

As part of the testing and validation of search pipelines and analysis
codes, simulated signals are added into the interferometer length
control system to produce mirror motions similar to what would be
generated if a gravitational wave signal were present.
Table~\ref{tab:S5R5PulsHardInjParams} shows the parameters of the set
of simulated CW signals injected into the LIGO detectors;
we shall often refer to these injections also as ``fake pulsars''. 
\begin{table*}
\begin{tabular}{ccrcrccrrcrcc} \hline \hline
Name & &$f_{P}$ (Hz) & & $\dot f$ ($\mathrm{Hz}~\mathrm{s}^{-1}$) & &$\alpha$ (rad) & $\delta$ (rad) & $\psi$ (rad) & $\phi_0$ (rad) & $\cos \iota$ (rad) & & $h_0$ \\
\hline
Fake pulsar~0 & & $265.5771052$ & &$-4.15\times 10^{-12}$ & &$1.248817$ & $-0.981180$ & $0.770087$ & $2.66$ & $0.794905$ & &$2.47 \times 10^{-25}$ \\
Fake pulsar~1 & & $849.0832962$ & &$-3.00\times 10^{-10}$ & &$0.652646$ & $-0.514042$ & $0.356036$ & $1.28$ & $0.463822$ & &$1.06 \times 10^{-24}$ \\
Fake pulsar~2 & & $575.163573$ & &$-1.37\times 10^{-13}$ & &$3.756929$ & $0.060109$ & $-0.221788$ & $4.03$ & $-0.928576$ & &$4.02 \times 10^{-24}$ \\
Fake pulsar~3 & & $108.8571594$ & &$-1.46\times 10^{-17}$ & &$3.113189$ & $-0.583579$ & $0.444280$ & $5.53$ & $-0.080666$ & &$1.63 \times 10^{-23}$ \\
Fake pulsar~4 & & $1403.163331$ & &$-2.54\times 10^{-8}$ & &$4.886707$ & $-0.217584$ & $-0.647939$ & $4.83$ & $0.277321$ & &$4.56 \times 10^{-23}$ \\
Fake pulsar~5 & & $52.80832436$ & &$-4.03\times 10^{-18}$ & &$5.281831$ & $-1.463269$ & $-0.363953$ & $2.23$ & $0.462967$ & &$4.85 \times 10^{-24}$ \\
Fake pulsar~6 & & $148.7190257$ & &$-6.73\times 10^{-9}$ & &$6.261385$ & $-1.141840$ & $0.470985$ & $0.97$ & $-0.153733$ & &$6.92 \times 10^{-25}$ \\
Fake pulsar~7 & & $1220.979581$ & &$-1.12\times 10^{-9}$ & &$3.899513$ & $-0.356931$ & $0.512323$ & $5.25$ & $0.756814$ & &$2.20 \times 10^{-24}$ \\
Fake pulsar~8 & & $194.3083185$ & &$-8.65\times 10^{-9}$ & &$6.132905$ & $-0.583263$ & $0.170471$ & $5.89$ & $0.073903$ & &$1.59 \times 10^{-23}$ \\
Fake pulsar~9 & & $763.8473165$ & &$-1.45\times 10^{-17}$ & &$3.471208$ & $1.321033$ & $-0.008560$ & $1.01$ & $-0.619187$ & &$8.13 \times 10^{-25}$ \\
\hline \hline
\end{tabular}
\caption{\label{tab:S5R5PulsHardInjParams} 
Simulated (``fake'') pulsar hardware injections during the S5 LIGO run, created with
the JPL DE405 Sun and Earth ephemeris files. The pulsar parameters are defined at the GPS reference time of 751\,680\,013~s in the SSB frame.
 These are the frequency $f_{P}$, the spindown $\dot f$, the sky position $(\alpha, \delta)$, the polarization angle $\psi$, the initial phase $\phi_0$,
 the inclination parameter $\cos \iota$ and the dimensionless strain amplitude $h_0$. These parameters correspond to the only set of hardware injections, injected into the S5 LIGO data, that fall within the GPS times of the S5R5 data.}
\end{table*}
 These injections were
active from the GPS epoch 829\,412\,600~s until 875\,301\,345~s. Of
these ten hardware-injected CW signals, eight had frequencies covered by
the S5R5 search frequency band: the fake pulsars 4 and 7 have
frequencies outside this band and thus have not been taken into
account during this analysis. The fake pulsars 6 and 8 have spindown
values outside the S5R5 search frequency derivative range.  

As a minor complication, the hardware injections were not active all
the time.  In the S5R5 data set, their duty cycle was $\sim 63 \%$ and $\sim 60 \%$ in L and H, respectively.
The hardware injections were active in 76 of the 121 S5R5
data segments, and they were completely absent in the remaining 45
segments. 
Their expected value of the Hough number count is thus:
\begin{equation}\label{eq:pulsarNc}
E[n_{\mathrm{c}}^{\mathrm{HI}}] = \sum_{j} \eta_{j} w_{j}+ p \, \sum_l w_{l},
\end{equation}
where the superscript ``HI'' refers to Hardware Injection, $j$ runs
over the number of data segments where the hardware injections were
active and $l$ runs over the remaining segments, and the $w_{j}$ are
the Hough weights given by Eq.~(\ref{eq:weights}); $\eta_{j}$ and $p$
are the probabilities that the estimated value of $2\mathcal{F}$ (for
a given data segment) crosses the threshold
$2\mathcal{F}_{\mathrm{th}}=5.2$ in the presence and absence of
a signal, respectively (expressions for $\eta_{j}$ and $p$ can be found
in \cite{JKSpaper,HierarchP2}).  

\begin{figure}[h]
\begin{center}
\includegraphics[width=8.9cm, angle=0, clip]{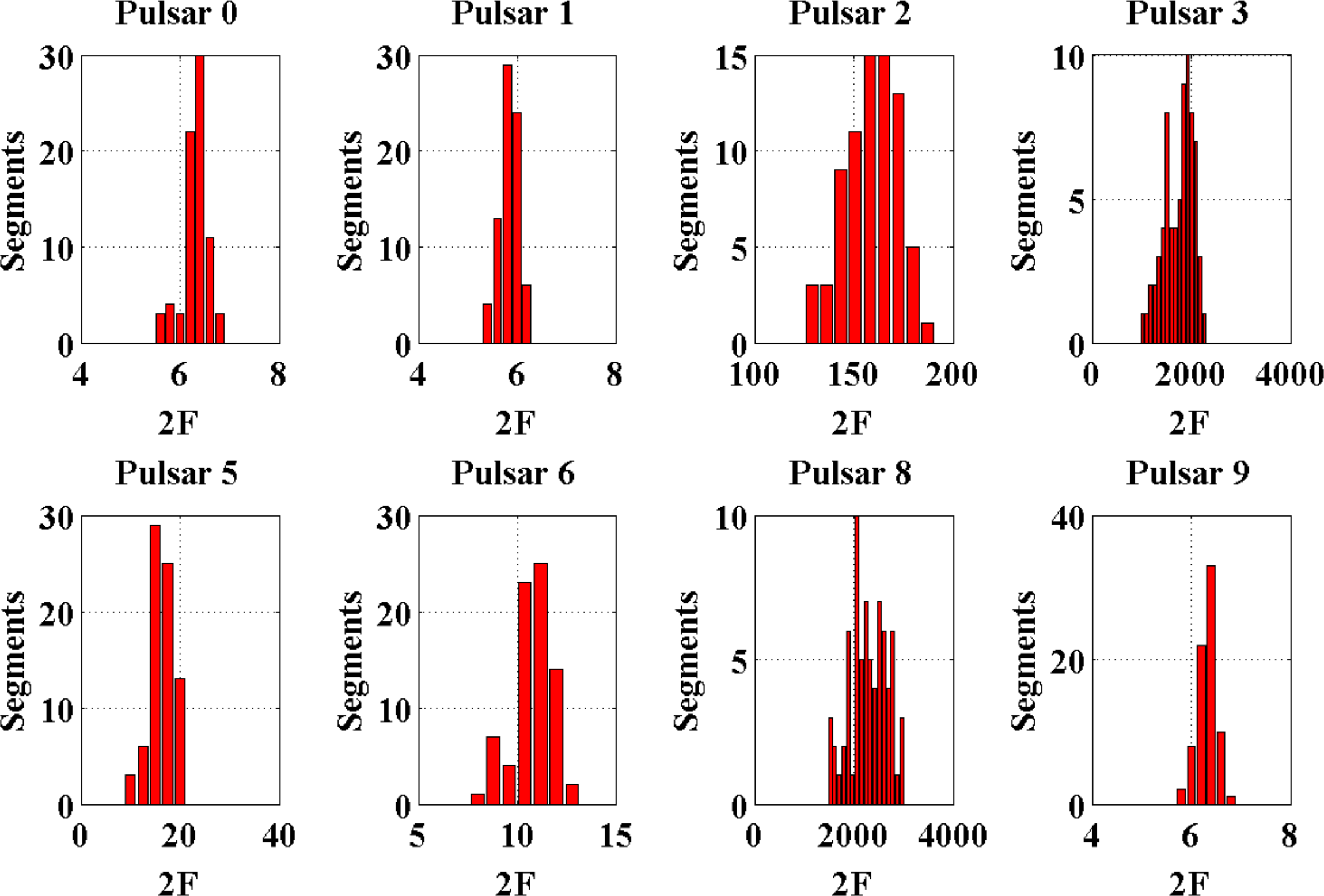}
\caption{\label{Fig:PulsHistsS5R5} Histograms of the $2\mathcal{F}$-values estimated, for every data segment, for all the S5 hardware injections covered by the investigated frequency range. The largest $2\mathcal{F}$-values are from the fake pulsars~2,~3,~5 and~8. In all the plots, the total number of data segments where the hardware injections were active is 76.}
\end{center}
\end{figure}

Figure~\ref{Fig:PulsHistsS5R5} shows, for each pulsar hardware injection, the
histograms of the expected $2\mathcal{F}$-values, computed for all the
segments where a particular hardware injection was active. From
this figure one can infer that the fake pulsars expected to be the
loudest in terms of $2\mathcal{F}$-values are pulsars 2, 3, 5 and 8.

The pulsar hardware injections go through the normal search and
post-processing pipelines in the usual way as described in the
previous sections. As discussed in the following, the fake pulsars recovered in the S5R5 search are pulsars 2, 3 and 5. 
On the other hand, fake pulsars 6 and 8 are missed because their
spindown values are outside our search range.
As expected, fake pulsars~0, 1 and~9 are not recovered because their
amplitudes are too weak and they do not pass the
$\left<2\mathcal{F_{\mathrm{HL}}}\right> > 6.5$ cut, as shown in
Table~\ref{tab:S5R5PulsHardInj}. In this table, the observed number counts of 69 to 72 are consistent with noise
fluctuations in those half-Hz bands (see Fig.~\ref{Fig:NcFrS5R5R6})
superseding the weak injected signals.
Note that the expected $2\mathcal{F}$-values shown in Fig.~\ref{Fig:PulsHistsS5R5} were computed assuming a search using exactly the correct signal parameters provided in Table~\ref{tab:S5R5PulsHardInjParams}, while the observed $\left<2\mathcal{F}_{\mathrm{HL}} \right>$-values in Table~\ref{tab:S5R5PulsHardInj} were obtained from our search which uses a grid of templates,
so significant mismatch is to be expected.
Table~\ref{tab:S5R5PulsHardInj} compares the expected and observed
values of the number counts associated with the S5 hardware injections
and the surviving S5R5 candidate events closest to them.  As in
Sec.~\ref{subsec:s5r3followup}, the measure of distance used here is a
Euclidean distance, expressed in bins, in the four dimensions
$(f,\dot{f},\alpha,\delta)$. 
\begin{table}[h!]
\caption{\label{tab:S5R5PulsHardInj} Comparison between the expected
   and observed number counts ($E[n_{\mathrm{c}}^{\mathrm{HI}}]$ and
   $N_{\mathrm{c}}^{\mathrm{S5R5}}$ respectively) associated with the
   hardware injections  and the recovered S5R5 candidates closest to
   these injections. The $\left<2\mathcal{F}_{\mathrm{HL}}
   \right>$-values for each of the candidates are also listed. 
The fake signals labeled pulsar~4 and~7 were not taken into account in this analysis since they have frequencies outside the S5R5 search frequency range.
The ``expected'' values marked by asterisks are not actually expected
to be obtained because the spindown rates for those signals lie
outside the range of this search.}
\begin{ruledtabular}
\begin{tabular}{lccc}
Name & $E[n_{\mathrm{c}}^{\mathrm{HI}}]$ & $N_{\mathrm{c}}^{\mathrm{S5R5}}$ & $\left<2\mathcal{F}_{\mathrm{HL}} \right>$      \\
\hline
Fake pulsar~0 & 53 & 69 & 5.6 \\
Fake pulsar~1 & 49 & 71 & 5.7 \\
Fake pulsar~2 & 88 & 80 & 22.3\\
Fake pulsar~3 & 89 & 96 & 197.1 \\
Fake pulsar~5 & 85 & 70 & 6.6 \\
Fake pulsar~6 & 76$^\ast$ & 71 & 5.3 \\
Fake pulsar~8 & 87$^\ast$ & 72 & 5.5 \\
Fake pulsar~9 & 53 & 72 & 5.8 \\
\end{tabular}
\end{ruledtabular}
\end{table}

Table~\ref{tab:PulsS5R5params} lists the
parameters of pulsars 2, 3 and 5 and the parameters of the
corresponding recovered candidates.  We successfully find candidates near the correct signal parameters.
The mismatch in spindown might seem large, but in fact the injections were
found at the nearest spindown template. The number count values show 
consistency within the $3\sigma$ range.
\begin{table}[h!]
\footnotesize
\caption{\label{tab:PulsS5R5params} Study of hardware injections in
  the S5R5 search. Values of $(f, \alpha, \delta, \dot{f})$ for the
  fake pulsar~2,~3,~5 and for the closest recovered candidates (denoted as
  Cand~2,~3,~5) are listed. The $n_c$-value in the last column
  represents the expectation value $E[n_{\mathrm{c}}^{\mathrm{HI}}]$ for the fake pulsars
  and the observed number count $N_{\mathrm{c}}^{\mathrm{S5R5}}$ for the corresponding recovered
  candidate. }
\begin{ruledtabular}
\begin{tabular}{lccccc}
Name & $f$~(Hz) & $\alpha$~(rad) & $\delta$~(rad) & $\dot f$~(Hz~s$^{-1}$) & $n_{\mathrm{c}}$ \\
\hline\\
Fake pulsar~2 & 575.163573 & 3.756929 & 0.060109 & $-1.37 \times 10^{-13}$ & 88 \\
Cand~2 & 575.163556 & 3.757514 & 0.065354 & $-1.64 \times 10^{-11}$ & 80 \\
\\
\hline \\
Fake pulsar~3 & 108.857159 & 3.113189 & -0.583579 & $-1.46 \times 10^{-17}$ & 89 \\
Cand~3 &  108.857158 & 3.09806 &   -0.5839483 & $-1.64 \times 10^{-11}$ & 96 \\
\\
\hline \\
Fake pulsar~5 & 52.8083243 & 5.281831 & -1.463269 & $-4.03 \times 10^{-18}$ & 85 \\
Cand~5 &  52.8082977 & 5.58845 &   -1.470972 & $-1.64 \times 10^{-11}$ & 70 \\
\\
\end{tabular}
\end{ruledtabular}
\end{table}

\section{\label{sec:conclus} CONCLUSIONS}
No evidence for continuous gravitational waves
 has been observed in
the search presented here. Upper limits on the intrinsic gravitational wave strain
have been derived using standard population-based methods, and are
shown in Fig.~\ref{Fig:SensEstimS5R5R6}.
These results are about a factor of $3$ more constraining than those
from the previous Einstein@Home search in early S5
data~\cite{EatHS5R1}. This improvement comes from using more data (a
year versus two months), from using a multi-detector coherent
statistic (versus a single-detector statistic), from a lower effective
threshold in the coherent detection stage, and from a more sensitive
incoherent method to combine the information from the coherently
analyzed segments. The largest effect comes from lowering the
effective threshold.  Indeed much of the improvement in sensitivity is
attributable to improved data analysis methods (as opposed to improved
detector sensitivity).  If we had used the much higher threshold of 25
on $2\mathcal{F}$, as in~\cite{EatHS5R1}, our
sensitivity would have been a factor of $\sim 2.5$ worse than our
final upper limits, thereby undoing almost all of the factor of 3
improvement mentioned above.  

We have not included second time-derivatives of the frequency
in our search. This could
be astrophysically significant in some regions of parameter space,
as discussed in
Sec.~\ref{subsec:spindownorder}.  It is important to keep this caveat in
mind while interpreting our results.  
 
This is the most sensitive wide-frequency-range, all-sky search for CW signals performed to date. The upper limit values are comparable to those obtained recently using the PowerFlux method~\cite{S4PSH,S5Powerflux2011} on the entire S5 data set (S5R3+S5R5). Ref.~\cite{S5Powerflux2011}
searched for CW signals over the whole sky, in a smaller frequency
band (up to 800 Hz versus 1\,190~Hz here), but a broader spindown range up to $-6$~nHz/s.
Strain upper limits were set at the 95\% confidence in
0.25-Hz wide sub-bands. In particular, near 152~Hz, the PowerFlux
strict, all-sky upper limit on worst-case linearly (best-case
circularly) polarized strain amplitude $h_{0}$ is $\sim 1 \times
10^{-24}$ ($3.5 \times 10^{-25}$). As a comparison, at the same
frequency, this search constrains the strain to $h_{0} \lesssim 7.6 \times
10^{-25}$ (as shown in Fig.~\ref{Fig:SensEstimS5R5R6}), $9.2 \times
10^{-25}$ and $3.2 \times 10^{-25}$ for the case of average, linear
and circular polarization, respectively, with a 90\% confidence level
in a 0.5-Hz wide band.

The most constraining upper limit obtained by this search is $h_0^{90\%} \sim 7.6 \times 10^{-25}$ at 152 Hz; the corresponding maximum reach is roughly 4~kpc, assuming $\varepsilon \sim 10^{-4}$.

It has long been expected that searching a large parameter space for CW signals
will require hierarchical semi-coherent searches. This analysis is a milestone towards that goal, and we expect
that future analyses will build on the tools developed here.

There are a number of areas where further improvements are possible.
In the latest round of analysis (an Einstein@Home processing run that
began in March 2012), some of the post-processing techniques developed
for this analysis have been ``moved upstream'' to the hosts. One
example is the generalized $\mathcal{F}$-statistic consistency test~\cite{PKPLS}. This continues the pattern of
moving analyses formerly carried out in the post-processing stage (for
example, the incoherent combination step) onto host machines.  Another
step forward is in the semi-coherent algorithm that combines the
coherent analyses from the different segments.  The Hough algorithm
described here turned out to be rather cumbersome, and does not combine the coarse and fine
grids in an optimal way. The latest round of Einstein@Home processing
makes use of a simpler optimal semi-coherent method, which allows
longer coherent time baselines to be used.  This method, based on a
detailed analysis of correlations in parameter space
\cite{GlobCorr}, is described in\cite{Pletsch:2009uu}. Looking farther forward, we expect
to use higher-order spindown parameters both to search for a broader class of signals as well as to be able to employ longer coherent time baselines in the analysis.

The Advanced LIGO and Advanced Virgo detectors are currently under
construction, and should begin operations around 2016.  In comparison
with the current generation, these instruments will provide an
order-of-magnitude improvement in strain sensitivity, increasing the
volume of space observed by a factor of a thousand.

These and other improvements in data analysis methods and
instrumentation make us optimistic that we will eventually be able to
make direct detections of CW signals.  Such
detection will provide new insights into the internal structure,
formation history and population statistics of neutron stars. 

\acknowledgments

The authors gratefully acknowledge the support of the United States
National Science Foundation for the construction and operation of the
LIGO Laboratory, the Science and Technology Facilities Council of the
United Kingdom, the Max-Planck-Society, and the State of
Niedersachsen/Germany for support of the construction and operation of
the GEO600 detector, and the Italian Istituto Nazionale di Fisica
Nucleare and the French Centre National de la Recherche Scientifique
for the construction and operation of the Virgo detector. The authors
also gratefully acknowledge the support of the research by these
agencies and by the Einstein@Home volunteers, by the Australian Research Council, 
the International Science Linkages program of the Commonwealth of Australia,
the Council of Scientific and Industrial Research of India, 
the Istituto Nazionale di Fisica Nucleare of Italy, 
the Spanish Ministerio de Econom\'ia y Competitividad,
the Conselleria d'Economia Hisenda i Innovaci\'o of the
Govern de les Illes Balears, the Foundation for Fundamental Research
on Matter supported by the Netherlands Organisation for Scientific Research, 
the Polish Ministry of Science and Higher Education, the FOCUS
Programme of Foundation for Polish Science,
the Royal Society, the Scottish Funding Council, the
Scottish Universities Physics Alliance, The National Aeronautics and
Space Administration, the Carnegie Trust, the Leverhulme Trust, the
David and Lucile Packard Foundation, the Research Corporation, and the
Alfred P. Sloan Foundation. This document has been assigned LIGO
Laboratory Document No. LIGO-P1200026.


\appendix

\section{Problems in calculating the weights} 
\label{sec:bugs}

In this section we describe two software errors
that affected the main
hierarchical search code. While these errors did not invalidate
the search results and were dealt with adequately, we document them
here for completeness.

The first issue is connected with the choice of weights used to
construct the number count defined in Eqs.~\eqref{eq:nc} and
\eqref{eq:weights}.  The weights for each segment are computed
following Eq.~\eqref{eq:noisewt-harmonic} which uses the harmonic mean
of the noise spectra for each SFT.  However, the original method for
computing the weights used the arithmetic mean of the contributions
instead of the harmonic mean, which turned out to have the perverse effect that segments with a few
noisy SFTs got a disproportionately large weight.  This led to a
much larger value for the variance given in Eq.~\eqref{eq:sigmaW} and
a correspondingly anomalously small value of the significance defined
in Eq.~\eqref{eq:NcSign}.  All WUs were originally run using the
arithmetic mean which led to anomalously low values of the
significance CR due to non-stationary noise for a small number of
frequency bands: $[50, 202]$~Hz, $[328.5, 329]$~Hz, $[995.5,
1\,010]$~Hz and $[1\,069, 1\,075]$~Hz. The WUs for these bands were
re-computed with the weights given by 
Eq.~\eqref{eq:noisewt-harmonic}.  All other frequencies are unaffected
by this problem.

A second issue, which interacts with the problematic calculation of the
weights described above, is floating point inaccuracy in our
implementation of the Hough-transform algorithm.  A single threshold
crossing for the $\mathcal{F}$-statistic leads to a +1 in number count
for possibly a large number of points in parameter space
\cite{HierarchP2}, and it is not necessary to step through parameter
space point-by-point to calculate the final number count.  For the
vast majority of cases, our implementation of the Hough-transform
agrees with the brute force approach for calculating the number count,
but the two can occasionally differ.  If we were not using weights,
these differences would have a minor effect on the number count.
Occasionally, however, these floating point errors coincide with the
cases when we assign excessively large weights to particular segments
as discussed above.  In these cases, the discrepancies in the number
count can be large and in some cases may even exceed the number of
segments, which is in principle a strict upper bound on the number
count.  Note that our upper limits remain valid because they consistently use
the same search code, and any candidates are followed-up by
independent codes thereby avoiding spurious false alarms.
Using the modified weights based on
Eq.~\eqref{eq:noisewt-harmonic} fixes this problem as well.

\section{\label{sec:S5R3pp} S5R3 Post-Processing}

\begin{figure}[h!]
\begin{center}
\includegraphics[width=8.6cm, angle=0, clip]{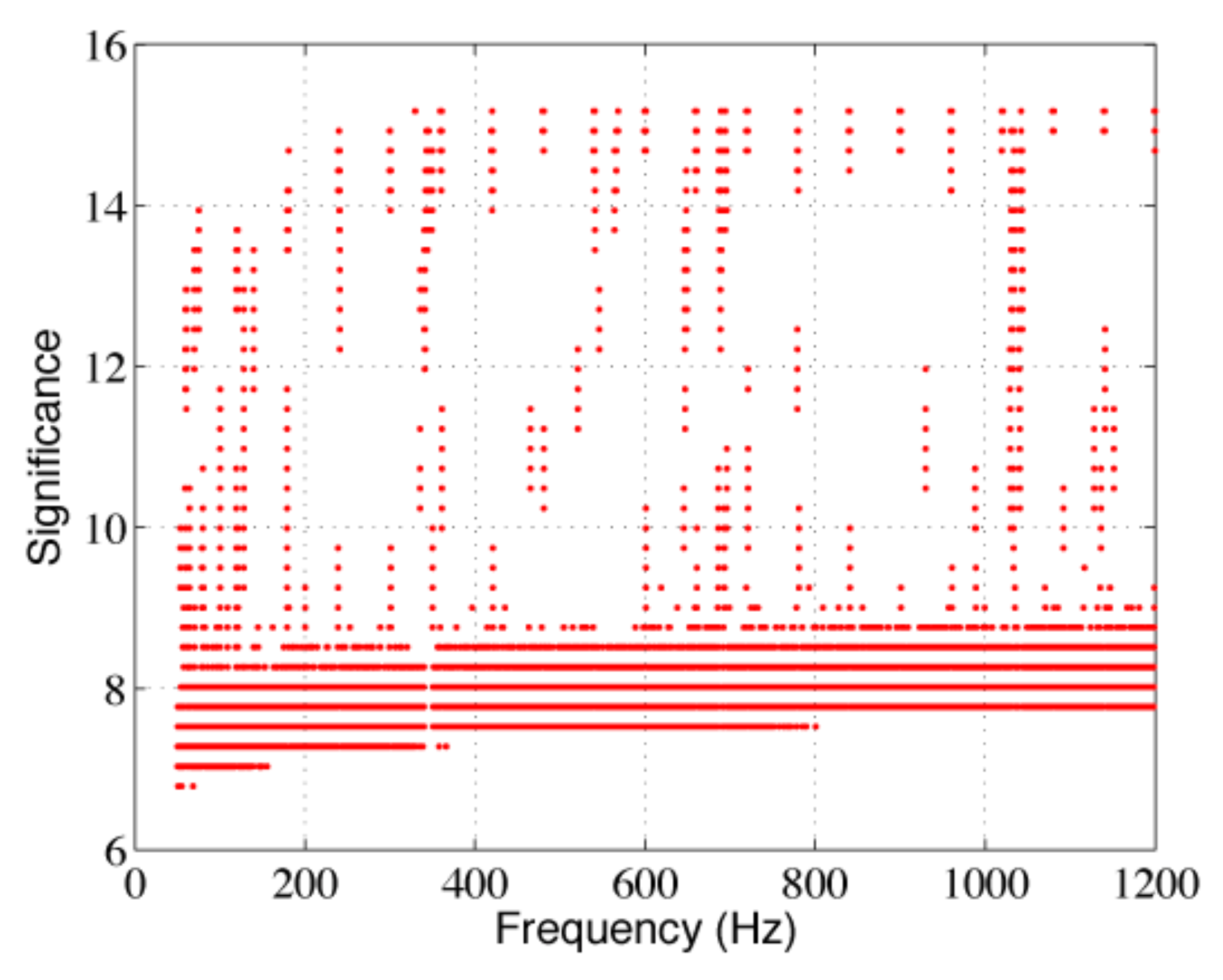} 
\caption{\label{Fig:NcFreq} Significance of 230\,000 loudest candidates selected in 0.5~Hz-wide frequency bands as a function of the frequency.}
\end{center}
\end{figure}

\begin{figure}[h]
\begin{center}
\includegraphics[width=8.5cm, angle=0, clip]{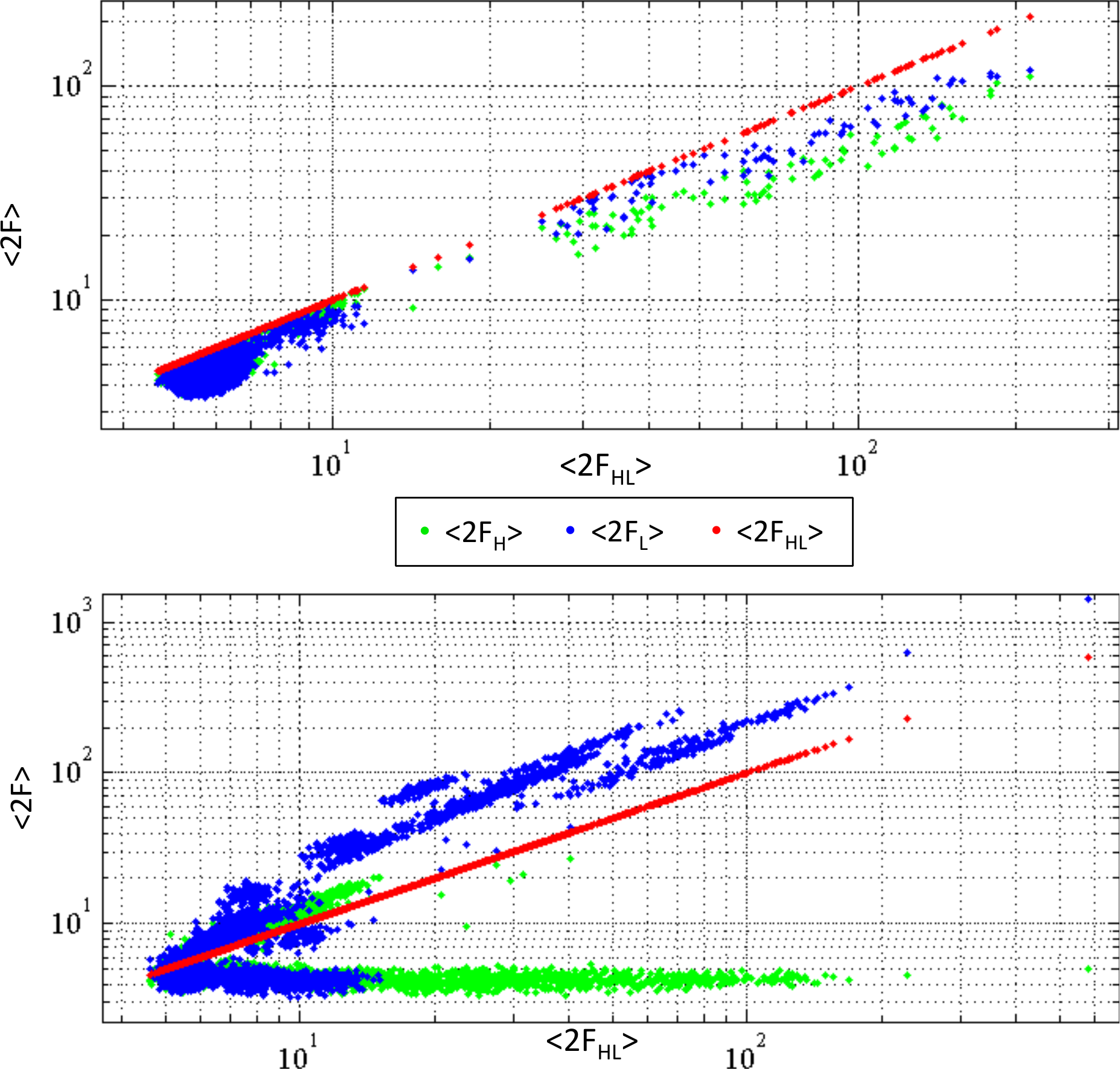}
\caption{\label{Fig:FstatVeto} Values of $2\mathcal{F}$ averaged over $84$ data segments for the single-detector case, $\left<2\mathcal{F}_{\mathrm{H}} \right>$ (green dots), $\left<2\mathcal{F}_{\mathrm{L}} \right>$ (blue dots), the multi-detector case ($\left<2\mathcal{F}_{\mathrm{HL}} \right>$, red dots) against those for the combined multi-detector statistic ($\left<2\mathcal{F}_{\mathrm{HL}} \right>$). The top (bottom) plot shows such values for 161\,785 (5\,994) surviving (vetoed) candidates such that $\left<2\mathcal{F}_{\mathrm{H}} \right>$ and (or) $\left<2\mathcal{F}_{\mathrm{L}} \right>$ is less (greater) than $\left<2\mathcal{F}_{\mathrm{HL}} \right>$.}
\end{center}
\end{figure}

The S5R3 run was launched on September 23, 2007 and ended on September
25, 2008. Like S5R5, it was an all-sky search. It used 7\,237 S5 LIGO
SFTs, collected between the GPS times of 818\,845\,553~s (Sat Dec 17
09:05:40 GMT 2005) and 851\,765\,191~s (Tue Jan 02 09:26:17 GMT
2007). The data analyzed consisted of 3\,781 SFTs from H and 3\,456
SFTs from L.  The number of data segments used for the S5R3 run was
84, with duration each $T_{\mathrm{seg}}=25$~hours. The search
frequency range was $[50, 1\,200]$~Hz, with a frequency resolution
$\delta f \sim 6.7 \times 10^{-6}$~Hz and a mismatch value $m=0.3$,
leading to the spindown resolution given by
Eq.~(\ref{eq:wrongdfdot}). In the S5R3 run, each WU analyzed a
constant frequency bandwidth $B \simeq 16$~mHz, the full spindown
interval, ranging roughly from $-1.6$~nHz~s$^{-1}$ to $-3.1 \times
10^{-11}$~Hz s$^{-1}$ in steps of $3.8 \times 10^{-10}$~Hz s$^{-1}$,
and a certain region of the sky, as already discussed in
Sec.~\ref{sec:EatHWUdesign}.  The sky-grids and output file formats are
identical to those used in S5R5.  A major difference with S5R5 are the
weights: no weighting scheme was used in S5R3 and all the weights
$w_i$ appearing in Eq.~(\ref{eq:nc}) were set to unity. In total, $\sim 7 \times 10^{10}$ S5R3 candidates were sent back to
the Einstein@Home server. 
The S5R3 post-processing was performed before that for S5R5, and it was used as a ``test-bed'' for the latter.
It consists of the same series of steps described previously in
Sec.~\ref{sec:S5R5pp}.  The significance values of the 100 loudest
candidates in 0.5~Hz-wide frequency bands are plotted in
Fig.~\ref{Fig:NcFreq} as a function of the frequency. 
These represent a total of 230\,000 candidates. This set is reduced by $\sim 27 \%$
via the removal of instrumental noise artifacts (listed in
Tables~\ref{tab:HS5linesRS} and~\ref{tab:LS5linesRS}) and of the
candidates from search frequency bands close to the fake noise
(according to Table~\ref{tab:whiteNoiseBandsH}).  For all the
surviving 167\,779 candidates, the $\mathcal{F}$-statistic consistency
veto described in Sec.~\ref{subsec:fstatconsistency} removes an
additional 3.6\%.  Figure~\ref{Fig:FstatVeto} shows the values of
$\left<2\mathcal{F}_{\mathrm{H}} \right>$,
$\left<2\mathcal{F}_{\mathrm{L}} \right>$ and
$\left<2\mathcal{F}_{\mathrm{HL}} \right>$ versus
$\left<2\mathcal{F}_{\mathrm{HL}} \right>$ for 161\,785 candidate
events that survived (top plot) and 5\,994 that were excluded (bottom
plot).

\begin{figure}[h]
\begin{center}
\includegraphics[width=8.3cm, angle=0, clip]{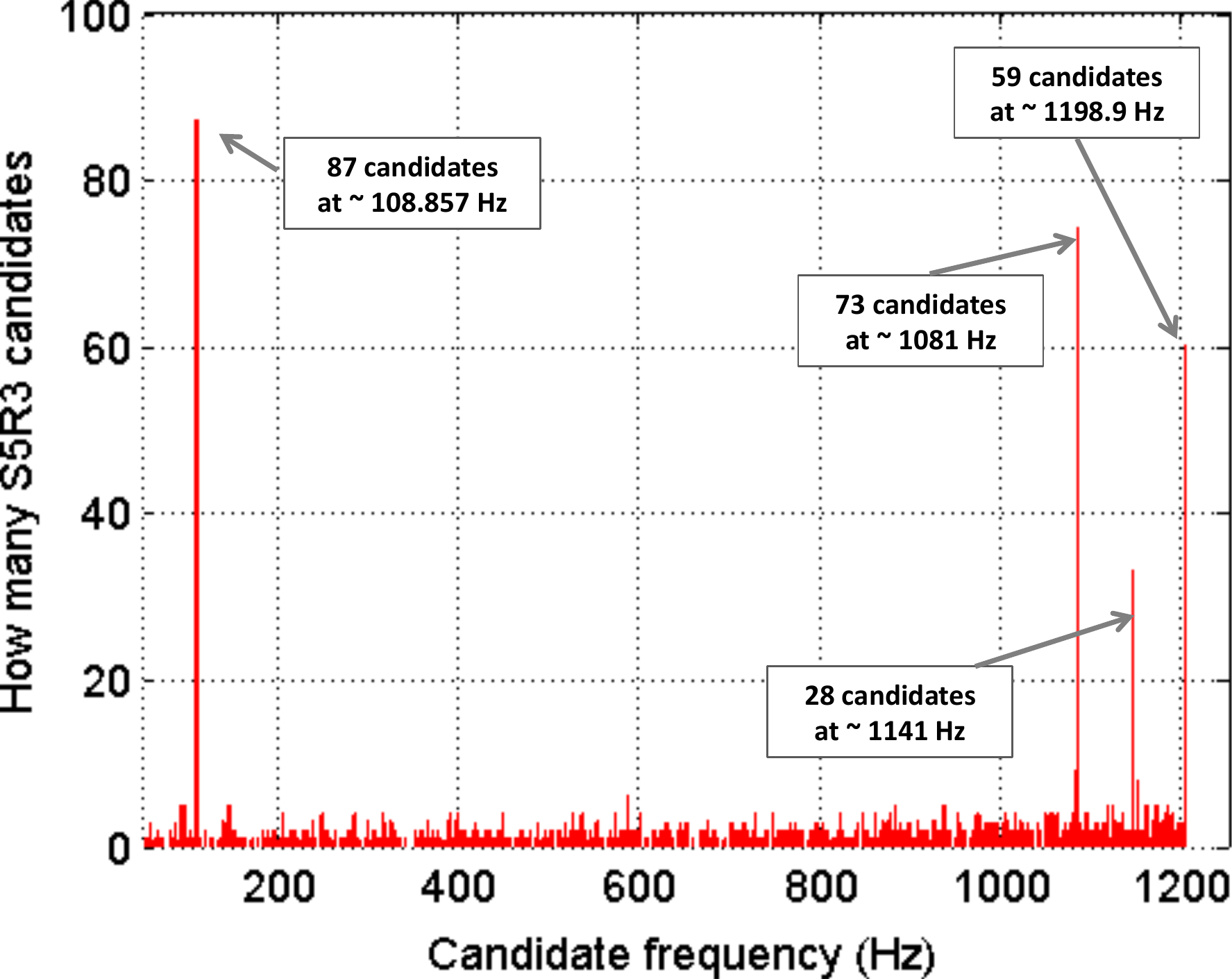} 
\caption{\label{Fig:S5R3FUcands} Histogram of $1\,465 $ S5R3 candidates that have been further investigated through follow-up study using the S5R5 data set.}
\end{center}
\end{figure}

Candidates whose $\left<2\mathcal{F}_{\mathrm{HL}} \right>$-value is greater than 6.5, i.e.\ 1\,465 out of 161\,785, are followed-up by performing a hierarchical search using the S5R5 data set. This run, whose details are described in Section~\ref{sec:S5R5pp}, consists of $\sim 46 \%$ more data than was used for S5R3 and is thus more sensitive than S5R5.
The followed-up candidates are plotted in Fig.~\ref{Fig:S5R3FUcands}
as a function of the frequency: 87 candidates are clustered at $\sim 108.857$~Hz and represent the
contribution of the hardware injection pulsar~3~\cite{S5R3HIs}; $73$ 
candidates have frequencies peaked around $\sim 1081$~Hz, $59$
candidates at $\sim 1198.9$~Hz and $28$ at $\sim 1141$~Hz. 
Except for the candidates that represent the contribution from the hardware injected signal, none of the S5R5 candidates is consistent with a signal at a level consistent with the observed S5R3 excess. We then conclude that no gravitational wave signal is observed in the S5R3 data.

\section{\label{sec:knownS5lines} Instrumental noise artifacts}
This appendix contains lists of the main known spectral lines of instrumental origin in the LIGO detectors during the S5 run. They were individually identified with a particular source, or were members of identified combs of lines found in many channels (e.g.\ 60~Hz combs), or the source was unknown, but they were found in frequency coincidence between the gravitational wave channel and auxiliary channels. In the latter case, to ensure that actual gravitational wave signals were not rejected, the SNR in the auxiliary channel had to be at least 5 times larger than in the gravitational wave channel and the density of lines in the auxiliary channel had to be low enough that accidental coincidence with a line in the gravitational wave channel was highly unlikely. The spectral lines and harmonics detected in H and L are listed in Tables~\ref{tab:HS5linesRS} and~\ref{tab:LS5linesRS}, respectively. As mentioned earlier, on the basis of these tables, about 22\% and 25\% of candidates have been excluded from the analysis in the S5R5 and S5R3 post-processing, respectively.

For each candidate with frequency $f_{\mathrm{c}}$ and derivative
$\dot{f}_{\mathrm{c}}$, the candidate was rejected if a band $\Delta
f_{\mathrm{c}} \simeq f_{\mathrm{c}} \times 10^{-4} +
|\dot{f_{\mathrm{c}}}| \times 10^{7}$~s on either side of the
signal had any overlap with an instrumental line band. This was done
in order to take into account the maximum possible Doppler shift due
to the Earth's orbital velocity, which is roughly $10^{-4}$ in units
of the speed of light, and the maximum frequency shift due to the
spindown of the source over the $\sim \pm 10^{7}~\mathrm{s}$ time span
relative to the reference time during 
the Einstein@Home run.

\begin{center} \footnotesize
\tablefirsthead{%
  \hline
  \hline 
  \multicolumn{1}{c}{$f_{\mathrm{L}}$~(Hz)} & 
  \multicolumn{1}{c}{Harmonics} &
  \multicolumn{1}{c}{LF} &
  \multicolumn{1}{c}{HF} &
  \multicolumn{1}{c}{Cause} \\
  \hline}
\tablehead{%
  \hline}
\tabletail{%
  \hline
  \multicolumn{5}{c}{\small\sl -continues-}\\
  \hline}
\tablelasttail{\hline}
\bottomcaption{\label{tab:HS5linesRS} Known S5 H spectral artifacts. The different columns represent: (I) the central frequency of the instrumental line; (II) the number of harmonics including the fundamental; (III) Low-Frequency (LF) bound of the knockout band; (IV) High-Frequency (HF) bound of the knockout band; (V) the cause of the line (see key below). When there were higher harmonics, the third and fourth columns were multiplied by the harmonic number to yield the proper LF and HF bounds. 
}
\begin{supertabular}{ l c c@{\hspace{2mm}} c@{\hspace{2mm}} c }
1.0 & 1000 & 0.9999194 & 1.0000806 & Electronics\\
46.70 & 1 & 46.6932 & 46.7068 & Calibration\\
48 & 1 & 47.96 & 48.04 & Pulsed-heating\\
51 & 1 & 50.96 & 51.04 & Pulsed-heating\\
54 & 1 & 53.96 & 54.04 & Pulsed-heating\\
57 & 1 & 56.96 & 57.04 & Pulsed-heating\\
60 & 121 & 59.96 & 60.04 & Mains \\
63 & 1 & 62.96 & 63.04 & Pulsed-heating\\
66 & 1 & 65.96 & 66.04 & Pulsed-heating\\
69 & 1 & 68.96 & 69.04 & Pulsed-heating\\
72 & 1 & 71.96 & 72.04 & Pulsed-heating\\
85.80 & 1 & 85.79 & 85.81 & Electronics\\
89.9 & 1 & 89.84 & 89.96 & Electronics\\
93.05 & 1 & 93.04 & 93.06 & Unknown\\
93.25 & 1 & 93.24 & 93.26 & Unknown\\
139.95 & 1 & 139.94 & 139.96 & Electronics\\
164.52 & 1 & 164.51 & 164.53 & Electronics\\
329.51 & 2 & 329.49 & 329.53 & Wire\\
329.58 & 1 & 329.56 & 329.59 & Electronics\\
329.59 & 2 & 329.57 & 329.61 & Wire\\
329.70 & 2 & 329.67 & 329.72 & Wire \\ 
329.78 & 2 & 329.75 & 329.8 & Wire\\
329.86 & 1 & 329.85 & 329.87 & Electronics\\
335.695 & 1 & 335.67 & 335.72 & Wire \\
335.7230 & 1 & 335.698 & 335.748 & Wire \\
335.7410 & 1 & 335.716 & 335.766 & Wire \\
335.8200 & 1 & 335.795 & 335.845 & Wire \\
343.2879 & 1 & 343.261 & 343.315 & Wire\\
343.4145 & 1 & 343.394 & 343.435 & Wire \\
343.9272 & 1 & 343.907 & 343.948 & Wire\\
344.0584 & 1 & 344.038 & 344.079 & Wire\\
344.5247 & 1 & 344.499 & 344.55 & Wire\\
344.6685 & 1 & 344.647 & 344.69 & Wire\\
344.7186 & 1 & 344.692 & 344.745 & Wire\\
344.8280 & 1 & 344.810 & 344.847 & Wire \\
347.1824 & 1 & 347.16 & 347.204 & Wire\\
347.3107 & 1 & 347.29 & 347.331 & Wire\\
347.3635 & 1 & 347.34 & 347.387 & Wire\\
347.5099 & 1 & 347.489 & 347.531 & Wire\\
347.5818 & 1 & 347.557 & 347.606 & Wire\\
347.6860 & 1 & 347.664 & 347.708 & Wire \\
347.7230 & 1 & 347.703 & 347.743 & Wire\\
393.1000 & 1 & 393.093 & 393.107 & Calibration\\
539.43 & 1 & 539.42 & 539.44 & Electronics\\
546.06 & 3 & 545.89 & 546.21 & Wire\\
548.36 & 1 & 548.35 & 548.37 & Electronics\\
564.07 & 3 & 563.9 & 564.22 & Wire\\
566.10 & 3 & 565.93 & 566.25 & Wire\\
568.10 & 3 & 567.93 & 568.25 & Wire\\
646.385 & 3 & 646.22 & 646.535 & Wire\\
648.835 & 3 & 648.67 & 648.985 & Wire\\
649.87 & 1 & 649.86 & 649.88 & Unknown\\
659.31 & 1 & 659.30 & 659.32 & Electronics\\
686.6615 & 1 & 686.634 & 686.689 & Wire\\
686.9176 & 1 & 686.896 & 686.939 & Wire \\
688.0224 & 1 & 688 & 688.044 & Wire\\
688.2825 & 1 & 688.26 & 688.305 & Wire\\
689.1301 & 1 & 689.108 & 689.152 & Wire\\
689.4262 & 1 & 689.404 & 689.449 & Wire\\
689.5036 & 1 & 689.482 & 689.525 & Wire\\
689.7361 & 1 & 689.715 & 689.758 & Wire \\
694.4720 & 1 & 694.447 & 694.497 & Wire\\
694.7292 & 1 & 694.703 & 694.755 & Wire\\
695.0220 & 1 & 694.999 & 695.045 & Wire\\
695.2091 & 1 & 695.185 & 695.233 & Wire\\
695.4274 & 1 & 695.404 & 695.451 & Wire\\
695.4814 & 1 & 695.457 & 695.506 & Wire\\
915.80 & 1 & 915.79 & 915.81 & Electronics\\
960 & 1 & 959.99 & 960.01 & Timing\\
961 & 1 & 960.99 & 961.01 & Timing\\
995.50 & 1 & 995.49 & 995.51 & Electronics\\
1009.70 & 1 & 1009.69 & 1009.71 & Electronics\\
1030.55 & 1 & 1030.48 & 1030.63 & Wire\\
1032.19 & 1 & 1032.16 & 1032.23 & Wire\\
1032.58 & 1 & 1032.56 & 1032.61 & Wire\\
1033.78 & 1 & 1033.77 & 1033.79 & Electronics\\
1033.8766 & 1 & 1033.84 & 1033.92 & Wire\\
1034.3294 & 1 & 1034.3 & 1034.36 & Wire\\
1034.4549 & 1 & 1034.42 & 1034.49 & Wire\\
1034.821 & 1 & 1034.78 & 1034.86 & Wire\\
1042.25 & 1 & 1042.18 & 1042.32 & Wire \\
1042.3785 & 1 & 1042.35 & 1042.41 & Wire\\
1042.8179 & 1 & 1042.8 & 1042.84 & Wire\\
1043.0272 & 1 & 1042.99 & 1043.06 & Wire\\
1043.3351 & 1 & 1043.31 & 1043.36 & Wire\\
1043.455 & 1 & 1043.38 & 1043.53 & Wire\\
1144.3 & 1 & 1144.29 & 1144.31 & Calibration\\
1374.4509 & 1 & 1374.43 & 1374.47 & Wire\\
1376.6139 & 1 & 1376.59 & 1376.64 & Wire\\
1377.1423 & 1 & 1377.12 & 1377.17 & Wire\\
1378.7493 & 1 & 1378.72 & 1378.78 & Wire\\
1379.3999 & 1 & 1379.37 & 1379.43 & Wire\\
1379.5062 & 1 & 1379.48 & 1379.53 & Wire\\
1380.0283 & 1 & 1380 & 1380.05 & Wire\\
1390.0061 & 1 & 1389.98 & 1390.03 & Wire\\
1390.6821 & 1 & 1390.66 & 1390.71 & Wire\\
1391.4240 & 1 & 1391.4 & 1391.45 & Wire\\
1391.5967 & 1 & 1391.57 & 1391.62 & Wire\\
1718.5697 & 1 & 1718.54 & 1718.6 & Wire\\
1721.9155 & 1 & 1721.89 & 1721.94 & Wire\\
1724.0104 & 1 & 1723.94 & 1724.08 & Wire\\
1724.9704 & 1 & 1724.95 & 1725 & Wire\\
1725.6181 & 1 & 1725.59 & 1725.64 & Wire\\
1737.9391 & 1 & 1737.92 & 1737.96 & Wire\\
1738.9907 & 1 & 1738.97 & 1739.01 & Wire\\
1739.8250 & 1 & 1739.8 & 1739.85 & Wire\\
1740.0280 & 1 & 1740 & 1740.05 & Wire\\
\end{supertabular}
\end{center}

\begin{center}\footnotesize
\tablefirsthead{%
  \hline
  \hline 
  \multicolumn{1}{c}{$f_{\mathrm{L}}$~(Hz)} & 
  \multicolumn{1}{c}{Harmonics} &
  \multicolumn{1}{c}{LF} &
  \multicolumn{1}{c}{HF} &
  \multicolumn{1}{c}{Cause} \\
  \hline}
\tablehead{%
  \hline}
\tabletail{%
  \hline
  \multicolumn{5}{c}{\small\sl -continues-}\\
  \hline}
\tablelasttail{\hline}
\bottomcaption{\label{tab:LS5linesRS} Known S5 L spectral artifacts.
The columns are the same as in Table~\ref{tab:HS5linesRS}.
}
\begin{supertabular}{ l c c@{\hspace{2mm}} c@{\hspace{2mm}} c }
1.0 & 1000 & 0.9999194 & 1.0000806 & Electronics\\
54.7000 & 1 & 54.6932 & 54.7068 & Calibration\\
59.0683 & 1 & 58.9749 & 59.1617 & Pulsed-heating\\
59.3918 & 1 & 59.3146 & 59.469 & Pulsed-heating\\
59.7382 & 1 & 59.5942 & 59.8822 & Pulsed-heating\\
60 & 121 & 59.96 & 60.04 & Mains\\
60.2731 & 1 & 60.1556 & 60.3906 & Pulsed-heating\\
60.5918 & 1 & 60.5284 & 60.6552 & Pulsed-heating\\
60.9497 & 1 & 60.8609 & 61.0385 & Pulsed-heating\\
93.2903 & 1 & 93.2758 & 93.3048 & Electronics\\
96.7082 & 1 & 96.6959 & 96.7205 & Electronics\\
139.9387 & 1 & 139.92 & 139.958 & Electronics\\
145.0622 & 1 & 145.047 & 145.078 & Electronics\\
186.5874 & 1 & 186.565 & 186.61 & Electronics\\
193.4164 & 1 & 193.395 & 193.437 & Electronics\\
233.2314 & 1 & 233.185 & 233.277 & Electronics\\
241.7774 & 1 & 241.713 & 241.842 & Electronics\\
329.2339 & 2 & 329.216 & 329.252 & Wire\\
329.3409 & 2 & 329.323 & 329.359 & Wire\\
329.4025 & 2 & 329.379 & 329.426 & Wire\\
335.276 & 1 & 335.256 & 335.296 & Wire\\
335.4100 & 1 & 335.386 & 335.434 & Wire\\
335.5950 & 1 & 335.57 & 335.62 & Wire\\
335.7770 & 1 & 335.752 & 335.802 & Wire\\
342.9424 & 1 & 342.915 & 342.97 & Wire\\
343.0980 & 1 & 343.075 & 343.121 & Wire\\
343.355 & 1 & 343.335 & 343.375 & Wire\\
343.4726 & 1 & 343.451 & 343.494 & Wire\\
343.6231 & 1 & 343.6 & 343.647 & Wire\\
344.266 & 1 & 344.246 & 344.286 & Wire\\
344.4132 & 1 & 344.392 & 344.434 & Wire\\
346.6349 & 1 & 346.603 & 346.667 & Wire\\
346.8060 & 1 & 346.784 & 346.828 & Wire\\
346.8727 & 1 & 346.85 & 346.896 & Wire\\
346.9151 & 1 & 346.895 & 346.935 & Wire\\
346.9650 & 1 & 346.945 & 346.985 & Wire\\
347.0370 & 1 & 347.017 & 347.057 & Wire\\
396.7 & 1 & 396.693 & 396.707 & Calibration\\
685.9147 & 1 & 685.893 & 685.937 & Wire\\
686.2051 & 1 & 686.172 & 686.238 & Wire\\
686.8158 & 1 & 686.792 & 686.84 & Wire\\
687.0511 & 1 & 687.021 & 687.081 & Wire\\
687.3246 & 1 & 687.301 & 687.348 & Wire\\
688.8577 & 1 & 688.832 & 688.883 & Wire\\
693.4187 & 1 & 693.392 & 693.445 & Wire\\
693.6827 & 1 & 693.652 & 693.713 & Wire\\
693.7638 & 1 & 693.74 & 693.788 & Wire\\
693.9111 & 1 & 693.888 & 693.934 & Wire\\
693.9834 & 1 & 693.958 & 694.008 & Wire\\
694.0889 & 1 & 694.058 & 694.12 & Wire\\
960 & 1 & 959.99 & 960.01 & Timing\\
961 & 1 & 960.99 & 961.01 & Timing\\
1029.5578 & 1 & 1029.53 & 1029.58 & Wire\\
1030.7536 & 1 & 1030.73 & 1030.78 & Wire\\
1031.1536 & 1 & 1031.13 & 1031.18 & Wire\\
1033.5104 & 1 & 1033.49 & 1033.53 & Wire\\
1040.3507 & 1 & 1040.33 & 1040.37 & Wire\\
1040.6940 & 1 & 1040.67 & 1040.72 & Wire\\
1040.7343 & 1 & 1040.71 & 1040.76 & Wire\\
1040.7859 & 1 & 1040.76 & 1040.81 & Wire\\
1041.0204 & 1 & 1041 & 1041.04 & Wire\\
1041.1701 & 1 & 1041.15 & 1041.19 & Wire\\
1041.2731 & 1 & 1041.25 & 1041.29 & Wire\\
1150.0661 & 1 & 1149.15 & 1150.98 & Calibration\\
1151.9118 & 1 & 1151.56 & 1152.26 & Calibration\\
1372.9742 & 1 & 1372.95 & 1373 & Wire\\
1374.6601 & 1 & 1374.64 & 1374.68 & Wire\\
1375.2021 & 1 & 1375.18 & 1375.23 & Wire\\
1378.3695 & 1 & 1378.34 & 1378.39 & Wire\\
1387.3946 & 1 & 1387.37 & 1387.42 & Wire\\
1387.9327 & 1 & 1387.9 & 1387.96 & Wire\\
1387.9660 & 1 & 1387.92 & 1388.01 & Wire\\
1388.0561 & 1 & 1388.03 & 1388.08 & Wire\\
1388.3850 & 1 & 1388.35 & 1388.42 & Wire\\
1388.5530 & 1 & 1388.53 & 1388.58 & Wire\\
1388.7127 & 1 & 1388.69 & 1388.74 & Wire\\
1716.8006 & 1 & 1716.77 & 1716.83 & Wire\\
1718.8679 & 1 & 1718.84 & 1718.89 & Wire\\
1719.5480 & 1 & 1719.52 & 1719.57 & Wire\\
1723.4861 & 1 & 1723.46 & 1723.51 & Wire\\
1734.7999 & 1 & 1734.78 & 1734.82 & Wire\\
1735.9610 & 1 & 1735.94 & 1735.99 & Wire\\
1736.1977 & 1 & 1736.16 & 1736.23 & Wire\\
1736.4134 & 1 & 1736.39 & 1736.44 & Wire\\
1920.0000 & 1 & 1919.99 & 1920.01 & Timing\\
1921.0000 & 1 & 1920.99 & 1921.01 & Timing\\
1922.0009 & 1 & 1921.98 & 1922.02 & Timing\\
\end{supertabular}
\end{center}

A short explanation of the key to the line sources listed in the fifth
column of Tables~\ref{tab:HS5linesRS} and~\ref{tab:LS5linesRS}
follows.
\begin{description}
\item[Mains] lines at multiples of the 60~Hz electrical power system
  frequency; the dominant coupling mechanism at 60~Hz was from 
  magnetic fields generated by electric currents coupling to the permanent
  magnets mounted on the test masses. 

\item[Electronics] produced by either electronic circuit oscillations,
  or by slight data corruption associated with repetitive processes in
  the data acquisition computers. The line was identified in power
  supply voltage variation, magnetic fields from electronics or by
  direct measurements.

\item[Calibration] produced for calibration purposes by moving a test
  mass with the actuation system. 

\item[Timing] introduced by the timing verification
  system. 

\item[Wire] a vibrational resonance of a mirror
  suspension wire. 

\item[Pulsed-heating] produced by cyclically
  pulsed mains heating circuits, coupling to the test mass magnets via
  magnetic fields. 

\item[Unknown] a line of unknown source that
  appeared in auxiliary channels and met the rejection criteria noted
  in the text.

\end{description}

\newpage
\bibliography{EatH_S5R3P}

\begin{thebibliography}{56}%
\makeatletter
\providecommand \@ifxundefined [1]{%
 \@ifx{#1\undefined}
}%
\providecommand \@ifnum [1]{%
 \ifnum #1\expandafter \@firstoftwo
 \else \expandafter \@secondoftwo
 \fi
}%
\providecommand \@ifx [1]{%
 \ifx #1\expandafter \@firstoftwo
 \else \expandafter \@secondoftwo
 \fi
}%
\providecommand \natexlab [1]{#1}%
\providecommand \enquote  [1]{``#1''}%
\providecommand \bibnamefont  [1]{#1}%
\providecommand \bibfnamefont [1]{#1}%
\providecommand \citenamefont [1]{#1}%
\providecommand \href@noop [0]{\@secondoftwo}%
\providecommand \href [0]{\begingroup \@sanitize@url \@href}%
\providecommand \@href[1]{\@@startlink{#1}\@@href}%
\providecommand \@@href[1]{\endgroup#1\@@endlink}%
\providecommand \@sanitize@url [0]{\catcode `\\12\catcode `\$12\catcode
  `\&12\catcode `\#12\catcode `\^12\catcode `\_12\catcode `\%12\relax}%
\providecommand \@@startlink[1]{}%
\providecommand \@@endlink[0]{}%
\providecommand \url  [0]{\begingroup\@sanitize@url \@url }%
\providecommand \@url [1]{\endgroup\@href {#1}{\urlprefix }}%
\providecommand \urlprefix  [0]{URL }%
\providecommand \Eprint [0]{\href }%
\@ifxundefined \urlstyle {%
  \providecommand \doi  [0]{\begingroup \@sanitize@url \@doi}%
  \providecommand \@doi [1]{\endgroup \@@startlink {\doibase
  #1}doi:\discretionary {}{}{}#1\@@endlink }%
}{%
  \providecommand \doi  [0]{doi:\discretionary{}{}{}\begingroup
  \urlstyle{rm}\Url }%
}%
\providecommand \doibase [0]{http://dx.doi.org/}%
\providecommand \Doi [0]{\begingroup \@sanitize@url \@Doi }%
\providecommand \@Doi  [1]{\endgroup\@@startlink{\doibase#1}\@@Doi}%
\providecommand \@@Doi [1]{#1\@@endlink}%
\providecommand \selectlanguage [0]{\@gobble}%
\providecommand \bibinfo  [0]{\@secondoftwo}%
\providecommand \bibfield  [0]{\@secondoftwo}%
\providecommand \translation [1]{[#1]}%
\providecommand \BibitemOpen [0]{}%
\providecommand \bibitemStop [0]{}%
\providecommand \bibitemNoStop [0]{.\EOS\space}%
\providecommand \EOS [0]{\spacefactor3000\relax}%
\providecommand \BibitemShut  [1]{\csname bibitem#1\endcsname}%
\bibitem [{\citenamefont {Bildsten}(1998)}]{NonAxNS1}%
  \BibitemOpen
  \bibfield  {author} {\bibinfo {author} {\bibfnamefont {L.}~\bibnamefont
  {Bildsten}},\ }\href@noop {} {\bibfield  {journal} {\bibinfo  {journal}
  {Astrophys. J.},\ }\textbf {\bibinfo {volume} {501}},\ \bibinfo {pages} {L89}
  (\bibinfo {year} {1998})}\BibitemShut {NoStop}%
\bibitem [{\citenamefont {Ushomirsky}\ \emph {et~al.}(2000)\citenamefont
  {Ushomirsky}, \citenamefont {Cutler},\ and\ \citenamefont
  {Bildsten}}]{NonAxNS2}%
  \BibitemOpen
  \bibfield  {author} {\bibinfo {author} {\bibfnamefont {G.}~\bibnamefont
  {Ushomirsky}}, \bibinfo {author} {\bibfnamefont {C.}~\bibnamefont {Cutler}},
  \ and\ \bibinfo {author} {\bibfnamefont {L.}~\bibnamefont {Bildsten}},\ }\Doi
  {10.1046/j.1365-8711.2000.03938.x} {\bibfield  {journal} {\bibinfo  {journal}
  {Mon. Not. Roy. Astron. Soc.},\ }\textbf {\bibinfo {volume} {319}},\ \bibinfo
  {pages} {902} (\bibinfo {year} {2000})},\ \Eprint
  {http://arxiv.org/abs/astro-ph/0001136} {arXiv:astro-ph/0001136} \BibitemShut
  {NoStop}%
\bibitem [{\citenamefont {Cutler}(2002)}]{NonAxNS3}%
  \BibitemOpen
  \bibfield  {author} {\bibinfo {author} {\bibfnamefont {C.}~\bibnamefont
  {Cutler}},\ }\Doi {10.1046/j.1365-8711.2000.03938.x} {\bibfield  {journal}
  {\bibinfo  {journal} {Phys. Rev. D},\ }\textbf {\bibinfo {volume} {66}},\
  \bibinfo {pages} {084025} (\bibinfo {year} {2002})}\BibitemShut {NoStop}%
\bibitem [{\citenamefont {Melatos}\ and\ \citenamefont
  {Payne}(2005)}]{NonAxNS4}%
  \BibitemOpen
  \bibfield  {author} {\bibinfo {author} {\bibfnamefont {A.}~\bibnamefont
  {Melatos}}\ and\ \bibinfo {author} {\bibfnamefont {D.}~\bibnamefont
  {Payne}},\ }\Doi {10.1086/428600} {\bibfield  {journal} {\bibinfo  {journal}
  {Astrophys. J.},\ }\textbf {\bibinfo {volume} {623}},\ \bibinfo {pages}
  {1044} (\bibinfo {year} {2005})},\ \Eprint
  {http://arxiv.org/abs/astro-ph/0503287} {arXiv:astro-ph/0503287} \BibitemShut
  {NoStop}%
\bibitem [{\citenamefont {Owen}(2005)}]{NonAxNS5}%
  \BibitemOpen
  \bibfield  {author} {\bibinfo {author} {\bibfnamefont {B.~J.}\ \bibnamefont
  {Owen}},\ }\Doi {10.1103/PhysRevLett.95.211101} {\bibfield  {journal}
  {\bibinfo  {journal} {Phys. Rev. Lett.},\ }\textbf {\bibinfo {volume} {95}},\
  \bibinfo {pages} {211101} (\bibinfo {year} {2005})},\ \Eprint
  {http://arxiv.org/abs/astro-ph/0503399} {arXiv:astro-ph/0503399} \BibitemShut
  {NoStop}%
\bibitem [{\citenamefont {Abramovici}\ \emph {et~al.}(1992)\citenamefont
  {Abramovici}, \citenamefont {Althouse}, \citenamefont {Drever}, \citenamefont
  {Gursel}, \citenamefont {Kawamura} \emph {et~al.}}]{ligoref}%
  \BibitemOpen
  \bibfield  {author} {\bibinfo {author} {\bibfnamefont {A.}~\bibnamefont
  {Abramovici}}, \bibinfo {author} {\bibfnamefont {W.~E.}\ \bibnamefont
  {Althouse}}, \bibinfo {author} {\bibfnamefont {R.~W.}\ \bibnamefont
  {Drever}}, \bibinfo {author} {\bibfnamefont {Y.}~\bibnamefont {Gursel}},
  \bibinfo {author} {\bibfnamefont {S.}~\bibnamefont {Kawamura}},  \emph
  {et~al.},\ }\href@noop {} {\bibfield  {journal} {\bibinfo  {journal}
  {Science},\ }\textbf {\bibinfo {volume} {256}},\ \bibinfo {pages} {325}
  (\bibinfo {year} {1992})}\BibitemShut {NoStop}%
\bibitem [{\citenamefont {Abbott}\ \emph
  {et~al.}(2009){\natexlab{a}}\citenamefont {Abbott} \emph
  {et~al.}}]{ligoref2}%
  \BibitemOpen
  \bibfield  {author} {\bibinfo {author} {\bibfnamefont {B.~P.}\ \bibnamefont
  {Abbott}} \emph {et~al.},\ }\href@noop {} {\bibfield  {journal} {\bibinfo
  {journal} {Rep. Prog. Phys.},\ }\textbf {\bibinfo {volume} {72}},\ \bibinfo
  {pages} {076901} (\bibinfo {year} {2009}{\natexlab{a}})}\BibitemShut
  {NoStop}%
\bibitem [{\citenamefont {{Bradaschia}}\ \emph {et~al.}(1991)\citenamefont
  {{Bradaschia}}, \citenamefont {{Calloni}}, \citenamefont {{Cobal}},
  \citenamefont {{Del Fabbro}}, \citenamefont {{di Virgilio}} \emph
  {et~al.}}]{Virgo}%
  \BibitemOpen
  \bibfield  {author} {\bibinfo {author} {\bibfnamefont {C.}~\bibnamefont
  {{Bradaschia}}}, \bibinfo {author} {\bibfnamefont {E.}~\bibnamefont
  {{Calloni}}}, \bibinfo {author} {\bibfnamefont {M.}~\bibnamefont {{Cobal}}},
  \bibinfo {author} {\bibfnamefont {R.}~\bibnamefont {{Del Fabbro}}}, \bibinfo
  {author} {\bibfnamefont {A.}~\bibnamefont {{di Virgilio}}},  \emph {et~al.},\
  }in\ \href@noop {} {\emph {\bibinfo {booktitle} {Banff Summer Institute on
  Gravitation}}},\ \bibinfo {editor} {edited by\ \bibinfo {editor}
  {\bibnamefont {{R.~Mann \& P.~Wesson}}}}\ (\bibinfo {year} {1991})\ pp.\
  \bibinfo {pages} {499--514}\BibitemShut {NoStop}%
\bibitem [{\citenamefont {Accadia}\ \emph {et~al.}(2012)\citenamefont {Accadia}
  \emph {et~al.}}]{virgo2}%
  \BibitemOpen
  \bibfield  {author} {\bibinfo {author} {\bibfnamefont {T.}~\bibnamefont
  {Accadia}} \emph {et~al.},\ }\href@noop {} {\bibfield  {journal} {\bibinfo
  {journal} {Journal of Instrumentation},\ }\textbf {\bibinfo {volume} {7}},\
  \bibinfo {pages} {P03012} (\bibinfo {year} {2012})}\BibitemShut {NoStop}%
\bibitem [{\citenamefont {{Danzmann}}(1995)}]{Geo}%
  \BibitemOpen
  \bibfield  {author} {\bibinfo {author} {\bibfnamefont {K.}~\bibnamefont
  {{Danzmann}}},\ }in\ \href@noop {} {\emph {\bibinfo {booktitle} {First
  Edoardo Amaldi conference on gravitational wave experiments}}},\ \bibinfo
  {editor} {edited by\ \bibinfo {editor} {\bibnamefont {{E.~Coccia, G.~Pizzella
  \& F.~Ronga (World Scientific, Singapore)}}}}\ (\bibinfo {year} {1995})\ pp.\
  \bibinfo {pages} {100--111}\BibitemShut {NoStop}%
\bibitem [{\citenamefont {{Tsubono}}(1995)}]{TAMA}%
  \BibitemOpen
  \bibfield  {author} {\bibinfo {author} {\bibfnamefont {K.}~\bibnamefont
  {{Tsubono}}},\ }in\ \href@noop {} {\emph {\bibinfo {booktitle} {First Edoardo
  Amaldi conference on gravitational wave experiments}}},\ \bibinfo {editor}
  {edited by\ \bibinfo {editor} {\bibnamefont {{E.~Coccia, G.~Pizzella \&
  F.~Ronga (World Scientific, Singapore)}}}}\ (\bibinfo {year} {1995})\ pp.\
  \bibinfo {pages} {100--111}\BibitemShut {NoStop}%
\bibitem [{\citenamefont {Jaranowski}\ \emph {et~al.}(1998)\citenamefont
  {Jaranowski}, \citenamefont {Kr{\'o}lak},\ and\ \citenamefont
  {Schutz}}]{JKSpaper}%
  \BibitemOpen
  \bibfield  {author} {\bibinfo {author} {\bibfnamefont {P.}~\bibnamefont
  {Jaranowski}}, \bibinfo {author} {\bibfnamefont {A.}~\bibnamefont
  {Kr{\'o}lak}}, \ and\ \bibinfo {author} {\bibfnamefont {B.~F.}\ \bibnamefont
  {Schutz}},\ }\Doi {10.1103/PhysRevD.58.063001} {\bibfield  {journal}
  {\bibinfo  {journal} {Phys. Rev. D},\ }\textbf {\bibinfo {volume} {58}},\
  \bibinfo {pages} {063001} (\bibinfo {year} {1998})},\ \Eprint
  {http://arxiv.org/abs/gr-qc/9804014} {arXiv:gr-qc/9804014} \BibitemShut
  {NoStop}%
\bibitem [{\citenamefont {Cutler}\ and\ \citenamefont
  {Schutz}(2005)}]{MultiIfoFstat}%
  \BibitemOpen
  \bibfield  {author} {\bibinfo {author} {\bibfnamefont {C.}~\bibnamefont
  {Cutler}}\ and\ \bibinfo {author} {\bibfnamefont {B.~F.}\ \bibnamefont
  {Schutz}},\ }\Doi {10.1103/PhysRevD.72.063006} {\bibfield  {journal}
  {\bibinfo  {journal} {Phys. Rev. D},\ }\textbf {\bibinfo {volume} {72}},\
  \bibinfo {pages} {063006} (\bibinfo {year} {2005})},\ \Eprint
  {http://arxiv.org/abs/gr-qc/0504011} {arXiv:gr-qc/0504011} \BibitemShut
  {NoStop}%
\bibitem [{\citenamefont {Prix}(2007)}]{MultiFstat2}%
  \BibitemOpen
  \bibfield  {author} {\bibinfo {author} {\bibfnamefont {R.}~\bibnamefont
  {Prix}},\ }\Doi {10.1103/PhysRevD.75.023004, 10.1103/PhysRevD.75.069901}
  {\bibfield  {journal} {\bibinfo  {journal} {Phys. Rev. D},\ }\textbf
  {\bibinfo {volume} {75}},\ \bibinfo {pages} {023004} (\bibinfo {year}
  {2007})},\ \Eprint {http://arxiv.org/abs/gr-qc/0606088} {arXiv:gr-qc/0606088}
  \BibitemShut {NoStop}%
\bibitem [{\citenamefont {Prix}\ and\ \citenamefont {Krishnan}(2009)}]{Bstat}%
  \BibitemOpen
  \bibfield  {author} {\bibinfo {author} {\bibfnamefont {R.}~\bibnamefont
  {Prix}}\ and\ \bibinfo {author} {\bibfnamefont {B.}~\bibnamefont
  {Krishnan}},\ }\Doi {10.1088/0264-9381/26/20/204013} {\bibfield  {journal}
  {\bibinfo  {journal} {Class. Quant. Grav.},\ }\textbf {\bibinfo {volume}
  {26}},\ \bibinfo {pages} {204013} (\bibinfo {year} {2009})},\ \Eprint
  {http://arxiv.org/abs/0907.2569} {arXiv:0907.2569 [gr-qc]} \BibitemShut
  {NoStop}%
\bibitem [{\citenamefont {Abbott}\ \emph {et~al.}(2007)\citenamefont {Abbott}
  \emph {et~al.}}]{S2ScoX1}%
  \BibitemOpen
  \bibfield  {author} {\bibinfo {author} {\bibfnamefont {B.}~\bibnamefont
  {Abbott}} \emph {et~al.} (\bibinfo {collaboration} {LIGO Scientific
  Collaboration}),\ }\Doi {10.1103/PhysRevD.76.082001} {\bibfield  {journal}
  {\bibinfo  {journal} {Phys. Rev. D},\ }\textbf {\bibinfo {volume} {76}},\
  \bibinfo {pages} {082001} (\bibinfo {year} {2007})},\ \Eprint
  {http://arxiv.org/abs/gr-qc/0605028} {arXiv:gr-qc/0605028} \BibitemShut
  {NoStop}%
\bibitem [{\citenamefont {Abadie}\ \emph
  {et~al.}(2010){\natexlab{a}}\citenamefont {Abadie} \emph {et~al.}}]{S5CasA}%
  \BibitemOpen
  \bibfield  {author} {\bibinfo {author} {\bibfnamefont {J.}~\bibnamefont
  {Abadie}} \emph {et~al.} (\bibinfo {collaboration} {LIGO Scientific
  Collaboration}),\ }\Doi {10.1088/0004-637X/722/2/1504} {\bibfield  {journal}
  {\bibinfo  {journal} {Astrophys. J.},\ }\textbf {\bibinfo {volume} {722}},\
  \bibinfo {pages} {1504} (\bibinfo {year} {2010}{\natexlab{a}})},\ \Eprint
  {http://arxiv.org/abs/1006.2535} {arXiv:1006.2535 [gr-qc]} \BibitemShut
  {NoStop}%
\bibitem [{\citenamefont {Brady}\ and\ \citenamefont
  {Creighton}(2000)}]{HierarchP1}%
  \BibitemOpen
  \bibfield  {author} {\bibinfo {author} {\bibfnamefont {P.~R.}\ \bibnamefont
  {Brady}}\ and\ \bibinfo {author} {\bibfnamefont {T.}~\bibnamefont
  {Creighton}},\ }\Doi {10.1103/PhysRevD.61.082001} {\bibfield  {journal}
  {\bibinfo  {journal} {Phys. Rev. D},\ }\textbf {\bibinfo {volume} {61}},\
  \bibinfo {pages} {082001} (\bibinfo {year} {2000})},\ \Eprint
  {http://arxiv.org/abs/gr-qc/9812014} {arXiv:gr-qc/9812014} \BibitemShut
  {NoStop}%
\bibitem [{\citenamefont {Krishnan}\ \emph {et~al.}(2004)\citenamefont
  {Krishnan}, \citenamefont {Sintes}, \citenamefont {Papa}, \citenamefont
  {Schutz}, \citenamefont {Frasca} \emph {et~al.}}]{HierarchP2}%
  \BibitemOpen
  \bibfield  {author} {\bibinfo {author} {\bibfnamefont {B.}~\bibnamefont
  {Krishnan}}, \bibinfo {author} {\bibfnamefont {A.~M.}\ \bibnamefont
  {Sintes}}, \bibinfo {author} {\bibfnamefont {M.~A.}\ \bibnamefont {Papa}},
  \bibinfo {author} {\bibfnamefont {B.~F.}\ \bibnamefont {Schutz}}, \bibinfo
  {author} {\bibfnamefont {S.}~\bibnamefont {Frasca}},  \emph {et~al.},\ }\Doi
  {10.1103/PhysRevD.70.082001} {\bibfield  {journal} {\bibinfo  {journal}
  {Phys. Rev. D},\ }\textbf {\bibinfo {volume} {70}},\ \bibinfo {pages}
  {082001} (\bibinfo {year} {2004})},\ \Eprint
  {http://arxiv.org/abs/gr-qc/0407001} {arXiv:gr-qc/0407001} \BibitemShut
  {NoStop}%
\bibitem [{\citenamefont {Cutler}\ \emph {et~al.}(2005)\citenamefont {Cutler},
  \citenamefont {Gholami},\ and\ \citenamefont {Krishnan}}]{HierarchP3}%
  \BibitemOpen
  \bibfield  {author} {\bibinfo {author} {\bibfnamefont {C.}~\bibnamefont
  {Cutler}}, \bibinfo {author} {\bibfnamefont {I.}~\bibnamefont {Gholami}}, \
  and\ \bibinfo {author} {\bibfnamefont {B.}~\bibnamefont {Krishnan}},\ }\Doi
  {10.1103/PhysRevD.72.042004} {\bibfield  {journal} {\bibinfo  {journal}
  {Phys. Rev. D},\ }\textbf {\bibinfo {volume} {72}},\ \bibinfo {pages}
  {042004} (\bibinfo {year} {2005})},\ \Eprint
  {http://arxiv.org/abs/gr-qc/0505082} {arXiv:gr-qc/0505082} \BibitemShut
  {NoStop}%
\bibitem [{\citenamefont {Abbott}\ \emph {et~al.}(2005)\citenamefont {Abbott}
  \emph {et~al.}}]{S2Hough}%
  \BibitemOpen
  \bibfield  {author} {\bibinfo {author} {\bibfnamefont {B.}~\bibnamefont
  {Abbott}} \emph {et~al.} (\bibinfo {collaboration} {LIGO Scientific
  Collaboration}),\ }\Doi {10.1103/PhysRevD.72.102004} {\bibfield  {journal}
  {\bibinfo  {journal} {Phys. Rev. D},\ }\textbf {\bibinfo {volume} {72}},\
  \bibinfo {pages} {102004} (\bibinfo {year} {2005})},\ \Eprint
  {http://arxiv.org/abs/gr-qc/0508065} {arXiv:gr-qc/0508065} \BibitemShut
  {NoStop}%
\bibitem [{\citenamefont {Abbott}\ \emph {et~al.}(2008)\citenamefont {Abbott}
  \emph {et~al.}}]{S4PSH}%
  \BibitemOpen
  \bibfield  {author} {\bibinfo {author} {\bibfnamefont {B.}~\bibnamefont
  {Abbott}} \emph {et~al.} (\bibinfo {collaboration} {LIGO Scientific
  Collaboration}),\ }\Doi {10.1103/PhysRevD.77.022001,
  10.1103/PhysRevD.80.129904, 10.1103/PhysRevD.77.022001,
  10.1103/PhysRevD.80.129904} {\bibfield  {journal} {\bibinfo  {journal} {Phys.
  Rev. D},\ }\textbf {\bibinfo {volume} {77}},\ \bibinfo {pages} {022001}
  (\bibinfo {year} {2008})},\ \Eprint {http://arxiv.org/abs/0708.3818}
  {arXiv:0708.3818 [gr-qc]} \BibitemShut {NoStop}%
\bibitem [{\citenamefont {Abbott}\ \emph
  {et~al.}(2009){\natexlab{b}}\citenamefont {Abbott} \emph
  {et~al.}}]{EatHS4R2}%
  \BibitemOpen
  \bibfield  {author} {\bibinfo {author} {\bibfnamefont {B.}~\bibnamefont
  {Abbott}} \emph {et~al.} (\bibinfo {collaboration} {LIGO Scientific
  Collaboration}),\ }\Doi {10.1103/PhysRevD.79.022001} {\bibfield  {journal}
  {\bibinfo  {journal} {Phys. Rev. D},\ }\textbf {\bibinfo {volume} {79}},\
  \bibinfo {pages} {022001} (\bibinfo {year} {2009}{\natexlab{b}})},\ \Eprint
  {http://arxiv.org/abs/0804.1747} {arXiv:0804.1747 [gr-qc]} \BibitemShut
  {NoStop}%
\bibitem [{\citenamefont {Abbott}\ \emph
  {et~al.}(2009){\natexlab{c}}\citenamefont {Abbott} \emph
  {et~al.}}]{EatHS5R1}%
  \BibitemOpen
  \bibfield  {author} {\bibinfo {author} {\bibfnamefont {B.}~\bibnamefont
  {Abbott}} \emph {et~al.} (\bibinfo {collaboration} {LIGO Scientific
  Collaboration}),\ }\Doi {10.1103/PhysRevD.80.042003} {\bibfield  {journal}
  {\bibinfo  {journal} {Phys. Rev. D},\ }\textbf {\bibinfo {volume} {80}},\
  \bibinfo {pages} {042003} (\bibinfo {year} {2009}{\natexlab{c}})},\ \Eprint
  {http://arxiv.org/abs/0905.1705} {arXiv:0905.1705 [gr-qc]} \BibitemShut
  {NoStop}%
\bibitem [{\citenamefont {Abbott}\ \emph
  {et~al.}(2009){\natexlab{d}}\citenamefont {Abbott} \emph
  {et~al.}}]{S5Powerflux2009}%
  \BibitemOpen
  \bibfield  {author} {\bibinfo {author} {\bibfnamefont {B.}~\bibnamefont
  {Abbott}} \emph {et~al.} (\bibinfo {collaboration} {LIGO Scientific
  Collaboration}),\ }\Doi {10.1103/PhysRevLett.102.111102} {\bibfield
  {journal} {\bibinfo  {journal} {Phys. Rev. Lett.},\ }\textbf {\bibinfo
  {volume} {102}},\ \bibinfo {pages} {111102} (\bibinfo {year}
  {2009}{\natexlab{d}})},\ \Eprint {http://arxiv.org/abs/0810.0283}
  {arXiv:0810.0283 [gr-qc]} \BibitemShut {NoStop}%
\bibitem [{\citenamefont {Abadie}\ \emph {et~al.}(2012)\citenamefont {Abadie}
  \emph {et~al.}}]{S5Powerflux2011}%
  \BibitemOpen
  \bibfield  {author} {\bibinfo {author} {\bibfnamefont {J.}~\bibnamefont
  {Abadie}} \emph {et~al.} (\bibinfo {collaboration} {LIGO Scientific
  Collaboration and Virgo Collaboration}),\ }\href@noop {} {\bibfield
  {journal} {\bibinfo  {journal} {Phys. Rev. D},\ }\textbf {\bibinfo {volume}
  {85}},\ \bibinfo {pages} {022001} (\bibinfo {year} {2012})},\ \Eprint
  {http://arxiv.org/abs/1110.0208} {arXiv:1110.0208 [gr-qc]} \BibitemShut
  {NoStop}%
\bibitem [{\citenamefont {Krishnan}(2005)}]{HoughP2}%
  \BibitemOpen
  \bibfield  {author} {\bibinfo {author} {\bibfnamefont {B.}~\bibnamefont
  {Krishnan}} (\bibinfo {collaboration} {LIGO Scientific Collaboration}),\
  }\Doi {10.1088/0264-9381/22/18/S40} {\bibfield  {journal} {\bibinfo
  {journal} {Class. Quant. Grav.},\ }\textbf {\bibinfo {volume} {22}},\
  \bibinfo {pages} {S1265} (\bibinfo {year} {2005})},\ \Eprint
  {http://arxiv.org/abs/gr-qc/0506109} {arXiv:gr-qc/0506109} \BibitemShut
  {NoStop}%
\bibitem [{Ein()}]{Einstweb}%
  \BibitemOpen
  \href@noop {} {}\bibinfo {note}
  {\url{http://einstein.phys.uwm.edu}.}\BibitemShut {Stop}%
\bibitem [{\citenamefont {Smoluchowski}(1970)}]{Smoluchowski}%
  \BibitemOpen
  \bibfield  {author} {\bibinfo {author} {\bibfnamefont {R.}~\bibnamefont
  {Smoluchowski}},\ }\href@noop {} {\bibfield  {journal} {\bibinfo  {journal}
  {Phys. Rev. Lett},\ }\textbf {\bibinfo {volume} {24}},\ \bibinfo {pages}
  {923} (\bibinfo {year} {1970})}\BibitemShut {NoStop}%
\bibitem [{\citenamefont {{Ruderman}}(1991)}]{Ruderman1991}%
  \BibitemOpen
  \bibfield  {author} {\bibinfo {author} {\bibfnamefont {M.}~\bibnamefont
  {{Ruderman}}},\ }\Doi {10.1086/170745} {\bibfield  {journal} {\bibinfo
  {journal} {\apj},\ }\textbf {\bibinfo {volume} {382}},\ \bibinfo {pages}
  {587} (\bibinfo {year} {1991})}\BibitemShut {NoStop}%
\bibitem [{\citenamefont {Horowitz}\ and\ \citenamefont
  {Kadau}(2009)}]{Horowitz}%
  \BibitemOpen
  \bibfield  {author} {\bibinfo {author} {\bibfnamefont {C.}~\bibnamefont
  {Horowitz}}\ and\ \bibinfo {author} {\bibfnamefont {K.}~\bibnamefont
  {Kadau}},\ }\Doi {10.1103/PhysRevLett.102.191102} {\bibfield  {journal}
  {\bibinfo  {journal} {Phys. Rev. Lett.},\ }\textbf {\bibinfo {volume}
  {102}},\ \bibinfo {pages} {191102} (\bibinfo {year} {2009})},\ \Eprint
  {http://arxiv.org/abs/0904.1986} {arXiv:0904.1986 [astro-ph.SR]} \BibitemShut
  {NoStop}%
\bibitem [{\citenamefont {Taylor}\ and\ \citenamefont
  {Weisberg}(1989)}]{Taylor:1989sw}%
  \BibitemOpen
  \bibfield  {author} {\bibinfo {author} {\bibfnamefont {J.~H.}\ \bibnamefont
  {Taylor}}\ and\ \bibinfo {author} {\bibfnamefont {J.}~\bibnamefont
  {Weisberg}},\ }\Doi {10.1086/167917} {\bibfield  {journal} {\bibinfo
  {journal} {Astrophys. J.},\ }\textbf {\bibinfo {volume} {345}},\ \bibinfo
  {pages} {434} (\bibinfo {year} {1989})}\BibitemShut {NoStop}%
\bibitem [{\citenamefont {Bonazzola}\ and\ \citenamefont
  {Gourgoulhon}(1996)}]{Bonazzola:1995rb}%
  \BibitemOpen
  \bibfield  {author} {\bibinfo {author} {\bibfnamefont {S.}~\bibnamefont
  {Bonazzola}}\ and\ \bibinfo {author} {\bibfnamefont {E.}~\bibnamefont
  {Gourgoulhon}},\ }\href@noop {} {\bibfield  {journal} {\bibinfo  {journal}
  {Astron. Astrophys.},\ }\textbf {\bibinfo {volume} {312}},\ \bibinfo {pages}
  {675} (\bibinfo {year} {1996})},\ \Eprint
  {http://arxiv.org/abs/astro-ph/9602107} {arXiv:astro-ph/9602107} \BibitemShut
  {NoStop}%
\bibitem [{\citenamefont {Balasubramanian}\ \emph {et~al.}(1996)\citenamefont
  {Balasubramanian}, \citenamefont {Sathyaprakash},\ and\ \citenamefont
  {Dhurandhar}}]{MetricBSD}%
  \BibitemOpen
  \bibfield  {author} {\bibinfo {author} {\bibfnamefont {R.}~\bibnamefont
  {Balasubramanian}}, \bibinfo {author} {\bibfnamefont {B.}~\bibnamefont
  {Sathyaprakash}}, \ and\ \bibinfo {author} {\bibfnamefont {S.}~\bibnamefont
  {Dhurandhar}},\ }\Doi {10.1103/PhysRevD.53.3033, 10.1103/PhysRevD.54.1860}
  {\bibfield  {journal} {\bibinfo  {journal} {Phys. Rev. D},\ }\textbf
  {\bibinfo {volume} {53}},\ \bibinfo {pages} {3033} (\bibinfo {year}
  {1996})},\ \Eprint {http://arxiv.org/abs/gr-qc/9508011} {arXiv:gr-qc/9508011}
  \BibitemShut {NoStop}%
\bibitem [{\citenamefont {Owen}(1996)}]{MetricBen}%
  \BibitemOpen
  \bibfield  {author} {\bibinfo {author} {\bibfnamefont {B.~J.}\ \bibnamefont
  {Owen}},\ }\Doi {10.1103/PhysRevD.53.6749} {\bibfield  {journal} {\bibinfo
  {journal} {Phys. Rev. D},\ }\textbf {\bibinfo {volume} {53}},\ \bibinfo
  {pages} {6749} (\bibinfo {year} {1996})},\ \Eprint
  {http://arxiv.org/abs/gr-qc/9511032} {arXiv:gr-qc/9511032} \BibitemShut
  {NoStop}%
\bibitem [{\citenamefont {Brady}\ \emph {et~al.}(1998)\citenamefont {Brady},
  \citenamefont {Creighton}, \citenamefont {Cutler},\ and\ \citenamefont
  {Schutz}}]{ForMetricExprRes}%
  \BibitemOpen
  \bibfield  {author} {\bibinfo {author} {\bibfnamefont {P.~R.}\ \bibnamefont
  {Brady}}, \bibinfo {author} {\bibfnamefont {T.}~\bibnamefont {Creighton}},
  \bibinfo {author} {\bibfnamefont {C.}~\bibnamefont {Cutler}}, \ and\ \bibinfo
  {author} {\bibfnamefont {B.~F.}\ \bibnamefont {Schutz}},\ }\Doi
  {10.1103/PhysRevD.57.2101} {\bibfield  {journal} {\bibinfo  {journal} {Phys.
  Rev. D},\ }\textbf {\bibinfo {volume} {57}},\ \bibinfo {pages} {2101}
  (\bibinfo {year} {1998})},\ \Eprint {http://arxiv.org/abs/gr-qc/9702050}
  {arXiv:gr-qc/9702050} \BibitemShut {NoStop}%
\bibitem [{\citenamefont {Palomba}\ \emph {et~al.}(2005)\citenamefont
  {Palomba}, \citenamefont {Astone},\ and\ \citenamefont
  {Frasca}}]{Palomba2005}%
  \BibitemOpen
  \bibfield  {author} {\bibinfo {author} {\bibfnamefont {C.}~\bibnamefont
  {Palomba}}, \bibinfo {author} {\bibfnamefont {P.}~\bibnamefont {Astone}}, \
  and\ \bibinfo {author} {\bibfnamefont {S.}~\bibnamefont {Frasca}},\ }\Doi
  {10.1088/0264-9381/22/18/S39} {\bibfield  {journal} {\bibinfo  {journal}
  {Class. Quant. Grav.},\ }\textbf {\bibinfo {volume} {22}},\ \bibinfo {pages}
  {S1255} (\bibinfo {year} {2005})}\BibitemShut {NoStop}%
\bibitem [{Boi()}]{Boinc1}%
  \BibitemOpen
  \href@noop {} {}\bibinfo {note}
  {\url{http://boinc.berkeley.edu/}.}\BibitemShut {Stop}%
\bibitem [{\citenamefont {{Anderson}}(2004)}]{Boinc2}%
  \BibitemOpen
  \bibfield  {author} {\bibinfo {author} {\bibfnamefont {D.~P.}\ \bibnamefont
  {{Anderson}}},\ }in\ \href@noop {} {\emph {\bibinfo {booktitle} {Proceedings
  of the Fifth IEEE/ACM International Workshop on Grid Computing (GRIDÕ04)}}},\
  \bibinfo {editor} {edited by\ \bibinfo {editor} {\bibnamefont {{IEEE Computer
  Society, Washington, DC (USA)}}}}\ (\bibinfo {year} {2004})\ pp.\ \bibinfo
  {pages} {4--10}\BibitemShut {NoStop}%
\bibitem [{\citenamefont {{Anderson}}\ \emph {et~al.}(2006)\citenamefont
  {{Anderson}}, \citenamefont {{Christensen}},\ and\ \citenamefont
  {{Allen}}}]{Boinc3}%
  \BibitemOpen
  \bibfield  {author} {\bibinfo {author} {\bibfnamefont {D.~P.}\ \bibnamefont
  {{Anderson}}}, \bibinfo {author} {\bibfnamefont {C.}~\bibnamefont
  {{Christensen}}}, \ and\ \bibinfo {author} {\bibfnamefont {B.}~\bibnamefont
  {{Allen}}},\ }in\ \href@noop {} {\emph {\bibinfo {booktitle} {Proceedings of
  the 2006 ACM/IEEE conference on Supercomputing}}},\ \bibinfo {editor} {edited
  by\ \bibinfo {editor} {\bibnamefont {{IEEE Computer Society, Tampa, Florida
  (USA)}}}}\ (\bibinfo {year} {2006})\ pp.\ \bibinfo {pages}
  {126--136}\BibitemShut {NoStop}%
\bibitem [{\citenamefont {Ostriker}\ and\ \citenamefont
  {Gunn}(1969)}]{Ostriker:1969if}%
  \BibitemOpen
  \bibfield  {author} {\bibinfo {author} {\bibfnamefont {J.}~\bibnamefont
  {Ostriker}}\ and\ \bibinfo {author} {\bibfnamefont {J.}~\bibnamefont
  {Gunn}},\ }\Doi {10.1086/150160} {\bibfield  {journal} {\bibinfo  {journal}
  {Astrophys.J.},\ }\textbf {\bibinfo {volume} {157}},\ \bibinfo {pages} {1395}
  (\bibinfo {year} {1969})}\BibitemShut {NoStop}%
\bibitem [{\citenamefont {Lyne}\ \emph {et~al.}(1996)\citenamefont {Lyne} \emph
  {et~al.}}]{VelaBrakInd}%
  \BibitemOpen
  \bibfield  {author} {\bibinfo {author} {\bibfnamefont {A.~G.}\ \bibnamefont
  {Lyne}} \emph {et~al.},\ }\href@noop {} {\bibfield  {journal} {\bibinfo
  {journal} {Nature},\ }\textbf {\bibinfo {volume} {381}},\ \bibinfo {pages}
  {497} (\bibinfo {year} {1996})}\BibitemShut {NoStop}%
\bibitem [{\citenamefont {Lyne}\ \emph {et~al.}(1993)\citenamefont {Lyne} \emph
  {et~al.}}]{CrabBrakInd}%
  \BibitemOpen
  \bibfield  {author} {\bibinfo {author} {\bibfnamefont {A.~G.}\ \bibnamefont
  {Lyne}} \emph {et~al.},\ }\href@noop {} {\bibfield  {journal} {\bibinfo
  {journal} {MNRAS},\ }\textbf {\bibinfo {volume} {265}},\ \bibinfo {pages}
  {1003} (\bibinfo {year} {1993})}\BibitemShut {NoStop}%
\bibitem [{\citenamefont {Aulbert}\ and\ \citenamefont
  {Fehrmann}(2008)}]{atlas}%
  \BibitemOpen
  \bibfield  {author} {\bibinfo {author} {\bibfnamefont {C.}~\bibnamefont
  {Aulbert}}\ and\ \bibinfo {author} {\bibfnamefont {H.}~\bibnamefont
  {Fehrmann}},\ }\href@noop {} {\bibfield  {journal} {\bibinfo  {journal}
  {T\"{a}tigkeitsbericht 2008, Max-Planck-Gesellschaft}} (\bibinfo {year}
  {2008})}\BibitemShut {NoStop}%
\bibitem [{\citenamefont {{P.~Leaci, M.~A.~Papa and
  H.~Pletsch}}(tion)}]{PaolaMAPpaper}%
  \BibitemOpen
  \bibfield  {author} {\bibinfo {author} {\bibnamefont {{P.~Leaci, M.~A.~Papa
  and H.~Pletsch}}},\ }\href@noop {} {\bibfield  {journal} {\bibinfo  {journal}
  {New methods to filter out spurious disturbances in continuous wave search
  analyses from gravitational wave detectors}} (\bibinfo {year} {in
  preparation})}\BibitemShut {NoStop}%
\bibitem [{\citenamefont {Audet}\ and\ \citenamefont {{Dennis,
  Jr.}}(2004)}]{Audet04meshadaptive}%
  \BibitemOpen
  \bibfield  {author} {\bibinfo {author} {\bibfnamefont {C.}~\bibnamefont
  {Audet}}\ and\ \bibinfo {author} {\bibfnamefont {J.~E.}\ \bibnamefont
  {{Dennis, Jr.}}},\ }\href@noop {} {\bibfield  {journal} {\bibinfo  {journal}
  {SIAM Journal on Optimization},\ }\textbf {\bibinfo {volume} {17}},\ \bibinfo
  {pages} {188} (\bibinfo {year} {2004})}\BibitemShut {NoStop}%
\bibitem [{\citenamefont {Le~Digabel}(2011)}]{LeDigabel2011}%
  \BibitemOpen
  \bibfield  {author} {\bibinfo {author} {\bibfnamefont {S.}~\bibnamefont
  {Le~Digabel}},\ }\Doi {10.1145/1916461.1916468} {\bibfield  {journal}
  {\bibinfo  {journal} {ACM Trans. Math. Softw.},\ }\textbf {\bibinfo {volume}
  {37}},\ \bibinfo {pages} {44:1} (\bibinfo {year} {2011})}\BibitemShut
  {NoStop}%
\bibitem [{\citenamefont {Shaltev}(ress)}]{MirosProc}%
  \BibitemOpen
  \bibfield  {author} {\bibinfo {author} {\bibfnamefont {M.}~\bibnamefont
  {Shaltev}},\ }\href@noop {} {\bibfield  {journal} {\bibinfo  {journal}
  {Journal of Physics: Conf. Series}} (\bibinfo {year}
  {http://arxiv.org/abs/1201.4656, in press})}\BibitemShut {NoStop}%
\bibitem [{\citenamefont {Abramson}\ \emph {et~al.}(2009)\citenamefont
  {Abramson}, \citenamefont {Audet}, \citenamefont {{Dennis, Jr.}},\ and\
  \citenamefont {{Le Digabel}}}]{AbramsonADD09}%
  \BibitemOpen
  \bibfield  {author} {\bibinfo {author} {\bibfnamefont {M.~A.}\ \bibnamefont
  {Abramson}}, \bibinfo {author} {\bibfnamefont {C.}~\bibnamefont {Audet}},
  \bibinfo {author} {\bibfnamefont {J.~E.}\ \bibnamefont {{Dennis, Jr.}}}, \
  and\ \bibinfo {author} {\bibfnamefont {S.}~\bibnamefont {{Le Digabel}}},\
  }\href
  {http://dblp.uni-trier.de/db/journals/siamjo/siamjo20.html#AbramsonADD09}
  {\bibfield  {journal} {\bibinfo  {journal} {SIAM Journal on Optimization},\
  }\textbf {\bibinfo {volume} {20}},\ \bibinfo {pages} {948} (\bibinfo {year}
  {2009})}\BibitemShut {NoStop}%
\bibitem [{\citenamefont {Abadie}\ \emph
  {et~al.}(2010){\natexlab{b}}\citenamefont {Abadie} \emph {et~al.}}]{Calib}%
  \BibitemOpen
  \bibfield  {author} {\bibinfo {author} {\bibfnamefont {J.}~\bibnamefont
  {Abadie}} \emph {et~al.} (\bibinfo {collaboration} {LIGO Scientific
  Collaboration}),\ }\Doi {10.1016/j.nima.2010.07.089} {\bibfield  {journal}
  {\bibinfo  {journal} {Nucl. Instrum. Meth.},\ }\textbf {\bibinfo {volume}
  {A624}},\ \bibinfo {pages} {223} (\bibinfo {year} {2010}{\natexlab{b}})},\
  \Eprint {http://arxiv.org/abs/1007.3973} {arXiv:1007.3973 [gr-qc]}
  \BibitemShut {NoStop}%
\bibitem [{\citenamefont {Wette}(2012)}]{Wette.2012}%
  \BibitemOpen
  \bibfield  {author} {\bibinfo {author} {\bibfnamefont {K.}~\bibnamefont
  {Wette}},\ }\Doi {10.1103/PhysRevD.85.042003} {\bibfield  {journal} {\bibinfo
   {journal} {Phys.\ Rev.\ D},\ }\textbf {\bibinfo {volume} {85}},\ \bibinfo
  {pages} {042003} (\bibinfo {year} {2012})},\ \Eprint
  {http://arxiv.org/abs/1111.5650} {arXiv:1111.5650 [gr-qc]} \BibitemShut
  {NoStop}%
\bibitem [{\citenamefont {Prix}\ and\ \citenamefont
  {Wette}(2012)}]{Prix.Wette.2012}%
  \BibitemOpen
  \bibfield  {author} {\bibinfo {author} {\bibfnamefont {R.}~\bibnamefont
  {Prix}}\ and\ \bibinfo {author} {\bibfnamefont {K.}~\bibnamefont {Wette}},\
  }\href@noop {} {\emph {\bibinfo {title} {{Estimating sensitivity of the
  Einstein@Home search S5R5}}}},\ \bibinfo {type} {Tech. Rep.}\ \bibinfo
  {number} {{\bf T1200272}}\ (\bibinfo  {institution} {LIGO},\ \bibinfo {year}
  {2012})\ \bibinfo {note}
  {\url{https://dcc.ligo.org/cgi-bin/DocDB/ShowDocument?docid=T1200272}}\BibitemShut
  {NoStop}%
\bibitem [{\citenamefont {{R.~Prix, D.~Keitel, M.~A.~Papa, P.~Leaci and
  M.~Siddiqi}}(tion)}]{PKPLS}%
  \BibitemOpen
  \bibfield  {author} {\bibinfo {author} {\bibnamefont {{R.~Prix, D.~Keitel,
  M.~A.~Papa, P.~Leaci and M.~Siddiqi}}},\ }\href@noop {} {\bibfield  {journal}
  {\bibinfo  {journal} {An F-statistic based multi-detector veto for detector
  artifacts}} (\bibinfo {year} {in preparation})}\BibitemShut {NoStop}%
\bibitem [{\citenamefont {Pletsch}(2008)}]{GlobCorr}%
  \BibitemOpen
  \bibfield  {author} {\bibinfo {author} {\bibfnamefont {H.~J.}\ \bibnamefont
  {Pletsch}},\ }\Doi {10.1103/PhysRevD.78.102005} {\bibfield  {journal}
  {\bibinfo  {journal} {Phys. Rev. D},\ }\textbf {\bibinfo {volume} {78}},\
  \bibinfo {pages} {102005} (\bibinfo {year} {2008})},\ \Eprint
  {http://arxiv.org/abs/0807.1324} {arXiv:0807.1324 [gr-qc]} \BibitemShut
  {NoStop}%
\bibitem [{\citenamefont {Pletsch}\ and\ \citenamefont
  {Allen}(2009)}]{Pletsch:2009uu}%
  \BibitemOpen
  \bibfield  {author} {\bibinfo {author} {\bibfnamefont {H.~J.}\ \bibnamefont
  {Pletsch}}\ and\ \bibinfo {author} {\bibfnamefont {B.}~\bibnamefont
  {Allen}},\ }\Doi {10.1103/PhysRevLett.103.181102} {\bibfield  {journal}
  {\bibinfo  {journal} {Phys. Rev. Lett.},\ }\textbf {\bibinfo {volume}
  {103}},\ \bibinfo {pages} {181102} (\bibinfo {year} {2009})},\ \Eprint
  {http://arxiv.org/abs/0906.0023} {arXiv:0906.0023 [gr-qc]} \BibitemShut
  {NoStop}%
\bibitem [{\citenamefont {{The LIGO Scientific Collaboration and the Virgo
  Collaboration}}(2012)}]{S5R3HIs}%
  \BibitemOpen
  \bibfield  {author} {\bibinfo {author} {\bibnamefont {{The LIGO Scientific
  Collaboration and the Virgo Collaboration}}},\ }\href@noop {} {\emph
  {\bibinfo {title} {Recovery of hardware injections in the S5R3 Einstein@Home
  run for continuous wave searches}}},\ \bibinfo {type} {Tech. Rep.}\ \bibinfo
  {number} {{\bf T1200278}}\ (\bibinfo  {institution} {LSC and Virgo},\
  \bibinfo {year} {2012})\ \bibinfo {note}
  {\url{https://dcc.ligo.org/cgi-bin/DocDB/ShowDocument?docid=92307}}\BibitemShut
  {NoStop}%
\end{thebibliography}%

\end{document}